\pdfoutput=1
\documentclass[12pt]{article} 
\usepackage{amsfonts,amssymb,amsmath,amsthm,dsfont,graphicx,hyperref,mathrsfs,euscript,enumitem,framed, bm, mathtools, color}
\usepackage[title, titletoc]{appendix}
\usepackage[top=1in,bottom=1in,left=1in,right=1in]{geometry} 
\usepackage[onehalfspacing]{setspace}
\usepackage{ifthen,caption,subcaption,lscape}
\hypersetup{pdfborder = {0 0 0},colorlinks=true,linkcolor=blue,urlcolor=blue,citecolor=blue,filecolor=blue}
\usepackage{chngcntr} 
\usepackage{natbib}
\usepackage{longtable}
\usepackage{placeins}
\usepackage{threeparttable}
\newtheorem{thm}{Theorem}
\newtheorem{coro}{Corollary}
\newtheorem{lem}{Lemma}
\newtheorem{assumption}{Assumption}

\theoremstyle{definition}

\theoremstyle{definition}
\newtheorem{remark_tmp}{Remark}[section]
\newenvironment{remark}
{ \begin{remark_tmp} 	}
	{ 
		\medskip\hfill{\LARGE$\lrcorner$}
	\end{remark_tmp} 
}

\newtheorem{exmp}{Example}[section]

\allowdisplaybreaks[1]

\DeclareMathOperator*{\argmin}{arg\,min}
\DeclareMathOperator*{\argmax}{arg\,max}

\DeclareMathOperator*{\tr}{Tr}

\DeclareMathOperator*{\diag}{diag}

\newcommand\inde{\protect\mathpalette{\protect\independenT}{\perp}}
\def\independenT#1#2{\mathrel{\rlap{$#1#2$}\mkern2mu{#1#2}}}

\renewcommand{\P}{\mathbb{P}}
\newcommand{\E}{\mathbb{E}}
\newcommand{\V}{\mathbb{V}}
\newcommand{\Cov}{\mathbb{C}\mathrm{ov}}
\newcommand{\I}{\mathds{1}}

\newcommand{\bA}{\bm{A}}
\newcommand{\bH}{\bm{H}}

\newcommand{\bB}{\bm{B}}

\newcommand{\bF}{\bm{F}}
\newcommand{\bG}{\bm{G}}

\newcommand{\bI}{\bm{I}}

\newcommand{\bP}{\bm{P}}
\newcommand{\bQ}{\bm{Q}}
\newcommand{\bR}{\bm{R}}

\newcommand{\bU}{\bm{U}}

\newcommand{\bV}{\bm{V}}
\newcommand{\bW}{\bm{W}}
\newcommand{\bX}{\bm{X}}

\newcommand{\bZ}{\bm{Z}}

\newcommand{\ba}{\bm{a}}
\newcommand{\bb}{\bm{b}}

\newcommand{\bg}{\bm{g}}
\newcommand{\be}{\bm{e}}

\newcommand{\bq}{\bm{q}}

\newcommand{\bt}{\bm{t}}
\newcommand{\bu}{\bm{u}}
\newcommand{\bv}{\bm{v}}
\newcommand{\bx}{\bm{x}}

\newcommand{\bw}{\bm{w}}
\newcommand{\bz}{\bm{z}}

\newcommand{\balpha}{\boldsymbol{\alpha}}
\newcommand{\bepsilon}{\boldsymbol{\epsilon}}

\newcommand{\bbeta}{\boldsymbol{\beta}}

\newcommand{\bkappa}{\boldsymbol{\kappa}}

\newcommand{\bgamma}{\boldsymbol{\gamma}}

\newcommand{\biota}{\boldsymbol{\iota}}

\newcommand{\bvth}{\boldsymbol{\vartheta}}

\newcommand{\bpsi}{\boldsymbol{\psi}}
\newcommand{\bSigma}{\boldsymbol{\Sigma}}
\newcommand{\bUpsilon}{\boldsymbol{\Upsilon}}

\newcommand{\blambda}{\boldsymbol{\lambda}}
\newcommand{\bLambda}{\boldsymbol{\Lambda}}
\newcommand{\bmeta}{\boldsymbol{\eta}}
\newcommand{\bvrho}{\boldsymbol{\varrho}}

\newcommand{\bGamma}{\boldsymbol{\Gamma}}
\newcommand{\bXi}{\boldsymbol{\Xi}}
\newcommand{\bpi}{\boldsymbol{\pi}}
\newcommand{\bomega}{\boldsymbol{\omega}}

\newcommand{\ttr}{r}
\newcommand{\ttd}{\mathsf{d}}
\newcommand{\tts}{\mathsf{p}}
\newcommand{\tty}{\mathsf{y}}
\newcommand{\lvarsig}{\underline{\varsigma}}
\newcommand{\uvarsig}{\bar{\varsigma}}
\newcommand{\bfrb}{\bm{\mathfrak{b}}}
\newcommand{\bfrc}{\bm{\mathfrak{c}}}
\newcommand{\barrate}[1]{\bar{\mathfrak{r}}_{{#1},\subi}}

\newcommand{\subi}{\scriptscriptstyle\mathcal{N}_i}

\newcommand{\rate}[1]{\mathfrak{r}_{{#1},\subi}}
\newcommand{\biaserr}{\mathfrak{m}_p}
\newcommand{\barbiaserr}{\bar{\mathfrak{m}}_p}
\newcommand{\sam}{\delta_{Kp}}

\setenumerate[1]{label=(\alph*)}

\renewcommand{\thesection}{\arabic{section}}
\renewcommand{\theequation}{\arabic{equation}}
\renewcommand{\thefigure}{\arabic{figure}}
\renewcommand{\thetable}{\arabic{table}}
\renewcommand{\thelem}{\arabic{lemma}}
\renewcommand{\thecoro}{\arabic{coro}}
\renewcommand{\thethm}{\arabic{theorem}}
\renewcommand{\theassumption}{\arabic{assumption}}
\renewcommand{\thedefinition}{\arabic{definition}}

\numberwithin{equation}{section}
\numberwithin{thm}{section}
\numberwithin{lem}{section}
\numberwithin{coro}{thm}

\newcommand{\forloop}[5][1]{\setcounter{#2}{#3}\ifthenelse{#4}{#5\addtocounter{#2}{#1}\forloop[#1]{#2}{\value{#2}}{#4}{#5}}{}}

\usepackage{tocloft}
\makeatletter
\let\arxivmaketitle\maketitle
\let\arxivinternalmaketitle\@maketitle
\let\arxivtitle\title
\let\arxivauthor\author
\let\arxivdate\date
\let\arxivthanks\thanks
\let\arxivand\and
\makeatother

\begin{document}
\hypersetup{pageanchor=false}

\title{\vspace{-0.25in} \Large Causal Inference in Possibly Nonlinear Factor Models
	\thanks{We thank the co-Editor, Peter Hull, and four reviewers for their valuable comments.
	We also thank Sebastian Calonico, Matias Cattaneo, Richard Crump, Jianqing Fan, Max Farrell, Andreas Hagemann, 
	Michael Jansson, Lutz Kilian, 
	Xinwei Ma, Kenichi Nagasawa, Roc\'{\i}o Titiunik, Gonzalo Vazquez-Bare, Jingshen Wang, and
	participants at various seminars, workshops, and conferences for helpful comments and discussions.
	Feng gratefully acknowledges financial support from the National Natural Science
	Foundation of China (NSFC) through grants 72203122 and 72522001.}
\bigskip }
\author{
	Yingjie Feng\thanks{School of Economics and Management, Tsinghua University, Beijing 100084, China. 
	}}
\maketitle

\vspace{.5em}

\begin{abstract}
This paper develops a causal inference method for treatment effects models with noisily measured confounders.
The key feature is that a large number of noisy proxies are available and linked with the underlying latent confounders through an unknown, possibly nonlinear factor structure. 
The main building block is a local principal subspace approximation procedure that combines $K$-nearest-neighbor matching and  principal component analysis. 
Estimators of many causal parameters, including average treatment effects and counterfactual distributions, are constructed based on doubly-robust score functions,
	and their large-sample properties are established.
These results require the collection of proxies to be jointly informative about the latent confounders relevant to the outcome and treatment, while allowing some proxies to be uninformative, their identities to be unknown, and measurement errors to be correlated with the outcome or treatment.
We also obtain uniformly consistent estimators of the conditional average treatment effect at each unit's confounder values.
The results are illustrated with an empirical application studying the effect of political connections on stock returns of financial firms and a Monte Carlo experiment.
\end{abstract}

\medskip

\thispagestyle{empty}

\clearpage

\doublespacing
\setcounter{page}{1}
\hypersetup{pageanchor=true}

\pagestyle{plain}

\section{Introduction} \label{sec: introduction}

Understanding effects of policy interventions is central in many disciplines.
When observational data are used, researchers usually confront the challenge that the treatment is nonrandomly assigned based on some characteristics that are \textit{not} directly observed, making it difficult to uncover the true causal relation between the outcome and the treatment.
Nowadays, with the advent of big data, a researcher may have access to a large number of \textit{noisy} proxies for these latent confounders, possibly collected over many time periods (large panel data), across many products or product categories (recommendation systems), or from interactions with many other units (large network data).
For instance, in finance applications one may be concerned about the confounding effect of a firm's unobserved entrepreneurial ability or management quality, which may be imperfectly captured by a firm's historical performance such as stock returns \citep{holmstrom1993market,pan2015learning}. 
It is tempting to ask in this context whether and how high-dimensional proxies can mitigate confounding in causal inference.

\begin{figure}[!htbp]\centering
	\caption{Sampling Distributions of ATT $t$-Statistics}
	\label{figure:selected proxies}
	\begin{threeparttable}
	\includegraphics[width=\textwidth]{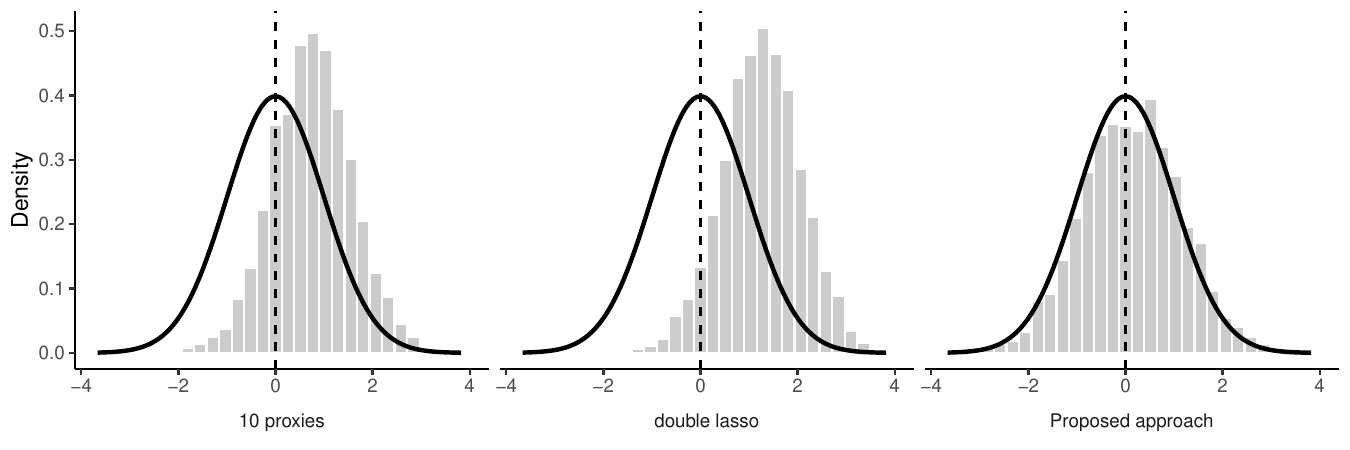}
	
	  \begin{tablenotes}[para,flushleft]
		\footnotesize
		\textit{Notes:} From left to right, the panels show sampling distributions of ATT \(t\)-statistics for estimators using the last 10 proxies, double Lasso, and the proposed method. Outcomes and treatments are generated as in Section \ref{sec: simulation}. The design includes 250 noisy proxies $x_{il}=\alpha_i+u_{il}$, with $\alpha_i$ a scalar latent confounder and $u_{il}$ a standard normal error. Dashed vertical lines mark zero; solid curves are standard normal densities.
	\end{tablenotes}
	\end{threeparttable}
\end{figure}

Unfortunately, due to the presence of measurement error, usual selection-on-observables approaches that naively control for a number of selected noisy proxies are generally invalid. This remains true even if modern machine learning techniques such as post double Lasso \citep{Belloni-Chernozhukov-Hansen_2014_RES} are employed for control variable selection.
Figure \ref{figure:selected proxies} shows a simple Monte Carlo experiment with 250 noisy proxies. The left panel shows the sampling distribution of the
$t$-statistic for the conventional doubly robust estimator of the average treatment effect on the treated (ATT) when we control for the last 10 noisy proxies, and 
the center panel shows the corresponding distribution for the ATT estimator using double Lasso for variable selection; both deviate substantially from the standard normal. 
This occurs because, in this context, sparsity-based techniques select a relatively small number of noisy proxies, leading to a non-vanishing bias caused by measurement error. A more detailed explanation is available in Section \ref{subsec: structure of LD measurements} below.

Moreover, the unclear, possibly nonlinear relationship between noisy proxies and latent confounders makes this problem even more challenging. 
Traditional solutions to measurement error in the literature often rely on strong functional form assumptions such as linearity \citep[see, e.g.,][for a review]{Schennach_2016_ARE}. However, in many applications confounders such as ability or quality are unobserved and often loosely defined; it is difficult or even impossible to specify how each proxy (e.g., the stock return in each period) relates to the latent confounders. In addition, when proxies are discrete, censored, truncated, or otherwise observed with limitations, nonlinear models arise naturally; one example is the graphon model, which characterizes link formation in network studies \citep{Gao-Lu-Zhou_2015_AoS}.

This paper proposes a novel approach for treatment effects analysis that addresses these issues in settings with high-dimensional proxies. The right panel of Figure \ref{figure:selected proxies} illustrates its performance in the same Monte Carlo experiment: the sampling distribution of its $t$-statistic is much closer to the standard normal. Our key assumption is that the noisy proxies are linked with the latent confounders via an unknown, possibly nonlinear factor model. 
The proposed method builds on a local principal component analysis (PCA) method developed in \cite{Feng_2023_wp_Nonlinear}. We first find the $K$-nearest neighbors ($K$-NN) for each unit based on observed proxies (with $K$ diverging as the sample size grows).
If different values of latent confounders can induce non-negligible differences in many (but possibly not all) observed proxies,
then the $K$ nearest neighbors, appropriately measured by the noisy proxies, are also close in terms of the latent confounders.
Next, within each local neighborhood formed by the $K$ matches, the underlying possibly nonlinear factor structure can be approximated by a local linear factor structure, which can be estimated via PCA.

As in linear factor models \citep{Bai_2003_ECMA}, the values of the latent confounders  cannot be exactly recovered without additional normalizations.
Nevertheless,  the nearest neighbors and estimated local factor loadings from local PCA provide enough information to restore unconfoundedness in  our treatment effects analysis. Intuitively, the nearest neighbors tell us, for instance, which control units are close to a particular treated unit in terms of the latent confounders, and the local factor loadings further characterize their distance. 
Such information suffices for usual causal analysis where \textit{predictions} based on latent confounders are needed.
Importantly, this strategy requires the proxies to be jointly informative about every latent confounder and the resulting local factor loadings to provide sufficient approximation power for causal analysis. See Remarks \ref{remark: informativeness} and \ref{remark: het error} below for detailed discussion.

Using the nearest neighbors and estimated local factor loadings, we propose a general local quasi-maximum likelihood method to estimate the conditional means of potential outcomes and the conditional treatment probabilities (generalized propensity scores), which form the basis of regression imputation and propensity score weighting estimators of treatment effects. 
The local region used in such estimation is defined by nearest neighbors, and the extracted local factor loadings play the role of generated regressors that provide further approximation to unknown functions of the latent confounders. The number of nearest neighbors, the main tuning parameter of our procedure, implicitly governs the ``bandwidth'' of the estimation
and determines the consistency of final estimators, whereas the number of local factor loadings extracted is analogous to the degree of the basis in local polynomial regression and can be taken as a small fixed number.

In contrast with standard nonparametric regression, the conditioning variables in this context are \textit{indirectly} obtained from proxies, and the noise in their factor structure restricts one's ability to select a bandwidth. Using a small number of nearest neighbors does \textit{not} necessarily lead to a small bandwidth and thus is not helpful for further bias reduction, 
which differs from other matching techniques based on a fixed number of matches (e.g., \citealp{Abadie-Imbens_2006_ECMA}). Consequently, the possibly large smoothing bias of the nonparametric ingredients may render the final inference on causal parameters invalid. To deal with this issue, we follow the Neyman-orthogonalization strategy that has been extensively applied in the recent double/debiased machine learning literature \citep{Chernozhukov-et-al_2018_EJ,Chernozhukov-et-al_2020_wp}. In treatment effects models, the widely used doubly-robust scores   \citep{Robins-Rotnitzky_1995_JASA,Cattaneo_2010_JoE} are estimating equations constructed based on the efficient influence function and are  automatically Neyman orthogonal. Taking advantage of this property, 
we can conduct valid inference under rate restrictions on the intermediate nonparametric estimators comparable to those commonly imposed in this literature. On the other hand, we do \textit{not} require sample splitting across units that separates nuisance function estimation from final inference, which is similar to   \cite{Cattaneo_2010_JoE,Farrell_2015_JoE,farrell2021deep}, but differs from the common practice in double machine learning.
This is made possible by carefully investigating the specific structure and the asymptotic expansion of our nonparametric ingredients 
(nearest-neighbor-based local smoothing estimators).

Leveraging these ideas, we develop a novel estimation and inference procedure for a large class of estimands, including counterfactual distributions and functionals thereof, and provide the basis for analyzing many causal quantities of interest such as average, quantile, and distributional treatment effects. 
In addition to the robustness with respect to the relationship between noisy proxies and latent confounders discussed above, our method has other appealing features in theory and practice. 
First, as a dimension reduction technique, the proposed method allows users to obtain low-dimensional  information on latent confounders  from high-dimensional noisy proxies. It only requires some but \textit{not all} proxies to be informative about latent confounders, and it is \textit{unnecessary} to know their identities a priori (see Remark \ref{remark: informativeness} below).  
Second, our theory permits a generic choice of the ``distance'' to define nearest neighbors, thus encompassing and extending previous matching strategies in the panel and network data literature
\citep[e.g.,][]{Zhang-Levina-Zhu_2017_BIMA, Bonhomme-Lamadon-Manresa_2021_wp}. 
Our theoretical and numerical results also show that the proposed method provides a more flexible approximation to the nonlinear factor structure and 
delivers more robust inference results, compared to local constant smoothing  based solely on matching techniques.
Third, thanks to the availability of high-dimensional data, we can obtain a uniformly consistent estimator of the conditional average treatment effect (CATE) at each unit's confounder values, and also relax some restrictive assumptions on the independence between measurement error and outcomes (or treatment) that are common in the prior literature.
Finally, the output of local PCA can be readily taken as input to many classical econometric estimation procedures such as local polynomial kernel regression \citep{Fan-Gijbels_1996_book}, which may be useful in other contexts such as diffusion index forecasts \citep{Stock-Watson_2002_JASA, Bai-Ng_2006_ECMA}, where prediction based on noisily measured variables is of interest.

 The paper is organized as follows. The rest of this section reviews the related literature. Section \ref{sec: basic model} introduces a basic binary treatment effects model and the nonlinear factor structure of high-dimensional proxies for latent confounders. Section \ref{sec: estimation} describes the estimation procedure and provides a step-by-step empirical illustration using data from \cite{Acemoglu-et-al_2016_JFE}. Section \ref{sec: main results} presents the main theoretical results and some Monte Carlo evidence. Section \ref{sec: extensions} extends the theory to uniform inference on counterfactual distributions. Section \ref{sec: conclusion} concludes. The Supplemental Appendix (SA) contains further discussion of key assumptions in the paper, several extensions of our theory, and all proofs and additional numerical and technical results. Replication files for the simulation study and empirical illustration are available at \url{https://github.com/yingjieum/replication-Feng_2026}.

\subsection{Related Literature}

This paper contributes to the large literature on causal inference and program evaluation (see \citealp{Abadie-Cattaneo_2018_ARE} for a review). 
Our theoretical framework is closely related to nonlinear models with measurement error. Much effort has been devoted to the identification of such models (\citealp{Schennach_2016_ARE}). For example, factor models can be utilized to construct repeated measurements of unobserved variables, which allow for identification of their distribution under suitable normalizations. A  general treatment following this strategy is available in \cite{Cunha-Heckman-Schennach_2010_ECMA}, using and extending results in \cite{Hu-Schennach_2008_ECMA}. 
A related strand of work uses a fixed number of noisy proxies to construct control functions \citep{Altonji-Mansfield_2018_AER, Nagasawa_2021_wp} or confounding bridge functions \citep{Deaner_2023_wp_ProxyControls,miao2018identifying,singh2020kernel}, enabling identification of treatment effects provided that the proxies are sufficiently informative and certain conditional independence restrictions hold.
By comparison, this paper considers the alternative setting with many noisy proxies and illustrates the gains from  high-dimensional data that are  typically unavailable in prior studies. In particular, identification of causal effects of interest does not require recovering the exact values (or distributions) of latent confounders, and a mixture of informative and uninformative proxies with unknown identities is permitted. Our extended theory in the SA further allows potential correlation between measurement error and the outcome or treatment. We also obtain a uniformly consistent estimator of the conditional average treatment effect at each unit's confounder values. See Section \ref{SA section: comparison of identifying assumptions} of the SA for a more detailed comparison of the identifying assumptions across these approaches.

The idea of local PCA, the technical building block of the inference method in this paper, has precedents in the machine learning literature \citep[e.g.,][]{Zhang-Zha_2004_SIAM, Arias-Lerman-Zhang_2017_JMLR}. These studies focus on the reconstruction of a global latent nonlinear structure underlying the data for the purpose of classification, clustering or data visualization, but do not provide formal estimation and inference theory.
Recently, building on and extending results on large-dimensional factor models \citep{Bai-Wang_2016_ARE}, 
\cite{Feng_2023_wp_Nonlinear} has established the estimation theory for local PCA based on a general neighborhood formulation in the nonlinear factor model, and applied it to optimal matrix estimation problems. Unlike these studies, in this paper the nonlinear factor structure itself is not the object of interest; rather, we use proxy-based neighborhoods and estimated local principal subspaces to estimate the conditional means of potential outcomes and propensity scores, which form the basis for inference on a large class of causal parameters. We also address a challenge due to heterogeneous approximation errors, which is specific to this causal inference problem (see Remark \ref{remark: het error} below).
The latent heterogeneity may also be addressed via other machine learning techniques. For instance, \cite{Bonhomme-Lamadon-Manresa_2021_wp} develops two-step grouped fixed-effects estimators that discretize latent heterogeneity by $K$-means clustering, which relies on 
a specific injective moment condition to ensure informativeness of proxies. 
By contrast, this paper relies on a more general, weaker informativeness requirement (see Assumption \ref{Assumption: Matching} and Remark \ref{remark: informativeness} for details) and
can achieve more flexible approximation of smooth functions of latent features via local PCA. 
The intermediate result (Theorem \ref{thm: QMLE}) also characterizes the uniform convergence of nonparametric estimators of individual-specific features and may be of independent interest.

\section{Basic Framework}
\label{sec: basic model}

In this section we consider a binary treatment effects model with latent confounders that suffices to communicate the key ideas of the proposed method. A more general framework with several extensions, including multi-valued treatment effects, general likelihood functions, and covariate-adjusted nonlinear factor models, is presented in the SA.

Suppose that a random sample  $\{(y_{i}, t_{i}, \bx_{i}, \bz_i)\}_{i=1}^n$ is observed, where $y_{i}\in\mathbb{R}$ is the outcome of interest, $t_i\in\{0,1\}$ is a binary treatment, 
and  $\bx_{i}\in\mathbb{R}^p$ and $\bz_i\in\mathbb{R}^{\ttr_z}$ are vectors of covariates. 
$\bx_i$ and $\bz_i$ play different roles: $\bx_i$ is a vector of noisy proxies for a vector of \textit{unobserved} confounders $\balpha_i\in\mathbb{R}^{\ttr_\alpha}$, whereas $\bz_i$ itself is a vector of	\textit{observed} confounders that can be controlled for directly. 
Some covariates may serve both purposes and therefore be included in both $\bx_i$ and $\bz_i$. 
The asymptotic theory in this paper is developed assuming the sample size $n$ and the number of noisy proxies $p$ simultaneously increase to infinity, whereas the total dimension $\ttr_\alpha+\ttr_z$ of the confounders is fixed. 

We follow the standard potential outcomes framework. Let $y_i(1)$ and $y_i(0)$ denote unit $i$'s potential outcomes under treatment and no treatment, respectively. The observed outcome can then be written as 
$y_{i}=t_{i}y_{i}(1)+(1-t_i)y_i(0)$. For the moment, we focus on the analysis of the ATT parameter $\E[y_i(1)-y_i(0)|t_i=1]$, and defer extensions to other causal parameters, including quantile and distributional treatment effects, to Section \ref{sec: extensions}. 

For the treated group, the treated  outcome $y_i(1)$ is observed while the untreated one $y_i(0)$ is not. Thus, we focus on the key challenge of identifying $\E[y_i(0)|t_i=1]$. This missing data problem is common in causal inference and is often overcome by imposing an unconfoundedness condition---the treatment assignment becomes independent of potential outcomes after conditioning on a set of covariates. In this paper we assume that
\begin{equation}\label{eq: unconfoundedness}
(y_i(1), y_i(0))\inde t_i\,|\,\balpha_i, \bz_i.
\end{equation}
Thus, $\balpha_i$ and $\bz_i$ play the role of confounders, but recall that $\balpha_i$ is \textit{not} directly observed in our setup, making \eqref{eq: unconfoundedness} different from the standard selection on observables assumption.

We assume a semi-linear regression model to characterize the untreated outcome:
\begin{equation}\label{eq: outcome}
	y_{i}(0)=\mu(\balpha_i)+\bz_i'\bbeta+\epsilon_{i}, \qquad \E[\epsilon_{i}|\balpha_i, \bz_i]=0,
\end{equation}
where $\mu(\cdot)$ is some unknown function whose functional form is left unspecified. 
\eqref{eq: outcome} implies that $\zeta_i:=\E[y_i(0)|\balpha_i,\bz_i]=\mu(\balpha_i)+\bz_i'\bbeta$. 
On the other hand, we assume a logistic semi-linear model for the treatment:
\begin{equation}\label{eq: treatment}
	t_{i}=\psi(e(\balpha_i)+\bz_i'\bgamma)+v_{i},\qquad 
	\E[v_{i}|\balpha_i, \bz_i]=0, 
\end{equation}
where $\psi(a)=1/(1+\exp(-a))$ is the standard logistic function and $e(\cdot)$ is an unknown function left unspecified. \eqref{eq: treatment} implies that $\varrho_i:=\P(t_i=1|\balpha_i,\bz_i)=\psi(e(\balpha_i)+\bz_i'\bgamma)$, which would be the usual propensity score if $\balpha_i$ were observed. 
Note that $\balpha_i$ and $\bz_i$ enter both equations, reflecting their roles as confounders in \eqref{eq: unconfoundedness}.

Equations \eqref{eq: outcome} and \eqref{eq: treatment} can be viewed as special cases of generalized semi-linear models with an identity link and a logistic link, respectively. Such choices of link functions are common in practice when the outcome is continuous and the treatment is binary, and they also match the setting of our empirical application. Extensions to other links are allowed in the general theory presented in the SA. 
Notice that despite the specific links used, our model remains nonparametric with respect to the dependence of the outcome and treatment on latent $\balpha_i$, since $\mu(\cdot)$ and $e(\cdot)$ are unspecified. This is appealing because $\balpha_i$ is unobserved (and sometimes difficult to precisely define); allowing it to enter the model nonparametrically offers greater flexibility and robustness in confounding adjustment. On the other hand, incorporating observed $\bz_i$ (e.g., various types of fixed effects) linearly is a usual strategy in practice to maintain tractability of the full model.

\subsection{Structure of High-Dimensional Proxies} \label{subsec: structure of LD measurements}

Since $\balpha_i$ is  not observed, regression techniques cannot be applied to estimate $\E[y_i(0)|\balpha_i, \bz_i]$ or $\P(t_i=1|\balpha_i,\bz_i)$ directly, making many classical methods based on  \eqref{eq: unconfoundedness}, such as regression adjustment and inverse probability weighting, infeasible. However, we have a vector of noisy proxies $\bx_i=(x_{i1},\cdots, x_{ip})'$ linked to the unobserved $\balpha_i$ through a nonlinear factor model:
\begin{equation}\label{eq: HD covariate}
	\bx_{i}=\bmeta(\balpha_i)+\bu_{i}, \quad \E[\bu_{i}|\{\balpha_j\}_{j\in[n]}]=0,
\end{equation}
where $\bmeta(\cdot)=(\eta_1(\cdot), \cdots, \eta_p(\cdot))':
\mathbb{R}^{\ttr_\alpha}\to\mathbb{R}^p$ is a vector of latent functions possibly heterogeneous across proxies, and $\bu_{i}=(u_{i1}, \cdots, u_{ip})'$ is a vector of idiosyncratic errors. 
Define $n\times p$ matrices $\bX=(\bx_1,\cdots, \bx_n)'$, $\bH=(\bmeta(\balpha_1),\cdots, \bmeta(\balpha_n))'$ and $\bU=(\bu_1,\cdots, \bu_n)'$. Then, \eqref{eq: HD covariate} can be written in matrix form $\bX=\bH+\bU$. Compared to linear factor models, the latent mean matrix $\bH$ could have full rank due to the potential nonlinearity of $\bmeta$.

Conceptually, the latent functions $\bmeta$ can be either fixed or random, depending on the specific context. For instance, when $\bx_i$ contains proxies such as test scores for one's latent ability, it may be appropriate to assume each $\eta_l$ is a fixed ``production function'' that maps one's ability to a specific output; when $p$ proxies are collected over time, $\{\eta_l(\cdot)\}_{l=1}^p$ may be viewed as a functional time series with each $\eta_l(\cdot)$ corresponding to a random function at time $l$ \citep[see, e.g.,][for more details about functional data]{Ferraty-Vieu_2006_book}. In the literature on panel and network data \citep[e.g.,][]{Bai_2003_ECMA, Gao-Lu-Zhou_2015_AoS}, it is also common to assume a more specific setup: $\eta_l(\balpha_i)=\bm{\varpi}_l'\balpha_i$ (interactive fixed-effects model) or $\eta_l(\balpha_i)=\eta(\balpha_i, \bm{\varpi}_l)$ (graphon model), for some fixed bivariate function $\eta(\cdot,\cdot)$ and latent variables $\bm{\varpi}_l$. The heterogeneity (and potential randomness) of $\eta_l$ across $l$ is then driven by $\bm{\varpi}_l$. For ease of presentation, we treat $\bmeta(\cdot)$ as fixed (conditioned upon) in this paper and remain agnostic about its specific structure and the potential dependence over $l$.

The high-dimensional $\bx_i$ contains information about $\balpha_i$ if $\bmeta(\cdot)$ is nondegenerate in a proper sense, as formalized later. Conceptually, we may find some low-dimensional statistic $\mathcal{T}_i(\bX)$, as a function of observables, such that  $\E[y_i(0)|\balpha_i, \bz_i]$ and $\P[t_i=1|\balpha_i, \bz_i]$, the building blocks of many treatment effects estimators, are approximately the same as $\E[y_i(0)|\mathcal{T}_i(\bX), \bz_i]$ and $\P[t_i=1|\mathcal{T}_i(\bX), \bz_i]$, when $p$ is large. In other words, the challenge of selection on the unobserved $\balpha_i$ in this case is resolved by selection on a proper transformation of observables.

\begin{remark}[Conditioning on $\bx_i$]\label{remark: condition on x}
	When many noisy proxies  are available, it is tempting to ask whether $\E[y_i(0)|\bx_i]$ is (approximately) the same as $\E[y_i(0)|\balpha_i]$, justifying the use of estimation methods that directly control for $\bx_i$. Unfortunately, this alternative approach suffers from several issues and is usually invalid. To gain some intuition, consider a Gaussian error model:
	\begin{equation*}
		y_i(0)=\alpha_i+\epsilon_i,\quad x_{il}=\alpha_i+u_{il},\;\;l=1, \cdots, p,
	\end{equation*}
	where $(\alpha_i, \epsilon_i, \bu_i')'\in\mathbb{R}^{2+p}$ is multivariate normal with mean zero and unit marginal variances, and its components are mutually uncorrelated except that $\sigma_{p\epsilon}:=\Cov(u_{ip},\epsilon_{i})$ is possibly nonzero. In this special case, it can be shown that
	\begin{align*}
		\E[y_i(0)|\bx_i]
		=\frac{p+\sigma_{p\epsilon}}{p+1}\E[y_i(0)|\alpha_i]+\frac{(1-\sigma_{p\epsilon})}{p+1}\sum_{l=1}^pu_{il}+\sigma_{p\epsilon}u_{ip}.
	\end{align*}
	This calculation has several implications. First, the two estimands $\E[y_i(0)|\bx_i]$ and $\E[y_i(0)|\alpha_i]$ generally differ for finite $p$, and the last term $\sigma_{p\epsilon}u_{ip}$ does \textit{not} vanish as $p$ grows. Thus, methods directly controlling for $\bx_i$ are invalid for estimating the conditional mean of potential outcomes (or the CATE parameter $\E[y_i(1)-y_i(0)|\balpha_i]$), even if only one measurement error $u_{ip}$ is correlated with the outcome.
	Such correlation is practically relevant in, for example, panel data settings, where noisy proxies may be past outcomes and the most recent ones may be strongly correlated with the outcome in the current period of interest.
    
    Second,	when $\sigma_{p\epsilon}=0$, though $\E[y_i(0)|\bx_i]-\E[y_i(0)|\alpha_i]$ converges in probability to zero as $p$ grows, the coefficients on the conditioning variables $\bx_i$ are small but \textit{non-sparse}, which implies that many sparsity-based high-dimensional regression techniques such as Lasso are invalid for estimating quantities such as $\E[y_i(0)|\alpha_i]$. In other words, conditioning on a few selected coordinates of the high-dimensional $\bx_i$ cannot remove confounding due to $\alpha_i$. 
	   
	Third, the specific linear, homogeneous structure of $x_{il}$'s in this example permits approximate recovery of $\alpha_i$ by averaging all proxies. However, when the functions $\eta_l(\cdot)$ are unknown and heterogeneous, it is no longer straightforward which transformation is able to remove confounding. Our proposed method below can be viewed as a general strategy to effectively extract information on $\alpha_i$ in possibly nonlinear settings, while allowing for potential correlation of measurement errors with outcomes or treatment conditional on the latent confounders. See more discussions below Assumption \ref{Assumption: indep of errors} and in Section \ref{SA section: condition on x} of the SA.
\end{remark}

\subsection{Notation}
For any scalars $a$ and $b$, $a\vee b=\max\{a,b\}$ and $a\wedge b=\min\{a,b\}$. Define $\sam=(K^{1/2}\wedge p^{1/2})/\sqrt{\log (n\vee p)}$. 
For a finite set $\mathcal{S}$, $|\mathcal{S}|$ denotes its cardinality. 
For an integer $m>0$, $[m]=\{1, 2, \cdots, m\}$. 
For a vector $\bv\in\mathbb{R}^\mathsf{d}$, $\|\bv\|=\sqrt{\bv'\bv}$ is the Euclidean norm of $\bv$, and for an $m\times n$ matrix $\bA$,  $\|\bA\|_{\max}=\max_{i\in[m], j\in[n]}|a_{ij}|$ is the entrywise sup-norm of $\bA$. 
Moreover,  $\bA_{i\cdot}$ denotes the $i$-th row of $\bA$.

For sequences of numbers or random variables, $a_n\lesssim b_n$ or $a_n=O(b_n)$ means $\limsup_n|a_n/b_n|$ is finite, and $a_n\lesssim_\P b_n$ or $a_n=O_\P(b_n)$ means $\limsup_{\varepsilon\rightarrow\infty}\limsup_n\P[|a_n/b_n|\geq\varepsilon]=0$.  $a_n=o(b_n)$ implies $a_n/b_n\rightarrow 0$, and $a_n=o_\P(b_n)$ implies $a_n/b_n\rightarrow_\P 0$, where $\rightarrow_\P$ denotes convergence in probability. $a_n\asymp b_n$ implies that $a_n\lesssim b_n$ and $b_n\lesssim a_n$. $a_n\asymp_\P b_n$ implies that $a_n\lesssim_\P b_n$ and $b_n\lesssim_\P a_n$. $\rightsquigarrow$ denotes convergence in distribution.
Moreover, for a possibly random sequence $\{a_{i,n}\}_{i\in[n]}$ and 
a strictly positive sequence $\{r_{i,n}\}_{i\in[n]}$, 
we write $a_{i,n}=\bar{o}_\P(r_{i,n})$ if 
$\max_{i\in[n]}|a_{i,n}/r_{i,n}|\rightarrow_\P 0$.

\section{Outline of Estimation Procedure} \label{sec: estimation}

This section describes our proposed estimation procedure. 
First, given the structure \eqref{eq: HD covariate}, relevant information on $\balpha_i$ is extracted from $\bx_i$ via a local PCA method. Second, we estimate $\zeta_i=\E[y_i(0)|\balpha_i, \bz_i]$ and $\varrho_i=\P(t_i=1|\balpha_i, \bz_i)$ via a local quasi-maximum likelihood method,  where the extracted information from the first step plays the role of kernel functions and generated regressors. Finally, the estimator of $\E[y_i(0)|t_i=1]$ is constructed based on the doubly-robust score function. See Algorithm \hyperlink{t2}{1} for a short summary. 

In addition to methodological discussions, each step below will be accompanied by an empirical illustration using the data from \cite{Acemoglu-et-al_2016_JFE}, who analyze how the announcement of the appointment of Tim Geithner as Treasury Secretary on November 21, 2008, affected stock returns of financial firms that were connected to him.
Specifically, the treatment of interest is exposure to the appointment through a connection to Geithner, beginning at a particular date (referred to as ``event day $0$'' hereafter). All firms remain untreated prior to the appointment. Starting at event day $0$, a subgroup of firms that are connected to Geithner are treated ($t_i=1$), while the other group remains untreated ($t_i=0$). Variables used in this analysis and the parameter of interest  are listed below:
\begin{itemize}\setlength\itemsep{.1em}
	\item Potential outcomes $y_{i}(1)$ and $y_i(0)$: the cumulative stock returns of firm $i$ from day $0$ to day $1$ that would be observed with and without Geithner connections;
	\item Noisy proxies $\bx_{i}$: the daily stock returns of firm $i$ prior to the Geithner announcement;
	\item Additional controls $\bz_i$: the size (log of total assets), profitability (return on equity), and leverage (total debt to total capital) of firm $i$ as of 2008;
	\item ATT parameter $\E[y_i(1)-y_i(0)|t_i=1]$: the average cumulative abnormal return of connected firms from day $0$ to day $1$.
\end{itemize}
The sample consists of $583$ firms in total ($n=583$) and $22$ of them are treated (``connected to Geithner''). To be comparable with the results in \cite{Acemoglu-et-al_2016_JFE}, the observed proxies $\bx_i$ include the daily stock returns over a $250$-day period ending $30$ days prior to the Geithner announcement ($p=250$). In this application, one concern is that even after controlling for the observed covariates $\bz_i$, ``connected and non-connected firms could be different in other ways'' \cite[p.~19]{Acemoglu-et-al_2016_JFE}. For instance, the two groups of firms may be different in some unobserved characteristics $\balpha_i$ such as entrepreneurial ability or management quality, which could be implicitly captured by stock returns as discussed before.

\begin{table*}[hbt!]
	\raisebox{\ht\strutbox}{\hypertarget{t2}{}}
	\centering
	\small
	\renewcommand{\arraystretch}{1.7}
	\begin{tabular}{p{.98\textwidth}}
		\hline\hline
		\textbf{Algorithm 1} (Estimation and inference for $\E[y_i(0)|t_i=1]$) \\
		\hline
		\textbf{Step 1: Latent-Confounder Information Extraction} \\
		\textbf{Input:} covariate matrix $\bX\in\mathbb{R}^{n\times p}$, tuning parameters $K$, $\ttd_i$\\
		\textbf{Output:} $\{\mathcal{N}_i, \widehat{\bLambda}_{\subi}\}_{i=1}^n$\\
		Split $\bX$ proxy-wise (column-wise) into two submatrices $\bX^{\dagger}\in\mathbb{R}^{n\times p^\dagger}$ and $\bX^{\ddagger}\in\mathbb{R}^{n\times p^\ddagger}$\\
		For each $i\in[n]$,\\
		(1) use $\bX^{\dagger}$ to obtain the set $\mathcal{N}_i$ of the $K$ nearest neighbors of unit $i$ based on distance 
		$\rho(\cdot, \cdot)$:  
		\begin{footnotesize}
		\begin{equation*}
		\mathcal{N}_i=\Big\{j_k(i): \sum_{\ell=1}^{n}\I\Big(\rho(\bX^\dagger_{i\cdot},\bX^\dagger_{\ell\cdot})\leq\rho(\bX^\dagger_{i\cdot}, \bX^{\dagger}_{j_k(i)\cdot})\Big)\leq K, \; 1\leq k\leq K\Big\}
		\end{equation*}
	    \end{footnotesize}
	    \vspace{-.5em}
	    
		(2) use $\bX_{\subi}^\ddagger$ to obtain the local factor loadings $\widehat{\bLambda}_{\subi}$  by local PCA:
		\begin{footnotesize}
		\begin{equation*}
		(\widehat{\bF}_{\subi}, \widehat{\bLambda}_{\subi})=
		\underset{\scalebox{0.6}{ $\tilde\bF_{\subi}\in\mathbb{R}^{p^\ddagger\times \ttd_i},  \tilde\bLambda_{\subi}\in\mathbb{R}^{K\times \ttd_i}$}}
		{\argmin}\tr\Big[\Big(\bX_{\subi}^\ddagger-\tilde{\bLambda}_{\subi}\tilde{\bF}_{\subi}'\Big)\Big(\bX_{\subi}^\ddagger-\tilde{\bLambda}_{\subi}\tilde{\bF}_{\subi}'\Big)'\Big]
		\end{equation*}
		\end{footnotesize}

		\textbf{Step 2: Local Quasi-Maximum Likelihood Estimation}\\
		\textbf{Input:}
		$\{y_i, t_i, \bz_i, \widehat{\bLambda}_{\subi}, \mathcal{N}_i\}_{i=1}^n$\\
		\textbf{Output:}
		fitted values $\{\widehat{\zeta}_{i}\}_{i=1}^n$ and   $\{\widehat{\varrho}_{i}\}_{i=1}^n$\\
		For each $i\in[n]$,\\
		(1) estimate $\zeta_{i}=\E[y_i(0)|\balpha_i,\bz_i]$ by local least squares using control units in $\mathcal{N}_i$; see \eqref{eq: local reg of outcome}\\
		(2) estimate $\varrho_i=\P(t_i=1|\balpha_i, \bz_i)$ by local logistic regression using units in $\mathcal{N}_i$; see \eqref{eq: local logit of treatment}
		\\

		\textbf{Step 3: Counterfactual Analysis}\\
		\textbf{Input:} $\{y_i, t_i, \widehat{\zeta}_{i}, \widehat{\varrho}_{i}\}_{i=1}^n$
		\\
		\textbf{Output:}  estimate $\widehat{\theta}$ of $\theta=\E[y_i(0)|t_i=1]$ and asymptotic variance estimate $\widehat{\sigma}^2$
		\begin{footnotesize}
	    \begin{equation*}
    	\widehat{\theta}=
	    \frac{1}{n}\sum_{i=1}^{n}\left[
	    \frac{t_i\widehat{\zeta}_{i}}{\widehat{\varrho}}+
	    \frac{\widehat{\varrho}_{i}}{\widehat{\varrho}}\frac{(1-t_i)(y_{i}-\widehat{\zeta}_{i})}{1-\widehat{\varrho}_{i}}\right],\quad
	    \widehat{\sigma}^2=
	    \frac{1}{n}\sum_{i=1}^{n}\bigg[\frac{t_i(\widehat{\zeta}_{i}-\widehat{\theta})^2}{\widehat{\varrho}^2}+
	    \frac{\widehat{\varrho}_{i}^2(1-t_i)(y_i-\widehat{\zeta}_{i})^2}{\widehat{\varrho}^2(1-\widehat{\varrho}_{i})^2}\bigg]
	    \end{equation*}
        \end{footnotesize}
		\\
		\hline
	\end{tabular}
\end{table*}

\subsection{Step 1: Latent-Confounder Information Extraction}
\label{subsec: step 1, latent variable extraction}
The goal of this step is to extract information on latent confounders $\balpha_i$ by employing the structure \eqref{eq: HD covariate}. The key ideas are outlined below. 

\smallskip
\textbf{\textit{Proxy-wise Data Splitting}}. 
We recommend users separate the $K$-NN matching and PCA steps via proxy-wise data splitting. 
Specifically, conduct a proxy-wise (i.e., column-wise) split of the index set $\mathcal{C}=[p]$ of $\bX$ into two non-overlapping subsets: $\mathcal{C}=\mathcal{C}^\dagger\cup\mathcal{C}^\ddagger$ with $p^\dagger=|\mathcal{C}^\dagger|$,  $p^\ddagger=|\mathcal{C}^\ddagger|$ and $p^\dagger\asymp p^\ddagger\asymp p$. Accordingly, the data matrix $\bX$ is divided into two submatrices $\bX^{\dagger}$ and $\bX^{\ddagger}$ containing proxies indexed by $\mathcal{C}^\dagger$ and $\mathcal{C}^\ddagger$, respectively.  $\bH^\dagger$, $\bH^\ddagger$, $\bU^{\dagger}$ and $\bU^{\ddagger}$ are defined similarly.
 
This proxy-wise splitting is used to obtain desired convergence properties of the local PCA estimator. It differs from sample splitting across units that separates nuisance function estimation and final inference  in the double machine learning literature.
 
\begin{remark}[Implementation: proxy-wise data splitting]
	In principle, proxy-wise splitting is used to make the two portions of data $\bX^\dagger$ and $\bX^\ddagger$ approximately independent (conditionally on the latent variables). For independent errors across $l$ as specified in Assumption \ref{Assumption: indep of errors} below, this goal can be achieved by, e.g., using a random proxy-wise split. 
	When proxies are collected over time and exhibit weak dependence, one may instead, e.g., take the first half of periods for matching and the second half for PCA, thereby preserving the time series structure.
\end{remark}

\medskip
\textbf{\textit{$K$-NN Matching}}. This step makes use of the proxy block labeled by $\dagger$, i.e., the submatrix of $\bX$ containing proxies indexed by $\mathcal{C}^{\dagger}$.
Let $\rho: \mathbb{R}^{p^\dagger}\times \mathbb{R}^{p^\dagger}\to\mathbb{R}$ be a ``distance'' function (we use the term loosely, so $\rho$ need not satisfy all the axioms of a metric). For each unit $i$, identify the index set $\mathcal{N}_i$ of its $K$ nearest neighbors (including unit $i$ itself):
\begin{equation}\label{eq: knn}
\mathcal{N}_i=\Big\{j_k(i): \sum_{\ell=1}^{n}\I\Big(\rho(\bX^\dagger_{i\cdot},\bX^\dagger_{\ell\cdot})\leq\rho(\bX^\dagger_{i\cdot}, \bX^\dagger_{j_k(i)\cdot})\Big)\leq K, \; 1\leq k\leq K\Big\}.
\end{equation}
If distances are tied at the cutoff, we break the tie by the original unit indices and retain exactly \(K\) units.

\begin{remark}[Implementation: choice of distance]
	Some usual choices of distance in the literature include 
		(i) squared Euclidean distance  $\rho(\bX^\dagger_{i\cdot}, \bX^\dagger_{j\cdot})=
		\|\bX^\dagger_{i\cdot}-\bX^\dagger_{j\cdot}\|^2/p^\dagger$; 
		(ii) pseudo-max distance   
		$\rho(\bX^\dagger_{i\cdot}, \bX^\dagger_{j\cdot})=\max_{\ell\neq i, j} |(\bX^\dagger_{i\cdot}-\bX^\dagger_{j\cdot})'\bX^\dagger_{\ell\cdot}|/p^\dagger$;
		and (iii) distance of averages $\rho(\bX^\dagger_{i\cdot}, \bX^\dagger_{j\cdot})=|\sum_{l\in\mathcal{C}^\dagger}(x_{il}-x_{jl})|/p^\dagger$. 
	Indirect matching based on each of these choices can be justified under certain conditions on the data generating process, and generally, none strictly dominates the others in theory. However, matching based on the Euclidean distance requires an unrealistic conditional homoskedasticity assumption, and matching using the distance of averages needs a strong injectivity condition on the averaged proxies. By contrast, the pseudo-max distance can be justified under practically more plausible conditions and thus is our recommended choice among the three. See Remarks \ref{remark: noise reduction} and \ref{remark: informativeness} and Section \ref{SA section: discuss assumptions} of the SA for details.
	
	Moreover, when proxies differ in scale or importance for revealing information on the latent variables, one can rescale or reweight different proxies when searching for nearest neighbors. Such transformations can be viewed as particular choices of the distance $\rho(\cdot, \cdot)$. For instance, given a vector of weights $w_l$, $l\in\mathcal{C}^\dagger$, one can define 
	$\rho(\bX^\dagger_{i\cdot}, \bX^\dagger_{j\cdot})=\sum_{l\in\mathcal{C}^\dagger}(w_lx_{il}-w_lx_{jl})^2$, which is the squared Euclidean distance of the reweighted proxies.
\end{remark}

\begin{remark}[Implementation: choice of $K$]\label{remark: select K}
	Our simulation studies in Section \ref{sec: simulation} show strong robustness of our method with respect to the choice of $K$. Theoretically, by taking advantage of doubly-robust score functions, our main results rely on mild restrictions on $K$, which reduce to $n/K^2=o(1)$ (up to $\log n$ terms) in practically relevant cases (see the discussion below Theorem \ref{thm: pointwise inference}). 
	Thus, many feasible choices of $K$ are valid in theory. 
	For instance, one can consider the nearest-neighbors-based 
	local $(m-1)$th-order polynomial regression	of $y_i$ on any $r$ out of the $p$ proxies. Under standard regularity conditions, 
	the optimal $K$ in the MSE sense for this regression is $O(n^{\frac{2m}{2m+r}})$ if $m$ is even, and is $O(n^{\frac{2m+2}{2m+2+r}})$ if $m$ is odd, which grows faster than $\sqrt{n}$ if $m>r/2$ or $m>r/2-1$. 
	Usual plug-in or cross-validation methods in the nonparametric regression literature can be employed to select $K$, though such choices may not be optimal for estimating the target causal parameter.
\end{remark}

Using the data from \cite{Acemoglu-et-al_2016_JFE}, we implement $K$-NN matching for each unit based on the pseudo-max distance of daily stock returns in the first 125 days. Given the discussion in Remark \ref{remark: select K}, we set $K=99$ as a benchmark choice, which is selected by leave-one-out cross validation for local constant nearest-neighbor regression of cumulative stock returns from day $0$ to $1$ on the pre-treatment return on day $-31$.
To have a sense of the performance of $K$-NN matching, we report some summary statistics of matching discrepancies in Table \ref{table: summary KNN}. Specifically, we calculate for each unit the maximum distance of matched pairs  divided by the standard deviation of distances across all pairs, which can be viewed as a normalized matching discrepancy in terms of the observed returns. Matching performs relatively well for treated units, whereas some control units have relatively distant matches. In later analysis, we check the robustness of our results by varying the number of nearest neighbors or dropping control units with large discrepancies.

\begin{table}[h]
	\centering
	\small
	\caption{$K$-NN Matching: Maximum Distance of Matched Pairs}
	\label{table: summary KNN}
	\begin{threeparttable}
		\setlength{\tabcolsep}{13pt}
		\renewcommand{\arraystretch}{1.25}
		\begin{tabular}{lrrrrrr}
\hline\hline
\multicolumn{1}{l}{}&\multicolumn{1}{c}{Min.}&\multicolumn{1}{c}{1st Qu.}&\multicolumn{1}{c}{Median}&\multicolumn{1}{c}{Mean}&\multicolumn{1}{c}{3rd Qu.}&\multicolumn{1}{c}{Max.}\tabularnewline
\hline
Treated&$0.671$&$1.043$&$1.166$&$1.199$&$1.313$&$1.840$\tabularnewline
Control&$0.602$&$0.881$&$1.064$&$1.228$&$1.345$&$6.638$\tabularnewline
\hline
\end{tabular}

		\smallskip
		
		\begin{tablenotes}[para,flushleft]
			\footnotesize\textit{Notes}:
			Matching uses $K=99$. For each unit, we normalize the maximum matched-pair distance by dividing it by the standard deviation of distances across all pairs.
		\end{tablenotes}
	\end{threeparttable}
\end{table}

\textbf{\textit{Local PCA}}.
This step makes use of the proxy block labeled by $\ddagger$, i.e., the submatrix of $\bX$ containing proxies indexed by $\mathcal{C}^{\ddagger}$.
Given the nearest neighbors of $i$ found in the first step, define a $K\times p^\ddagger$ submatrix $\bX_{\subi}^\ddagger$ of $\bX^\ddagger$, with each row corresponding to a neighbor's proxies that belong to $\mathcal{C}^\ddagger$. 
Similarly, $\bH_{\subi}^\ddagger$ and $\bU_{\subi}^\ddagger$ denote submatrices of $\bH^\ddagger$ and $\bU^\ddagger$ respectively, formed by rows corresponding to the $K$ nearest neighbors of unit $i$.
Then, we write
\begin{equation}\label{eq: local factor structure}
	\bX_{\subi}^\ddagger=\bLambda_{\subi}\bF_{\subi}'+\bXi_{\subi}+\bU_{\subi}^\ddagger,
\end{equation} 
where the local factor matrix $\bF_{\subi}\in\mathbb{R}^{p^\ddagger\times \ttd_i}$ and the loading matrix $\bLambda_{\subi}\in\mathbb{R}^{K\times \ttd_i}$ are defined to be a solution to the noise-free local PCA problem:
\begin{equation*}
	(\bF_{\subi}, \bLambda_{\subi})=
	\underset{\scalebox{0.7}{ $\tilde\bF_{\subi}\in\mathbb{R}^{p^\ddagger\times \ttd_i},  \tilde\bLambda_{\subi}\in\mathbb{R}^{K\times \ttd_i}$}}
	{\argmin}\tr\Big[\Big(\bH_{\subi}^\ddagger-\tilde\bLambda_{\subi}\tilde\bF_{\subi}'\Big)\Big(\bH_{\subi}^\ddagger-\tilde\bLambda_{\subi}\tilde\bF_{\subi}')'\Big]
\end{equation*}
such that $\tilde{\bF}_{\subi}'\tilde{\bF}_{\subi}/p^\ddagger=\bI_{\ttd_i}$ and $\tilde\bLambda_{\subi}'\tilde\bLambda_{\subi}/K$ is diagonal. $\ttd_i$ denotes the number of local factors extracted, and $\bXi_{\subi}\in\mathbb{R}^{K\times p^\ddagger}$ is the corresponding approximation error for $\bH_{\subi}^\ddagger$.  
In practice, we estimate $\bF_{\subi}$ and $\bLambda_{\subi}$  by PCA:
\begin{equation}\label{eq: lpca}
(\widehat{\bF}_{\subi}, \widehat{\bLambda}_{\subi})=
\underset{\scalebox{0.7}{ $\tilde\bF_{\subi}\in\mathbb{R}^{p^\ddagger\times \ttd_i},  \tilde\bLambda_{\subi}\in\mathbb{R}^{K\times \ttd_i}$}}
	{\argmin}\tr\Big[\Big(\bX_{\subi}^\ddagger-\tilde\bLambda_{\subi}\tilde\bF_{\subi}'\Big)\Big(\bX_{\subi}^\ddagger-\tilde\bLambda_{\subi}\tilde\bF_{\subi}')'\Big]
\end{equation}
such that $\tilde{\bF}_{\subi}'\tilde{\bF}_{\subi}/p^\ddagger=\bI_{\ttd_i}$ and $\tilde\bLambda_{\subi}'\tilde\bLambda_{\subi}/K$ is diagonal.

\begin{remark}[Implementation: choice of $\ttd_i$]\label{remark: number of confounders}
In this paper we treat $\ttd_i$ as a fixed tuning parameter, possibly heterogeneous across neighborhoods, that governs the number of terms in the local approximation. In principle, $\ttd_i$ is set such that the extracted local factors have stronger signal strength relative to the noise $\bU_{\subi}^\ddagger$ (see the discussion below Assumption \ref{Assumption: LPCA}). Note that  although the number of latent confounders $\ttr_\alpha$ implicitly affects the approximation quality of the proposed method,  it is \textit{unnecessary} in practice to determine $\ttr_\alpha$ in the above procedure. 
Also, $\ttd_i$ need \textit{not} be greater than or equal to $\ttr_\alpha$ (e.g., $\ttd_i=1$ is conceptually analogous to local constant smoothing in nonparametrics). 

In practice, we recommend users select $\ttd_i$ in a data-dependent fashion. For instance, in each local neighborhood we extract all local factors from $\bX_{\subi}^\ddagger$ whose singular values exceed $C_{np}(\sqrt{K}+\sqrt{p^\ddagger})$, where $C_{np}>0$ is a slowly diverging sequence (e.g., $C_{np}=\log\log(n\vee p)$). Since the noise matrix $\bU_{\subi}^\ddagger$ is typically of order $\sqrt{K}+\sqrt{p^\ddagger}$, this rule ensures that the extracted factors are sufficiently stronger than the noise, while the slow divergence of $C_{np}$ helps retain useful approximation power. Refined strategies based on similar ideas are available in the literature on PCA (e.g., \citealp{johnstone2001distribution,dobriban2019deterministic}).
\end{remark}

\begin{figure}[!h]
		\centering
		\small
		\caption{Local Singular Values and Component Selection}
		\begin{subfigure}{0.47\textwidth}
			\includegraphics[width=\textwidth]{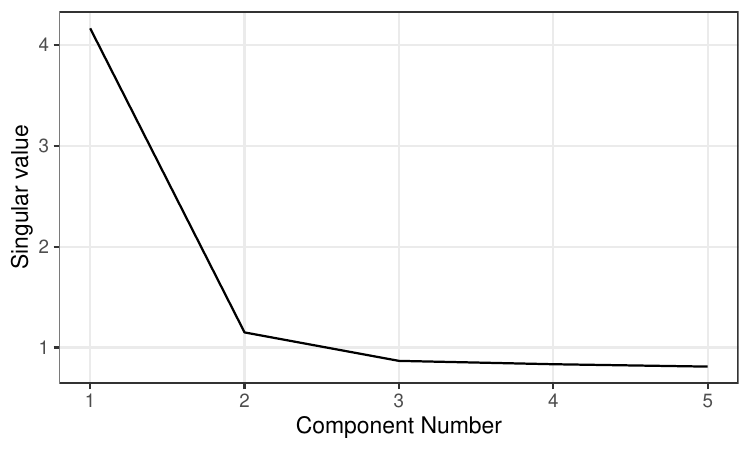}
			\caption{Singular values for one neighborhood}
		\end{subfigure}
		\begin{subfigure}{0.47\textwidth}
			\includegraphics[width=\textwidth]{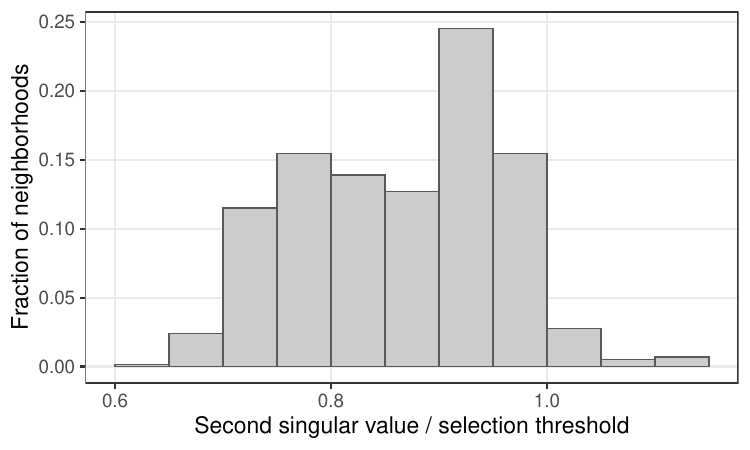}
			\caption{Second singular-value ratios}
		\end{subfigure}
		\label{figure:eigenvalue}
		{\footnotesize\noindent\parbox{0.95\linewidth}{%
			\textit{Notes:} Panel (a) reports the first five singular values for the neighborhood of American Express Co. Panel (b) reports the distribution across all full-sample neighborhoods of the second singular value divided by the selection threshold in Remark \ref{remark: number of confounders}.
		}}
\end{figure}

As an illustration, we implement local PCA for each unit using the data from \cite{Acemoglu-et-al_2016_JFE}. Recall that for each firm a set of nearest neighbors has been obtained using the daily stock returns in the first $125$ days. PCA is then conducted for this subgroup of firms using their daily stock returns in the next 125 days. In the subsequent empirical analysis, we employ the strategy described in Remark \ref{remark: number of confounders} to choose $\ttd_i$ in a data-dependent fashion. Figure \ref{figure:eigenvalue}(a) shows the leading singular values for the neighborhood of American Express Co., while Figure \ref{figure:eigenvalue}(b) summarizes the second singular value relative to this selection threshold across all full-sample neighborhoods. Overall, Figure \ref{figure:eigenvalue} shows that our rule selects one component in most neighborhoods and two in a few; Section \ref{SA subsection: application diagnostics} of the SA provides more detailed summaries and shows how changing the selection threshold affects the selected local dimensions $\ttd_i$ and the resulting ATT estimates.

\FloatBarrier
\FloatBarrier

\subsection{Step 2: Local Quasi-Maximum Likelihood Estimation}
\label{subsec: step 2, factor-augmented reg}

The goal of this step is to estimate $\zeta_i=\E[y_i(0)|\balpha_i, \bz_i]$ and $\varrho_i=\P(t_i=1|\balpha_i, \bz_i)$. Section \ref{SA sec: estimation} of the SA considers the more general quasi-maximum likelihood estimation (QMLE), but to be consistent with our basic setup in \eqref{eq: outcome} and \eqref{eq: treatment}, here we focus on the Gaussian likelihood for the outcome equation and the logit likelihood for the treatment equation.

A local least squares estimator of $\zeta_i$ is given by
\begin{equation}\label{eq: local reg of outcome}
\begin{split}
	\widehat{\zeta}_{i}&=
	\widehat{\mu}(\balpha_i)+\bz_{i}'\widehat{\bbeta}_{\subi},\quad \widehat{\mu}(\balpha_i)=
	\widehat{\blambda}_{i,\subi}'\widehat{\bb}_{\subi},\quad \text{where}\\
	(\widehat{\bb}_{\subi}',\;\widehat{\bbeta}_{\subi}')'&=
	\argmin_{(\bb_1',\; \bb_2')'\in\mathbb{R}^{\ttd_i+\ttr_z}}\;
	\sum_{j\in\mathcal{N}_i}
	(1-t_j)(y_j-\widehat{\blambda}_{j,\subi}'\bb_1-\bz_j'\bb_2)^2,
\end{split}
\end{equation}
where $\widehat{\blambda}_{j, \,\subi}$ is the estimated factor loading vector in $\widehat{\bLambda}_{\subi}$ corresponding to unit $j$.
For the treatment equation \eqref{eq: treatment}, the estimator of $\varrho_i=\P(t_i=1|\balpha_i,\bz_i)$ is given by
\begin{equation}\label{eq: local logit of treatment}
	\begin{split}
\widehat{\varrho}_i&=\psi(\widehat{e}(\balpha_i)+\bz_{i}'\widehat{\bgamma}_{\subi}), \quad \widehat{e}(\balpha_i)=\widehat{\blambda}_{i,\subi}'\widetilde{\bb}_{\subi}, \quad \text{where}\\
(\widetilde{\bb}_{\subi}',\;\widehat{\bgamma}_{\subi}')'&=
\argmax_{(\bb_1',\; \bb_2')'\in\mathbb{R}^{\ttd_i+\ttr_z}}\;
\sum_{j\in\mathcal{N}_i}
\log [\psi_j(\bb_1,\bb_2)^{t_j}(1-\psi_j(\bb_1,\bb_2))^{1-t_j}],
\end{split}
\end{equation}
where $\psi_j(\bb_1,\bb_2):=\psi(\widehat{\blambda}_{j,\subi}'\bb_1+\bz_j'\bb_2)$.

\begin{remark}[Implementation: QMLE]
	For local least squares we have closed-form solutions, while for more general quasi-MLE methods, the maximization is done numerically in practice.
	The local PCA outputs can be readily used as inputs to any existing fast MLE algorithms.
	Also, in practice, an intercept may be included alongside the estimated factor loadings in both local QMLE problems because PCA may not capture the constant basis term used for local approximations; see Example \ref{example: linear factor model} below. If the intercept lies in the span of the extracted loadings, one redundant column can be removed without changing the fitted values $\widehat{\zeta}_i$ and $\widehat{\varrho}_i$, which are the only quantities needed for causal inference.

	In \eqref{eq: local reg of outcome} and \eqref{eq: local logit of treatment}, we include the observed covariates $\bz_j$ as additional controls, without exploiting the semiparametric structure to estimate $\bbeta$ and $\bgamma$. Alternatively, these parameters could be estimated using semiparametric methods such as profiled quasi-maximum likelihood, although at greater computational cost. See, e.g., \cite{Hardle-Muller-Sperlich-Werwatz_2012_book}.
\end{remark}

\begin{figure}[!h]
	\small
	\caption{Local QMLE}
	\begin{center}
		\begin{subfigure}{0.47\textwidth}
			\includegraphics[width=\textwidth]{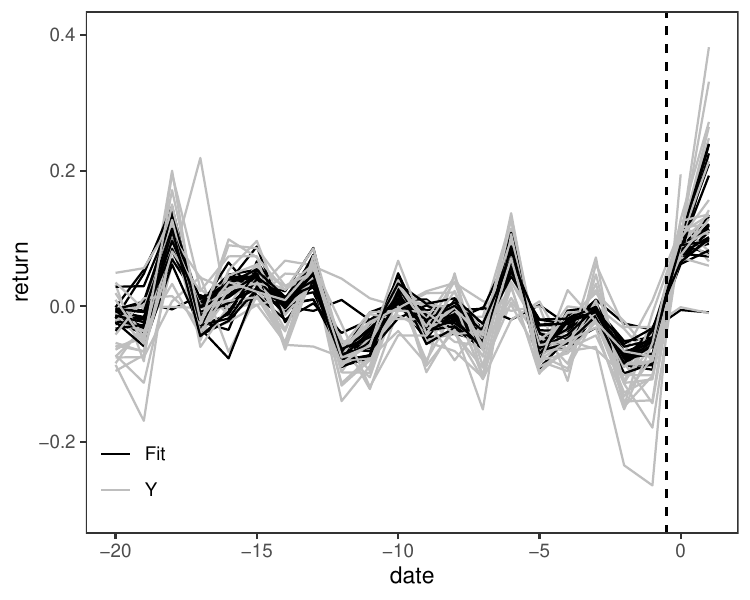}
			\caption{Conditional mean of the untreated outcome}
			\label{figure:qmle}
		\end{subfigure}
		\begin{subfigure}{0.47\textwidth}
			\includegraphics[width=\textwidth]{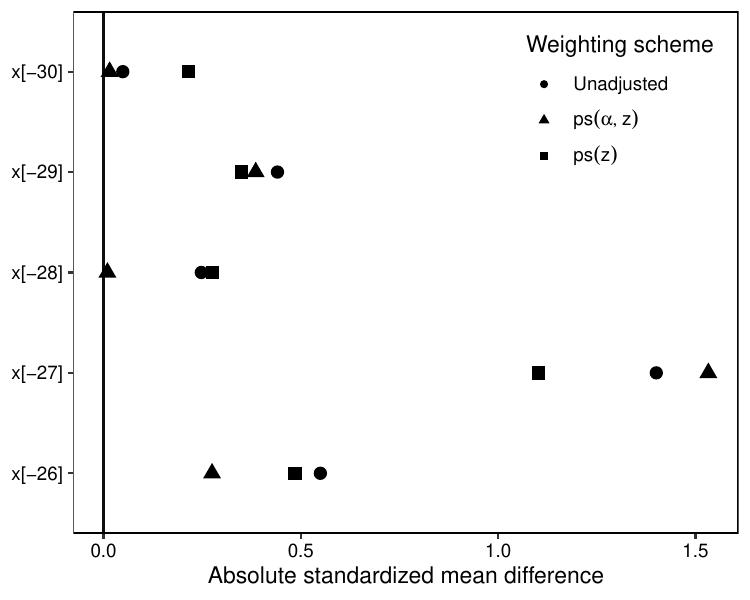}
			\caption{Balance in pre-treatment returns}
			\label{figure:balance}
		\end{subfigure}
	\end{center}
\vspace{-1em}
{\footnotesize\noindent\parbox{\linewidth}{%
	\textit{Notes:} The left panel shows the fitted values for the conditional mean of untreated returns for treated firms, along with the observed return sequences. The right panel shows the balance in returns from day $-30$ through day $-26$, where \textsf{ps}($\alpha$,\textsf{z}) and \textsf{ps}(\textsf{z}) denote propensity score estimates from  our approach and from a specification using $\bz_i$ only. 
}}
\end{figure}

As an illustration, for each treated firm, we find its nearest neighbors and obtain the local factor loadings. Then, for returns on each day from $-20$ through $+1$, we use the data from the treated firm's untreated neighbors to perform a local least squares regression on the extracted factor loadings. Figure \ref{figure:qmle} shows the fitted values in black and the observed daily returns in grey for the treated firms. 
Recall that the fitted values are the estimates of the conditional means of potential returns \textit{without} treatment given the latent variables. Relative to the fitted values, the observed returns of some treated firms decline as event day $0$ approaches, whereas those of most treated firms increase sharply after day $0$.
Given the propensity score estimates obtained in this step, we can also check the balance in pre-treatment returns that are not used for local PCA. The result is reported in Figure \ref{figure:balance}, which compares our approach with propensity score estimates based on $\bz_i$ only.
We remind readers that these diagnostics should be interpreted cautiously: even with perfectly balanced confounders $(\balpha_i,\bz_i)$, measurement errors related to treatment conditional on these confounders (as allowed in our more general model) may produce proxy imbalance.

\FloatBarrier
\FloatBarrier
\vspace{-2em}

\subsection{Step 3: Counterfactual Analysis}

The final step is to estimate $\theta=\E[y_{i}(0)|t_{i}=1]$. Let $\varrho=\P(t_{i}=1)$. 
Under our unconfoundedness assumption \eqref{eq: unconfoundedness}, 
\[
\theta=\E\left[\frac{t_i\zeta_{i}}{\varrho}+\frac{\varrho_{i}}{\varrho}\frac{(1-t_i)(y_{i}-\zeta_{i})}{1-\varrho_{i}}\right].
\]
Therefore, an estimator of $\theta$ is given by
\begin{equation}\label{eq: final estimator}
\widehat{\theta}=
\frac{1}{n}\sum_{i=1}^{n}\left[\frac{t_i\widehat{\zeta}_{i}}{\widehat{\varrho}}+\frac{\widehat{\varrho}_{i}}{\widehat{\varrho}}\frac{(1-t_i)(y_{i}-\widehat{\zeta}_{i})}{1-\widehat{\varrho}_{i}}\right],
\end{equation}
where $\widehat{\zeta}_{i}$ and $\widehat{\varrho}_{i}$ are obtained in the second step, and $\widehat{\varrho}=\sum_{i=1}^{n}t_i/n$.
For the purpose of inference, a simple plug-in estimator of the asymptotic variance is  
\begin{equation}\label{eq: standard error}
\widehat{\sigma}^2=
\frac{1}{n}\sum_{i=1}^{n}\bigg[\frac{t_i(\widehat{\zeta}_{i}-\widehat{\theta})^2}{\widehat{\varrho}^2}\bigg]+
\frac{1}{n}\sum_{i=1}^{n}\bigg[\frac{\widehat{\varrho}_{i}^2(1-t_i)(y_i-\widehat{\zeta}_{i})^2}{\widehat{\varrho}^2(1-\widehat{\varrho}_{i})^2}\bigg].		
\end{equation}
Under some regularity conditions, we show in Theorem \ref{thm: pointwise inference} below that
$\sqrt{n}\widehat{\sigma}^{-1}(\widehat{\theta}-\theta)\rightsquigarrow \mathsf{N}(0,1)$. 
Estimators of other causal parameters can be constructed accordingly. For instance, an estimator of ATT is simply $(\sum_{i=1}^nt_i)^{-1}\sum_{i=1}^nt_iy_i-\widehat{\theta}$.

\begin{remark}[Implementation: small denominators]
	It is well known in the literature that inverse probability weighting estimators may suffer from the ``small denominators'' issue (propensity scores too close to $0$ or $1$). In such cases some principled trimming methods may be needed to ensure more robust performance of the estimator. See, e.g., \cite{crump2009dealing,Busso-DiNardo-McCrary_2014_Restat,ma2020robust} for  discussions and possible solutions. 
\end{remark}

As an illustration, we estimate the average treatment effect of Geithner connections on cumulative returns from day 0 to day 1 (CAR[0,1]) for firms with connections. For the outcome equation, we implement a local least squares regression of cumulative stock returns of firms with no connections on the local factor loadings extracted previously; for the treatment equation, we perform a local logit estimation. We consider  several choices of $K=C\widehat{K}_{\tt CV}$ with $C=0.5, 1, 2$, where $\widehat{K}_{\tt CV}$ is the cross-validation benchmark defined above. We apply our procedure to the full sample and a base sample as defined in \cite{Acemoglu-et-al_2016_JFE} (the base sample excludes firms whose returns are highly correlated with Citigroup).

Results without covariates are reported in the first two columns of Table \ref{table:empirical illustration}, while the third and fourth columns give results that control for the three observed firm characteristics. For comparison, we also include the estimates based on double Lasso. Using the proposed method, the estimates of the average cumulative abnormal return for connected firms range from $0.049$ to $0.107$, and in most cases they are significant at the $0.05$ level. With covariates included, double Lasso produces larger point estimates than ours and larger standard errors. By contrast, \cite{Acemoglu-et-al_2016_JFE} uses a synthetic control method without covariate adjustment and finds a significant effect of 0.06 for the base sample.
We also examine robustness to excluding control firms with large normalized matching discrepancies; the results are reported in Section \ref{SA subsection: application diagnostics} of the SA.

\FloatBarrier
\begin{table}[h]
	\small
	\caption{Average Treatment Effect of Connections on the Treated}
	\label{table:empirical illustration}
	\centering
	\begin{minipage}{0.85\textwidth}
	\begin{threeparttable}
		\setlength{\tabcolsep}{6pt}
		\renewcommand{\arraystretch}{1.15}
		\resizebox{\linewidth}{!}{\begin{tabular}{lccccc}
\hline\hline
\multicolumn{1}{l}{\bfseries }&\multicolumn{2}{c}{\bfseries No Covariates}&\multicolumn{1}{c}{\bfseries }&\multicolumn{2}{c}{\bfseries Add Covariates}\tabularnewline
\cline{2-3} \cline{5-6}
\multicolumn{1}{l}{}&\multicolumn{1}{c}{Full Sample}&\multicolumn{1}{c}{Base Sample}&\multicolumn{1}{c}{}&\multicolumn{1}{c}{Full Sample}&\multicolumn{1}{c}{Base Sample}\tabularnewline
\hline
{\bfseries Local PCA, $K=$}&&&&&\tabularnewline
~~50&0.096&0.077&&0.060&0.057\tabularnewline
~~&(0.028)&(0.029)&&(0.024)&(0.03)\tabularnewline
~~99&0.105&0.087&&0.049&0.074\tabularnewline
~~&(0.028)&(0.026)&&(0.027)&(0.032)\tabularnewline
~~198&0.107&0.093&&0.064&0.080\tabularnewline
~~&(0.028)&(0.026)&&(0.029)&(0.03)\tabularnewline
\hline
{\bfseries Double Lasso}&&&&&\tabularnewline
~~&0.099&0.104&&0.092&0.109\tabularnewline
~~&(0.034)&(0.04)&&(0.033)&(0.042)\tabularnewline
\hline
\end{tabular}
}
		\smallskip
		
{\footnotesize\noindent\parbox{\linewidth}{%
		\textit{Notes}: The numbers in parentheses are standard errors.
		The base sample excludes firms highly correlated with Citigroup. 
}}
\end{threeparttable}
	\end{minipage}
\end{table}
\FloatBarrier

\subsection{Applicability and Diagnostic Checks}

Our method is intended for applications in which many proxies jointly carry information about the latent confounders. As formalized in Assumptions \ref{Assumption: Matching} and \ref{Assumption: LPCA} below, its applicability requires proxy-based matching to preserve proximity in the relevant latent confounders and the extracted local factor loadings to provide a sufficiently rich basis for approximating the outcome and treatment functions. In the present application, for example, these requirements mean that the return proxies are jointly informative about every relevant latent firm characteristic, their measurement errors are sufficiently weakly dependent over time, and the extracted local factor loadings capture variation related to these characteristics rather than measurement error.

Because the latent confounders are not observed, these requirements cannot be verified directly from the data. Nevertheless, the following empirical checks remain useful for evaluating the applicability of our method and the robustness of the conclusions:
\begin{enumerate}
	\item Repeat the analysis using several proxy-wise splits and compare nearest-neighbor sets, local loading spaces, and the resulting treatment-effect estimates.
	\item Use local factor loadings estimated from one proxy group to predict the proxies in a held-out group, and compare the prediction errors with a simple benchmark that does not use these factors. This check assesses whether different proxy blocks contain shared latent information that is useful for prediction.
	\item Report summaries of maximum within-neighborhood matching discrepancies, local singular values relative to the factor-selection threshold, selected numbers of local factors, and neighborhood composition.
	\item Examine propensity-score overlap and balance after adjustment and, when the application permits, report placebo effects for pre-treatment outcomes.
	\item Re-estimate the treatment effects across reasonable choices of $K$, the rule for selecting $\ttd_i$, the distance function, and the trimming threshold, and report whether the main conclusions are stable.
\end{enumerate}
Several of these checks are already part of our empirical illustration, including matching discrepancies in Table \ref{table: summary KNN}, a local singular-value diagnostic in Figure \ref{figure:eigenvalue}, balance in pre-treatment returns in Figure \ref{figure:balance}, and sensitivity to $K$ in Table \ref{table:empirical illustration}. For other diagnostics specific to our application, see Section \ref{SA subsection: application diagnostics} of the SA.

\section{Main Results} \label{sec: main results}

\subsection{Assumptions}

We begin with the unconfoundedness and overlap conditions commonly used in the causal inference literature. Recall that the confounders  $\balpha_i$ are not directly observed.

\begin{assumption}[Unconfoundedness and Overlap] 
	\label{Assumption: ID-counterfactual}
	$(y_i(0),y_i(1),t_i,\balpha_i, \bz_i)$ is i.i.d. over $i\in[n]$ and satisfies (a) $(y_i(0), y_i(1))\inde t_i\,|\, \balpha_i, \bz_i$;
	and (b) $0<c_{\min}\leq \P(t_i=1|\balpha_i, \bz_i)\leq c_{\max}<1$ for some constants $c_{\min}$ and $c_{\max}$ almost surely.
\end{assumption}
The conditional independence between the potential outcomes and the treatment is a key feature of our problem and will always be maintained. The data are assumed to be i.i.d. over $i$ for simplicity; in the empirical application, this means that we have a random sample of firms whose stock returns are independent of each other. With additional technicalities, it is possible to extend the results to allow for dependence within small clusters (so firms within the same industry can be correlated).

We impose some regularity conditions on covariates, errors and latent functions.
Let $\bZ_{\subi}$ have rows $\bz_j'$, $j\in\mathcal{N}_i$, in the same order as $\bLambda_{\subi}$, and let $\bP_{\subi}$ be the orthogonal projection onto the column space of $\bLambda_{\subi}$.
\begin{assumption}[Regularities]
	\label{Assumption: Regularities}
	$\balpha_i$ has a compact convex support with a density bounded and bounded away from zero; $\bz_i$ has a compact support, and $K^{-1}\bZ_{\subi}'(\bI_K-\bP_{\subi})\bZ_{\subi}$ has eigenvalues uniformly bounded away from zero in probability over $i\in[n]$; for some $\nu>2$,
	$\max_{i\in [n], l\in [p]}
	\E[|u_{il}|^{2+\nu}|\{\balpha_j\}_{j\in[n]}]\lesssim_\P 1$ and $\max_{i\in [n]}
	\E[|\epsilon_{i}|^{2+\nu}|\balpha_i, \bz_i]\lesssim_\P 1$; $\eta_l(\cdot)$ is bounded by a universal constant for all $l\in[p]$.
\end{assumption}
Most conditions in Assumption \ref{Assumption: Regularities} are standard. The local Gram condition is the usual noncollinearity requirement that the observed covariates $\bz_i$ retain nondegenerate variation after removing components spanned by the local factor loadings; it is vacuous when $\bz_i$ is absent. Section \ref{SA section: local design} of the SA provides more primitive sufficient conditions for this requirement.

Next, we impose an assumption restricting the errors in our model.
\begin{assumption}[Independence of Errors]\label{Assumption: indep of errors}
	$u_{il}$ is independent across $i\in[n]$ and over $l\in[p]$, conditional on $\{\balpha_i\}_{i\in[n]}$; 
	$\{(\epsilon_i,v_i)\}_{i\in[n]}$ is independent of $\{\bu_i\}_{i\in[n]}$ conditional on $\{(\balpha_i,\bz_i)\}_{i\in[n]}$.
\end{assumption}
As discussed before, Assumption \ref{Assumption: indep of errors} is common in the literature, implying that proxies are conditionally independent and do not directly affect the outcome or treatment.
We assume it here for simplicity only. In this high-dimensional setting, the conditional independence of $u_{il}$ across $l$ can be relaxed to, e.g., $\beta$-mixing conditions to accommodate proxies collected over time, albeit with more technical complexity.
Note that the proxy observations may still be dependent unconditionally across both $i$ and $l$ under Assumption \ref{Assumption: indep of errors}, since the conditional distributions of the $u_{il}$'s may depend on the full latent sample and the relationship among the $\eta_l$'s is unrestricted.
The conditional independence between the measurement error and the outcome (or treatment) can also be relaxed. In the SA, we introduce a quantity $\biaserr$ to explicitly characterize the bias due to the correlation of $\bu_i$ with $\epsilon_i$ or $v_i$, reflected through the correlation of $(\epsilon_i, v_i)$ with the proxy transformations $\mathcal{N}_i$ and $\widehat{\bLambda}_{\subi}$. Our results below maintain Assumption \ref{Assumption: indep of errors}, under which $\biaserr=0$, whereas the corresponding theorems in the SA allow $\biaserr$ to be nonzero under their stated rate conditions. Since it is technically more involved, we defer detailed discussion to Section \ref{SA section: indep of errors} in the SA.

In our context, it is unnecessary to identify $\balpha_i$ in an exact sense, since only predictions based on $\balpha_i$ matter for the causal analysis.
Instead, we rely on the following two key assumptions: Assumption \ref{Assumption: Matching} requires proxy-based matching to preserve proximity in the space of latent confounders, and Assumption \ref{Assumption: LPCA} requires the local basis recovered from the proxies to approximate $\mu(\cdot)$ and $e(\cdot)$ with sufficient accuracy.
   
\begin{assumption}[Indirect Matching] 
	\label{Assumption: Matching}\leavevmode
	For some fixed constants $\rho_0\in\mathbb{R}$, $\lvarsig>0$, $\uvarsig>0$ and some positive sequence $a_p=o(1)$, the following conditions hold:
	\begin{enumerate}[label=(\alph*)]		
		\item $\underset{i,j\in[n]:\,i\neq j}{\max}\;|\rho(\bX_{i\cdot}^\dagger, \bX_{j\cdot}^\dagger)-
		\rho(\bH_{i\cdot}^\dagger, \bH_{j\cdot}^\dagger)
		-\rho_0|\lesssim_\P a_p$;
		
		\item 
		$\underset{i,j\in[n]:\,\balpha_i\neq \balpha_j}{\max}\;\frac{\rho(\bH_{i\cdot}^\dagger,\, \bH_{j\cdot }^\dagger)}{\|\balpha_i-\balpha_j\|^{\uvarsig}}
		\lesssim_\P 1$ and 
		$\underset{i,j\in[n]:\,\balpha_i\neq \balpha_j}{\min}\;\frac{\rho(\bH_{i\cdot}^\dagger,\, \bH_{j\cdot}^\dagger)}{\|\balpha_i-\balpha_j\|^{\lvarsig}}\gtrsim_\P 1$.
	\end{enumerate}
\end{assumption}
Assumption \ref{Assumption: Matching} is a high-level condition, accommodating a generic choice of the distance $\rho$. In the following we make several remarks on each part, and defer more details, including sufficient conditions for Assumption \ref{Assumption: Matching}, to Section \ref{SA section: matching} of the SA. 

Part (a) implies that the distance of the noisy proxies can be translated into that of the noise-free mean structure.
This is typically possible when the chosen distance aggregates many proxies that are independent (or weakly dependent) conditional on the latent confounders. The form of aggregation depends on the distance, as illustrated below.

\begin{remark}[Noise reduction]\label{remark: noise reduction}
	For the Euclidean distance, we average the squared proxy differences.
	Since $\E[\bu_i|\balpha_i]=0$ and $u_{il}$ is independent over $l$, we roughly have $\|\bX_{i\cdot}^\dagger-\bX_{j\cdot}^\dagger\|^2/p^\dagger\approx\|\bH_{i\cdot}^\dagger-\bH_{j\cdot}^\dagger\|^2/p^\dagger+\|\bU_{i\cdot}^\dagger\|^2/p^\dagger+\|\bU_{j\cdot}^\dagger\|^2/p^\dagger$.
	Then, part (a) holds if $\|\bU_{i\cdot}^\dagger\|^2$ is approximately the same across all units, which amounts to  a (usually unrealistic) homoskedasticity condition---$\E[u_{il}^2|\balpha_i]$ is a constant $c_u$.
	Then, we can take $\rho_0=2c_u$ and $a_p=(\frac{\log n}{p})^{1/2}$.
	
	By contrast, for the pseudo-max distance, we average products of the proxy differences between two units and the proxy values of a third unit. Since units are independent of each other, we have
	$(\bX_{i\cdot}^\dagger-\bX_{j\cdot}^\dagger)'\bX_{\ell\cdot}^\dagger/p^\dagger\approx
	(\bH_{i\cdot}^\dagger-\bH_{j\cdot}^\dagger)'\bH_{\ell\cdot}^\dagger/p^\dagger$. Homoskedasticity of errors is unnecessary, and we can take
	$\rho_0=0$ and $a_p=(\frac{\log n}{p})^{1/2}$.
	
	For the distance of averages, the proxy differences are directly averaged: $\sum_{l\in\mathcal{C}^\dagger}[x_{il}-x_{jl}]/p^\dagger\approx\sum_{l\in\mathcal{C}^\dagger}[\eta_{l}(\balpha_i)-\eta_{l}(\balpha_j)]/p^\dagger$ since $\E[\bu_i|\balpha_i]=0$. Thus, we can take $\rho_0=0$ and $a_p=(\frac{\log n}{p})^{1/2}$.
\end{remark}

On the other hand, part (b) precisely links the distance of the noise-free mean structures with that of the latent confounders.
The upper bound requirement is mild. For example, if the latent functions $\eta_l$'s are Lipschitz continuous, we can take $\uvarsig=2$ for (squared) Euclidean distance and $\uvarsig=1$ for pseudo-max distance and distance of averages. 
By contrast, the lower bound is the key requirement for proxy informativeness, 
which needs to be understood in context and is not guaranteed by many proxies or smooth latent functions alone. 
The following remark discusses this requirement in greater detail.

\begin{remark}[Informativeness] \label{remark: informativeness}
	When the distance of averages is used, we typically need (the limit of) $\frac{1}{p^\dagger}\sum_{l\in\mathcal{C}^\dagger}\eta_l(\cdot)$ to be strictly monotonic. Then, we can take $\lvarsig=1$. This is possible only for a scalar $\alpha_i$.
	When there is more than one confounder, one may need to assume that each component of $\balpha_i$ is separately measured by a set of proxies. The averaged information across these sets then can be combined to define the final nearest-neighbor set.

	In addition, simple averaging may also lose useful information about confounders.
	For instance, consider a linear factor model: 
	$\eta_l(\alpha_i)=\varpi_l\alpha_i$ for some $\varpi_l\in\mathbb{R}$. 
	Strict monotonicity of the averaged $\eta_l$'s is violated if $\frac{1}{p^\dagger}\sum_{l\in\mathcal{C}^\dagger}\varpi_l\approx 0$.
	That is, the averaged proxies can be approximately uninformative about the latent confounder, even if
	a non-negligible subset of proxies have nonzero $\varpi_l$'s and thus are individually informative. By contrast,
	the Euclidean distance or pseudo-max distance can avoid this drawback. Generally, as long as $\balpha$  can generate sufficient variation in many (but not necessarily all) proxies and thus in the entire sequence $\{\eta_l(\balpha)\}_{l\in\mathcal{C}^\dagger}$, the lower bound requirement can be satisfied for Euclidean distance ($\lvarsig=2$) and pseudo-max distance ($\lvarsig=1$).
	For nonlinear factor models, this requires the entire proxy map to distinguish different latent values globally and to be locally informative about every confounding direction, while individual functions \(\eta_l(\cdot)\) need \textit{not} be monotonic; Section \ref{SA section: matching} of the SA provides precise conditions and examples. Also, 
	it is unnecessary to know the identities of informative proxies a priori,
	but if users do have some prior knowledge, a corresponding weighting scheme can be
	incorporated into the definition of the distance, which is completely accommodated in our framework.
\end{remark}

Next, we impose a high-level condition on the local approximation \eqref{eq: local factor structure}. It ensures that the local PCA output from the noisy proxy matrix $\bX_{\subi}^\ddagger$  can aid the approximation of the unknown functions $\mu(\cdot)$ and $e(\cdot)$ in the outcome and treatment equations. 
Recall that, by construction, $\bUpsilon_{\subi}:=(\bLambda_{\subi}'\bLambda_{\subi}/K)^{1/2}$ is diagonal, so we can write $\bUpsilon_{\subi}=\diag\{\upsilon_{1,\subi}, \cdots, \upsilon_{\ttd_i,\subi}\}$, where $\upsilon_{1, \subi}\geq \upsilon_{2,\subi}\geq\cdots\geq \upsilon_{\ttd_i,\subi}>0$. Thus, $\upsilon_{j,\subi}^2$ is the $j$th leading eigenvalue of $(\bH_{\subi}^\ddagger)'\bH_{\subi}^\ddagger/(p^\ddagger K)$. Let $\blambda_{j,\subi}$ denote the loading vector in $\bLambda_{\subi}$ corresponding to unit $j$, and recall $\sam=(K^{1/2}\wedge p^{1/2})/\sqrt{\log (n\vee p)}$.

\begin{assumption}[Local Approximation] \label{Assumption: LPCA}
	For each $i\in[n]$, the following conditions hold:
	\begin{enumerate}[label=(\alph*),  ref={\theassumption(\alph*)}]
    \item \label{assumption lpca a}
		$\max_{i\in[n]}\|\bF_{\subi}\|_{\max}\lesssim_\P1$ and $\underset{i\in[n]}{\max}\|\bLambda_{\subi}\bUpsilon_{\subi}^{-1}\|_{\max}\lesssim_\P 1$.
		Also, either $\upsilon_{j,\subi}/\upsilon_{j+1,\subi}\lesssim_\P 1$ or $\upsilon_{j+1,\subi}/\upsilon_{j,\subi}=o_\P(1)$ for $j\in[\ttd_i-1]$, and
		$\max_{i\in[n]}\sam^{-1}/\upsilon_{\ttd_i,\subi}=o_\P(1)$.
		\item 	\label{assumption lpca b}
		 $\rate{\eta}:=\|\bXi_{\subi}\|_{\max}=\bar{o}_\P(\upsilon_{\ttd_i,\subi})$, and  
		there exist some $\bfrb_{\subi}, \bfrc_{\subi}\in\mathbb{R}^{\ttd_i}$, uniformly bounded in probability over $i$, such that
		$\rate{\mu}:=\max_{j\in\mathcal{N}_i}|\mu(\balpha_j)-\bfrb_{\subi}'\blambda_{j,\subi}|=\bar{o}_\P(1)$ and 
		$\rate{e}:=\max_{j\in\mathcal{N}_i}|e(\balpha_j)-\bfrc_{\subi}'\blambda_{j,\subi}|=\bar{o}_\P(1)$.
	\end{enumerate}
\end{assumption}

Assumption \ref{Assumption: LPCA} is a high-level condition delineating the key signal-strength and approximation requirements for the local factor loadings $\bLambda_{\subi}$. We make several remarks on each part, and more detailed discussions, including sufficient conditions for Assumption \ref{Assumption: LPCA}, are available in Section \ref{SA section: lpca} of the SA.

First, $\bLambda_{\subi}$ can be viewed as an approximation basis
that characterizes the relation of different units in terms of $\balpha_i$. Thus, Assumption \ref{assumption lpca a} is a mild condition that specifies the possibly heterogeneous magnitude of different components of the basis. The requirement that  $\upsilon_{\ttd_i,\subi}$ is much larger than $\sam^{-1}$ is standard in the factor model literature \citep{Wang-Fan_2017_AoS}, implying that the weakest factor to be extracted has a stronger signal strength than the noise and thus is consistently estimable.

Second, Assumption \ref{assumption lpca b} quantifies the approximation accuracy for the three nonparametric components in our model,  $\bmeta(\cdot)$, $\mu(\cdot)$ and $e(\cdot)$, through the corresponding error bounds $\rate{\eta}$, $\rate{\mu}$, and $\rate{e}$. 
The key requirement in part (b) is that the same basis $\bLambda_{\subi}$ extracted from the noise-free proxy structure $\bH_{\subi}^\ddagger$ can also approximate $\mu(\cdot)$ and $e(\cdot)$ with vanishing errors.
The use of this common local basis is motivated by regular cases in which $\eta_l(\cdot)$, $\mu(\cdot)$, and $e(\cdot)$ are smooth functions of the same low-dimensional $\balpha_i$, so their local expansions share the same polynomial basis.
Together, parts (a) and (b) require the local basis to have sufficient signal strength for consistent recovery and sufficient approximation power for the outcome and treatment functions; the examples below illustrate these requirements.

\begin{exmp}[Linear factor model]\label{example: linear factor model}
	To fix ideas, consider a linear model $\eta_l(\balpha_i)=(\bm{C}\balpha_i)'\bm{\varpi}_l$ for some $\bm{\varpi}_l\in\mathbb{R}^{\ttr_\alpha}$. We introduce a diagonal matrix $\bm{C}=\diag\{c_1,\cdots, c_{\ttr_\alpha}\}$ to explicitly characterize the signal strength of factors. Assume that $\sum_{l\in\mathcal{C}^\ddagger}\bm{\varpi}_l\bm{\varpi}_l'/p^\ddagger$ and $\sum_{i=1}^n\balpha_i\balpha_i'/n$ are nondegenerate. As will be formalized below, when $K$-NN matching is primarily determined by the latent confounders, the neighborhood bandwidth $h_n$ is typically of order $(K/n)^{1/\ttr_\alpha}$.
	If $\sam^{-1}=o(h_n\min_j |c_{j}|)$ and $\ttd_i=\ttr_\alpha$, Assumption \ref{assumption lpca a} holds. Moreover, the linear factor representation is exact, so $\rate{\eta}$ in Assumption \ref{assumption lpca b} is exactly zero. Note that the extracted factor loadings capture local linear variation in $\balpha_i$ but need not include an exact constant. In practice, we can include an intercept in the local quasi-MLE; the intercept absorbs the common local level, while the extracted loadings continue to capture variation across neighboring units. Thus, if $\mu(\cdot)$ and $e(\cdot)$ are twice continuously differentiable, we can achieve local linear approximations of $\mu(\cdot)$ and  $e(\cdot)$, with $\rate{\mu}$ and $\rate{e}$ of order $(K/n)^{2/\ttr_\alpha}$.
	\hfill{\LARGE$\lrcorner$}
\end{exmp}

\begin{exmp}[Nonlinear models]\label{remark: monomial}
	Suppose that $\max_{j\in\mathcal{N}_i}\|\balpha_j-\balpha_i\|\lesssim_\P h_n$ for some  ``bandwidth'' $h_n$, and  $\mu(\cdot)$, $e(\cdot)$ and $\eta_l(\cdot)$'s are twice continuously differentiable with bounded derivatives. Let $\nabla\eta_l(\cdot)$ denote the gradient vector of $\eta_l(\cdot)$. Then, $|\eta_l(\balpha_j)-\eta_l(\balpha_i)- \nabla\eta_l(\balpha_i)'(\balpha_j-\balpha_i)|\lesssim_\P h_n^2$ for all $l\in[p]$.
	If $\frac{1}{p^\ddagger}\sum_{l\in\mathcal{C}^\ddagger}
	(\eta_l(\balpha_i), \nabla\eta_l(\balpha_i)')'(\eta_l(\balpha_i), \nabla\eta_l(\balpha_i)')$ is nondegenerate (i.e., the minimum eigenvalue is bounded away from zero), then we can let $\ttd_i=1+\ttr_\alpha$, and Assumption \ref{assumption lpca a} holds with $\upsilon_{1, \subi}\asymp_\P 1$ and $\upsilon_{j, \subi}\asymp_\P h_n$ for $j=2, \cdots, 1+\ttr_\alpha$. Accordingly, Assumption \ref{assumption lpca b} holds with $\rate{\eta}, \rate{\mu}$, and $\rate{e}$ of order $h_n^2$.
	The nondegeneracy condition on partial derivatives of $\eta_l$'s has an intuitive interpretation: the tangent plane  of
	the ``surface'' generated by $\bmeta(\cdot)$ does not collapse in any direction at the point $\balpha_i$ (so the proxies are indeed locally informative about the $\ttr_\alpha$ latent confounders).
	This nondegeneracy requirement is imposed jointly across proxies, and each individual $\eta_l(\cdot)$ need \textit{not} be monotonic.

	This idea extends to higher-order approximations. If each $\eta_l$ is $(m+1)$-times continuously differentiable and the local factors formed by the partial derivatives of $\eta_l$ up to order $m$ are nondegenerate, then the approximation errors $\rate{\eta}$, $\rate{\mu}$, and $\rate{e}$ are  $O(h_n^{m+1})$. The higher-order nondegeneracy condition may be restrictive for large $m$, since it requires sufficient nonlinearities of $\bmeta(\cdot)$; by contrast, it is less stringent when $m=0$ or $m=1$. See discussions in Section \ref{SA section: lpca} of the SA.
	\hfill{\LARGE$\lrcorner$}
\end{exmp}

\begin{remark}[Heterogeneous approximation power]\label{remark: het error}
	We allow the number of factors extracted $\ttd_i$ to vary across local neighborhoods. It accommodates the possibility that the nonlinear surface induced by $\bmeta(\cdot)$ may exhibit varying degrees of nonlinearity at different points. Relative to the noise, some approximation terms may yield signals too weak to be recovered from the proxies in specific regions.
	Accordingly, the ability of the local basis $\bLambda_{\subi}$ to approximate $\mu(\balpha_i)$ or $e(\balpha_i)$ may also be heterogeneous across units. 	For instance, if $\bmeta(\balpha)$ is almost linear at $\balpha=\balpha_i$, then the basis $\bLambda_{\subi}$ extracted from the proxies only captures the information about local constant and linear basis terms, and thus we cannot achieve higher-order approximation of $\mu(\balpha)$ at $\balpha=\balpha_i$. This feature is more specific to this causal inference problem. If the goal is to estimate the mean structure $\bH$ only, the recommended strategy in Remark \ref{remark: number of confounders} (extracting all principal components whose singular values exceed the noise edge) may allow us to approximately achieve $\rate{\eta}=O_\P(\sam^{-1})$ across all units.
	   	
	Two features of our method can alleviate the concern about such heterogeneity: (i) when $p$ is large, a mixture of less nonlinear proxies and highly nonlinear ones may still permit the characterization of higher-order basis terms; and (ii)  our main inference results only need the approximation errors to be sufficiently small in an  \textit{average} sense rather than uniformly over all units, so weak approximation power of $\bLambda_{\subi}$ is permitted in some (or even many) neighborhoods. See more discussion  in Section \ref{SA section: lpca} of the SA. 
\end{remark}

\subsection{Theoretical Results}
Recall that $K$ is the number of nearest neighbors, $p$ is the number of proxies, and
$\ttd_i$ is the number of leading local principal components extracted in the neighborhood of unit $i$. The asymptotic analysis is conducted assuming both $K$ and $p$ diverge as $n\rightarrow\infty$.

The indirect matching and local PCA are used to extract information on the latent variables $\balpha_i$. Specifically, under Assumptions \ref{Assumption: ID-counterfactual}--\ref{Assumption: Matching}, if $\frac{\log (n/K)}{K}=o(1)$ and $\frac{K\log n}{n}=o(1)$, it can be shown that
\[
\max_{i\in[n]}\max_{k\in[K]}
\|\balpha_i-\balpha_{j_k(i)}\|
\lesssim_\P (K/n)^{\uvarsig/(\lvarsig \ttr_\alpha)}+a_p^{1/\lvarsig}.
\]
This formalizes the idea that matching on the noisy proxies approximately matches the latent confounders as well. In particular, the implied matching discrepancy of nearest neighbors shrinks as $K/n\rightarrow0$ and $a_p\rightarrow 0$.
The first term, $(K/n)^{\uvarsig/(\lvarsig \ttr_\alpha)}$, is the usual discrepancy that would have existed even if the matching could be done
directly on $\balpha_i$. When $\lvarsig=\uvarsig$ (applicable to all three distances discussed above), the first term reduces to $(K/n)^{1/\ttr_\alpha}$, which is the implicit ``bandwidth'' for subsequent local estimation.
By contrast, the second term $a_p^{1/\lvarsig}$ arises from the fact that the matching can only be conducted on the noisy proxies rather than the latent confounders. 
As described in Remark \ref{remark: noise reduction}, typically, $a_p\rightarrow 0$   as $p\rightarrow\infty$ (more proxies).

	If Assumption \ref{Assumption: LPCA} also holds and $(np)^{\frac{2}{\nu}}\sam^{-2}\lesssim 1$, it can be further shown that
	\[
	\max_{i\in[n]}\frac{\|\widehat{\bLambda}_{\subi}-\bLambda_{\subi}\bR_{\subi}\|_{\max}}
	{\sam^{-1}+\rate{\eta}}\lesssim_\P 1,
	\]
for some rotation matrix $\bR_{\subi}$. This formalizes the idea that the extracted loading matrix $\widehat{\bLambda}_{\subi}$ from the noisy proxies ``consistently'' estimates the loading matrix $\bLambda_{\subi}$ that has approximation power for functions of the latent confounders.
The error bound shrinks as both $K$ and $p$ grow large and the approximation error $\rate{\eta}$ diminishes.  
Although not needed for our causal analysis, the same local PCA method can also recover the evaluated values $\{\eta_l(\balpha_i)\}_{i,l}$, i.e., the entries of the noise-free proxy matrix, but not the entire functions $\eta_l(\cdot)$ over their domains. Notice also that indirect matching and local PCA do not involve the outcome or treatment variables and thus, strictly speaking, the conditions on their errors in Assumptions \ref{Assumption: Regularities} and \ref{Assumption: indep of errors} are unnecessary for these results (in particular, the measure \(\mathfrak m_p\) of dependence-induced bias introduced in the SA plays no role).

Building on these properties of local PCA, we obtain our  first main result, which shows that the conditional means of potential outcomes and the propensity scores can be consistently estimated by the proposed local QMLE method.

\begin{thm}[Local QMLE]\label{thm: QMLE}
	Suppose that Assumptions \ref{Assumption: ID-counterfactual}--\ref{Assumption: LPCA} hold.
	If $(np)^{\frac{2}{\nu}}\sam^{-2}\lesssim 1$, then  
	$$
	\max_{i\in[n]}\frac{|\widehat{\zeta}_{i}-\zeta_{i}|}
	{\sam^{-1}+\rate{\eta}+\rate{\mu}}\lesssim_\P 1\quad \text{and}\quad
	\max_{i\in[n]}\frac{|\widehat{\varrho}_{i}-\varrho_{i}|}
	{\sam^{-1}+\rate{\eta}+\rate{e}}\lesssim_\P 1. 
	$$
	A detailed asymptotic expansion is given in Equation \eqref{eq: asymptotic expansion} in the SA.
\end{thm}
Since the functional forms of $\mu(\cdot)$ and $e(\cdot)$ are unspecified, Theorem \ref{thm: QMLE} exhibits the familiar bias–variance structure of nonparametric convergence rates: 
$\sam^{-1}$ is the variance term, and 
$\rate{\mu}$ and $\rate{e}$ reflect the smoothing bias. 
The major difference here is that the convergence rate also depends on the number of proxies $p$ and the approximation error $\rate{\eta}$, reflecting the fact that the information on $\balpha_i$ is indirectly obtained via local PCA.
Also, the convergence above is uniform over all the data points indexed by $i$, which respects the fact that $\balpha_i$ is not directly observed and we obtain information on it for the $n$ units in the dataset. Such a sample-point uniform convergence result differs from the whole-support uniformity derived (or assumed directly) in classical nonparametric or semiparametric estimation and is stronger than the $L_2$ convergence rates commonly established or imposed for nuisance estimators in high-dimensional and double/debiased machine learning analyses \citep[e.g.,][]{Chernozhukov-et-al_2018_EJ}.

\begin{remark}[CATE estimator]\label{remark: CATE}
	In this setup, a conditional average treatment effect given the confounders is defined as $\E[y_i(1)-y_i(0)|\balpha_i, \bz_i]$. Theorem \ref{thm: QMLE} shows that $\widehat{\zeta}_i$ is a uniformly consistent estimator of $\E[y_i(0)|\balpha_i, \bz_i]$. The same method can be applied to estimate $\E[y_i(1)|\balpha_i, \bz_i]$. This allows us to construct a valid CATE estimator evaluated at each individual's confounder values, with convergence uniform over the observed units $i\in[n]$, and to explore treatment effect heterogeneity along different dimensions of the confounders.
\end{remark}

Now, we are ready to apply the previous results to inference on the counterfactual means. The following theorem, our second main result, 
establishes the asymptotic normality of the proposed estimator. 

\begin{thm}[Causal Inference] 
	\label{thm: pointwise inference}
	Suppose that Assumptions \ref{Assumption: ID-counterfactual}--\ref{Assumption: LPCA} hold. If 
	$(np)^{\frac{2}{\nu}}\sam^{-2}\lesssim 1$ and 
	$\sqrt{n\log n}(\sam^{-2}+
	\sum_{i\in[n]}[\rate{\eta}^2+\rate{\mu}^2+\rate{e}^2]/n
	)=o_\P(1)$, then
	\begin{enumerate}
		\item 
		$\sqrt{n}(\widehat{\theta}-\theta)=
		\frac{1}{\sqrt{n}}\sum_{i=1}^{n}\varphi_{i}
		+o_\P(1)$, 
		where $\varphi_{i}=
		\frac{t_i(\zeta_{i}-\theta)}{\varrho}+
		\frac{\varrho_{i}}{\varrho}\frac{(1-t_i)(y_{i}-\zeta_{i})}{1-\varrho_{i}}$;
		
		\item 	If $\sigma^2:=\V[\varphi_i]>0$, then $\sqrt{n}(\widehat{\theta}-\theta)/\widehat{\sigma}\rightsquigarrow\mathsf{N}(0,1)$.
		
	\end{enumerate}	
\end{thm}

Theorem \ref{thm: pointwise inference} takes advantage of the doubly-robust score function and the properties of our first-step estimator to relax  conditions on the convergence rates of $\widehat{\zeta}_i$ and $\widehat{\varrho}_i$, while not requiring sample splitting across units. 
Specifically, the second rate restriction essentially requires 
the squared estimation errors---and hence their product---to be of a smaller order than $n^{-1/2}$,
which is consistent with results in the double/debiased machine learning literature.
Importantly, the approximation errors in our model, characterized by $\rate{\eta}$, $\rate{\mu}$ and $\rate{e}$, 
need \textit{not} be small uniformly over all units.
Instead, we only require that the \textit{average} squared errors be small, 
allowing for cases in which the nearest neighbors and local factor loadings from the first step may not be  
highly informative about the latent confounders and yield poor approximations for some units.
 
To get a better sense of the rate restrictions imposed, consider a simple but practically relevant case 
with $K\lesssim p$, $\nu$ sufficiently large (e.g., sub-Gaussian errors), and all the unknown functions $\eta_l$'s, $\mu$ and $e$ sufficiently smooth. Then, $\sqrt{n\log n}\sam^{-2}=o(1)$ reduces to $n/K^2=o(1)$, up to $\log n$ terms.
On the other hand, in the case of local linear approximation in Example \ref{remark: monomial},
$\rate{\eta}$, $\rate{\mu}$ and $\rate{e}$ are typically $O((K/n)^{2/\ttr_\alpha})$, and thus we roughly need
$\sqrt{n}(K/n)^{4/\ttr_\alpha}=o(1)$. If we use the recommended strategy in Remark \ref{remark: number of confounders} and extract all approximation terms with signal strength exceeding the noise level, then in the best case $\rate{\eta}, \rate{\mu}$ and $\rate{e}$ can be as small as $O(\sam^{-1})$. 
Then, all restrictions imposed in Theorem \ref{thm: pointwise inference} reduce to $n/K^2=o(1)$, up to $\log n$ terms.

Note that this paper focuses on large-$K$ asymptotics, which is analogous to a large bandwidth in kernel estimation or a small number of approximation terms in series estimation. If $K$ is small relative to the sample size, a non-negligible undersmoothing bias may arise in the distributional approximation, and bias-robust inference may be needed. See  \cite*{Cattaneo-Jansson_2018_ECMA,Cattaneo-Jansson-Ma_2019_REStud,Matsushita-Otsu_2019_wp} for further  discussion of undersmoothing bias and possible solutions.

\subsection{Numerical Results} 
\label{sec: simulation}

We conducted Monte Carlo experiments to check the finite sample performance of our proposed method. The SA contains complete results, and here we give a short summary. We consider (i) three specifications for $\bmeta(\cdot)$ with one latent confounder ($\ttr_\alpha=1$) and three with two latent confounders ($\ttr_\alpha=2$), with varying degrees of nonlinearity and smoothness; (ii) different sample size  combinations $(n,p)=(500, 125), (500, 250), (500, 500), (500, 750)$;  (iii) three specifications for the measurement errors $u_{il}$: i.i.d., conditionally heteroskedastic, and autocorrelated across $l$ (and correlated with $\epsilon_i$); and (iv) different choices of the tuning parameter $K$. 
We also compare our method with three alternatives in this setting: (i) the usual doubly robust estimator that controls for two $x_{il}$'s directly, which mimics the practice of naively treating the noisy proxies as the true confounders (``DR''); (ii) the doubly robust estimator that employs double Lasso to select controls among the high-dimensional $\bx_i$ (``DR-DL''); and (iii) the doubly robust estimator that uses local average smoothing to estimate conditional means of outcomes and propensity scores (i.e., taking local averages over nearest neighbors without the PCA step).
The idea of local average smoothing has been proposed in the panel and network data literature for different purposes \citep[e.g.,][]{Zhang-Levina-Zhu_2017_BIMA,Bonhomme-Lamadon-Manresa_2021_wp}.

To conserve space, we only summarize results for one setup. As shown by our theory, the key feature that determines the performance of our method is the signal strength in the nonlinear factor model, reflected by the effective rank of the noise-free matrix $\bH$.
Therefore, we consider the Gaussian kernel $\eta_l(\alpha_i)=2\exp(-(\alpha_i-\varpi_l)^2/\gamma)$, for which we can conveniently change the smoothness of $\bmeta(\cdot)$ by varying $\gamma$. Typically, decreasing $\gamma$ makes $\bH$ more complex with more non-negligible singular values (``high-rank''). We set $\gamma=0.5$, $0.1$, and $0.02$, corresponding to the low-, mid-, and high-rank designs in Table \ref{table:simul}, respectively. The measurement error $u_{il}$, given $\alpha_i$ and  $\varpi_l$, is independent over $i$ and $l$ and follows a zero-mean normal distribution with the conditional variance depending on $\alpha_i$ and $\varpi_l$. 
Also, $\E[y_i(0)|\alpha_i]=\alpha_i+\alpha_i^2$ and  $\E[y_i(1)|\alpha_i]=2\alpha_i+\alpha_i^2+1$, and conditional on $\alpha_i$, the treatment $t_i$ is Bernoulli with success probability $\varrho_i=\psi((\alpha_i-0.5)+(\alpha_i-0.5)^2)$. $\alpha_i, \varpi_l\sim\mathsf{U}[0,1]$, and the error terms in $y_{i}(0)$ and $y_{i}(1)$ are standard normal.
We consider 2\,000 simulated datasets with $n=500$ and $p=250$, which approximately match the dimensions of our empirical application. 
For each simulated dataset, we obtain a point estimate of ATT, and report root mean squared error (RMSE), coverage rate (CR) of the nominal 95\% confidence interval and its average length (AL) in Table \ref{table:simul}. 

The results in the first five columns are obtained using our proposed method, using the same strategy as described in 
our empirical illustration. Specifically, we take the first half ($p^\dagger=125$) of proxies for matching 
and the second half ($p^\ddagger=125$) for extracting the local factor loadings. 
We let $K=Cn^{4/5}$ for $C\in\{0.5, 1, 1.5, 2\}$, which has the MSE-optimal rate for the (infeasible) cross-sectional local constant/linear nearest-neighbor regression of $y_i$ on $\alpha_i$ and satisfies the restriction for our inference theory. We also consider a data-dependent choice $\widehat{K}_{\tt CV}$, which is obtained by  20-fold cross-validation for the local nearest-neighbor regression of $y_i$ on one noisy proxy $x_{ip}$. 
We choose the largest $\ttd_i$ such that the extracted factors are associated with singular values greater than $C_{np}(\sqrt{K}+\sqrt{p^\ddagger})$ for some $C_{np}>0$. For comparison, we report results based on DR and DR-DL in the last two columns.

Our method performs relatively well throughout the three specifications and is robust to a wide range of choices of $K$. By contrast, DR and DR-DL have unstable performance across different specifications, do not achieve the nominal coverage rate and usually have wider confidence intervals. The results in the SA further show that (i) the local average smoothing has more unstable performance across choices of $K$ and suffers from more severe undercoverage in more complex models (e.g., with multiple confounders); and (ii) larger $p$ improves the performance of our method, while for small $p$, the proposed method may undercover but still have better performance compared to DR and DR-DL.

\begin{table}[h]
	\renewcommand{\arraystretch}{1.25}
	\centering
	\scriptsize
	\caption{Simulation Results (heteroskedastic error, $n=500$, $p=250$)}
	\label{table:simul}
	\resizebox{.8\textwidth}{1.17\height}{\scriptsize \begin{tabular}{lccccccccc}
\hline\hline
\multicolumn{1}{l}{\bfseries }&\multicolumn{5}{c}{\bfseries Local PCA, $K=$}&\multicolumn{1}{c}{\bfseries }&\multicolumn{1}{c}{\bfseries DR}&\multicolumn{1}{c}{\bfseries }&\multicolumn{1}{c}{\bfseries DR-DL}\tabularnewline
\cline{2-6} \cline{8-8} \cline{10-10}
\multicolumn{1}{l}{}&\multicolumn{1}{c}{73}&\multicolumn{1}{c}{145}&\multicolumn{1}{c}{217}&\multicolumn{1}{c}{289}&\multicolumn{1}{c}{$\widehat{K}_{\mathtt{CV}}$}&\multicolumn{1}{c}{}&\multicolumn{1}{c}{}&\multicolumn{1}{c}{}&\multicolumn{1}{c}{}\tabularnewline
\hline
{\bfseries low-rank}&&&&&&&&&\tabularnewline
~~RMSE&$0.100$&$0.098$&$0.098$&$0.097$&$0.099$&&$0.173$&&$0.187$\tabularnewline
~~CR&$0.943$&$0.942$&$0.949$&$0.950$&$0.944$&&$0.867$&&$0.838$\tabularnewline
~~AL&$0.378$&$0.375$&$0.374$&$0.373$&$0.375$&&$0.509$&&$0.509$\tabularnewline
\hline
{\bfseries mid-rank}&&&&&&&&&\tabularnewline
~~RMSE&$0.096$&$0.094$&$0.093$&$0.093$&$0.094$&&$0.156$&&$0.147$\tabularnewline
~~CR&$0.947$&$0.951$&$0.951$&$0.951$&$0.951$&&$0.916$&&$0.915$\tabularnewline
~~AL&$0.375$&$0.372$&$0.372$&$0.373$&$0.373$&&$0.501$&&$0.481$\tabularnewline
\hline
{\bfseries high-rank}&&&&&&&&&\tabularnewline
~~RMSE&$0.100$&$0.098$&$0.099$&$0.102$&$0.099$&&$0.177$&&$0.196$\tabularnewline
~~CR&$0.938$&$0.940$&$0.944$&$0.940$&$0.940$&&$0.861$&&$0.823$\tabularnewline
~~AL&$0.376$&$0.375$&$0.377$&$0.381$&$0.376$&&$0.511$&&$0.517$\tabularnewline
\hline
\end{tabular}
}
	\flushleft\footnotesize{\textit{Notes}:
		RMSE = root MSE of point estimator, CR = coverage rate of $95\%$ nominal confidence intervals, AL = average interval length of $95\%$ nominal confidence intervals. $\widehat{K}_{\mathtt{CV}}$= cross-validation choice of $K$.
	}\newline
\end{table}

\section{Extension: Uniform Inference} \label{sec: extensions}
In many applications, the outcome of interest is a certain transformation of  the original potential outcome via a function $g(\cdot)\in\mathcal{G}$, and uniform inference over the function class $\mathcal{G}$ is desired. In general, the goal can be achieved in two steps: (i) strengthen the asymptotic expansion in Theorem \ref{thm: pointwise inference}(a) to be uniform, i.e., the remainder needs to be negligible uniformly over $g\in\mathcal{G}$; and (ii) show that the empirical process of influence functions indexed by $\mathcal{G}$ weakly converges to a limiting process. The general treatment of such issues can be found in, e.g.,   \cite{Barrett-Donald_2003_ECMA, Chernozhukov_2013_ECMA,Donald-Hsu_2014_JoE}. 

We will focus on counterfactual distributions, the analysis of which relies on a
particular function class $\mathcal{G}=\{y\mapsto \I(y\leq \tau):\tau\in\mathcal{Y}\}$ for some compact set $\mathcal{Y}\subset\mathbb{R}$. Each $g(\cdot)\in\mathcal{G}$ corresponds to a particular value $\tau\in\mathcal{Y}$. Therefore, we write $y_{i,\tau}(t)=\I(y_i(t)\leq \tau)$, $t=0,1$, and $y_{i,\tau}=\I(y_i\leq \tau)$. 
One may be interested in 
$\E[\I(y_i(t)\leq \tau)|t_i=\tilde{t}\,]$, the counterfactual distribution function of the potential outcome $y_i(t)$ for the group with treatment status $\tilde{t}$, which is a parameter in $\ell^\infty(\mathcal{Y})$, a function space of bounded functions on $\mathcal{Y}$ equipped with the supremum norm.
For now, we focus on  $\theta(\tau)=\E[\I(y_i(0)\leq \tau)|t_i=1]$ in our binary treatment framework, and defer the general case to the SA. Now, Equation \eqref{eq: outcome} becomes 
$$
y_{i,\tau}(0)=\zeta_{i,\tau}+\epsilon_{i,\tau},\quad\zeta_{i,\tau}:=\P(y_i(0)\leq \tau|\balpha_i, \bz_i)=\mu_{\tau}(\balpha_i)+\bz_i'\bbeta_{\tau}.
$$
For each $\tau$, the second step of the estimation procedure in Section \ref{sec: estimation} is implemented to obtain an estimator $\widehat{\zeta}_{i,\tau}$ of $\zeta_{i,\tau}$. 

To establish the limiting distribution of the proposed estimator, we impose an additional assumption.

\begin{assumption}[Uniform Inference] \label{Assumption: QMLE, uniform}
$\mu_{\tau}(\balpha)$ is Lipschitz in $\tau$ uniformly over $\balpha$;
for all $\tau\in\mathcal{Y}$, there exists some $\bfrb_{\tau,\subi}\in\mathbb{R}^{\ttd_i}$, uniformly bounded in probability over $i$ and $\tau$, such that 
$\barrate{\mu}:=\sup_{\tau\in\mathcal{Y}}\max_{j\in\mathcal{N}_i}|\mu_{\tau}(\balpha_j)-\bfrb_{\tau,\subi}'\blambda_{j,\subi}|=\bar{o}_\P(1)$.
\end{assumption}

The following theorem shows that the (scaled) counterfactual distribution process weakly converges to a limiting Gaussian process indexed by $\tau\in\mathcal{Y}$, which forms the basis of uniform inference. See \cite{vandeVaart_book_1996} for underlying technical details.
\begin{thm}[Uniform Inference]\label{thm: uniform inference}
	Suppose that Assumptions \ref{Assumption: ID-counterfactual}--\ref{Assumption: QMLE, uniform} hold, $(np)^{\frac{2}{\nu}}\sam^{-2}\lesssim 1$, and   
	$\sqrt{n\log n}(\sam^{-2}+\allowbreak
	\sum_{i\in[n]}[\rate{\eta}^2+\barrate{\mu}^2+\rate{e}^2]/n
	)=o_\P(1)$. Then,
	\[
	\sqrt{n}\Big(\widehat{\theta}(\cdot)-\theta(\cdot)\Big)=
	\frac{1}{\sqrt{n}}\sum_{i=1}^n\varphi_{i}(\cdot)+o_\P(1)\rightsquigarrow \mathsf{Z}(\cdot) \quad \text{in } \; \ell^\infty(\mathcal{Y}),
	\]
where $\varphi_{i}(\cdot)=
\frac{t_i(\zeta_{i,\cdot}-\theta(\cdot))}{\varrho}+
\frac{\varrho_{i}}{\varrho}\frac{(1-t_i)(y_{i,\cdot}-\zeta_{i,\cdot})}{1-\varrho_{i}}$ and $\mathsf{Z}(\cdot)$ is a zero-mean Gaussian process with covariance kernel $\E[\varphi_{i}(\tau_1)\varphi_{i}(\tau_2)]$
for $\tau_1,\tau_2\in\mathcal{Y}$.
\end{thm}

Under proper regularity conditions, the weak convergence above can be applied to construct inference procedures for other quantities such as quantile treatment effects by the functional delta method, as shown in the following corollary.

\begin{coro}[Functional Delta Method]\label{coro: functional delta method}
	Let the conditions in Theorem \ref{thm: uniform inference} hold. Consider the parameter $\tilde\theta$ as an element of a
	parameter space $D_{\tilde\theta}\subseteq \ell^\infty(\mathcal{Y})$, a function space of bounded functions on $\mathcal{Y}$, with $D_{\tilde\theta}$ containing the true value $\theta$. Let a functional $\Psi(\tilde\theta)$ mapping $D_{\tilde\theta}$ to $\ell^\infty(\mathcal{Q})$ for some index set $\mathcal Q$ be Hadamard differentiable in $\tilde\theta$ at $\theta$ with derivative
	$\dot{\Psi}_\theta$. Then,
	$\sup_{q\in\mathcal Q}|\sqrt{n}\{\Psi(\widehat{\theta})(q)-\Psi(\theta)(q)\}-
	n^{-1/2}\sum_{i=1}^{n}\dot{\Psi}_\theta(\varphi_{i})(q)|=o_\P(1)$ and $\sqrt{n}(\Psi(\widehat{\theta})(\cdot)-\Psi(\theta)(\cdot))\rightsquigarrow\dot{\Psi}_\theta(\mathsf{Z})(\cdot)$ that is a Gaussian process  in $\ell^\infty(\mathcal{Q})$ with mean zero and covariance kernel defined by the limit of the second
	moment of $\dot{\Psi}_\theta(\varphi_{i})$.
\end{coro}

The limiting Gaussian process can be approximated based on a practically feasible multiplier bootstrap procedure widely used in the literature. To be specific, take an i.i.d. sequence of random variables $\{\omega_i\}_{i\in[n]}$ independent of the data with mean zero and variance one. Define a uniformly consistent estimator of $\varphi_{i}(\cdot)$:
$$\widehat{\varphi}_{i}(\cdot)=
\frac{t_i(\widehat{\zeta}_{i,\cdot}-\widehat{\theta}(\cdot))}{\widehat{\varrho}}+
\frac{\widehat{\varrho}_{i}}{\widehat{\varrho}}\frac{(1-t_i)(y_{i,\cdot}-\widehat{\zeta}_{i,\cdot})}{1-\widehat{\varrho}_{i}}.$$ 
The following corollary shows that conditional on the data, $\frac{1}{\sqrt{n}}\sum_{i=1}^n\omega_i\widehat{\varphi}_{i}(\cdot)$ weakly converges to the same limiting process $\mathsf{Z}(\cdot)$ as in Theorem \ref{thm: uniform inference}. In practice, one only needs to simulate this feasible approximation process
by taking random draws of $\{\omega_i\}_{i\in [n]}$.

\begin{coro}[Multiplier Bootstrap] \label{coro: multiplier bootstrap}
	Let the conditions of Theorem \ref{thm: uniform inference} hold. Then, conditional on the data, $n^{-1/2}\sum_{i=1}^{n}\omega_i\widehat{\varphi}_{i}(\cdot)\rightsquigarrow\mathsf{Z}(\cdot)$ that is the Gaussian process defined in Theorem \ref{thm: uniform inference}, with probability approaching one.
\end{coro}

To showcase the uniform inference procedure, we use the data from \cite{Acemoglu-et-al_2016_JFE} to check the (first-order) stochastic dominance (SD) of $\theta_1(\cdot)$ over $\theta_0(\cdot)$, where $\theta_1(\cdot)$ and $\theta_0(\cdot)$ denote, respectively, the cumulative distribution functions (CDFs) of potential stock returns of firms connected to Geithner under connection and no connection.
By definition of SD, the null hypothesis is $\theta_1(\tau)\leq \theta_0(\tau)$ for all $\tau\in\mathcal{Y}$. An intuitive test statistic is $\sqrt{n}\sup_{\tau\in\mathcal{Y}}(\widehat{\theta}_1(\tau)-\widehat{\theta}_0(\tau))$. The null hypothesis is rejected if the test statistic is greater than a certain critical value. Given the asymptotic expansions of  the corresponding estimators $\widehat{\theta}_1(\cdot)$ and $\widehat{\theta}_0(\cdot)$, the critical value can be obtained by simulating the supremum of the approximation process. In practice, the supremum over the whole support is simply replaced by the maximum over a set of user-specified evaluation points.

\begin{figure}[!h]
	\centering
	\small
	\caption{Estimated CDFs of Potential Outcomes for the Treated}
	\includegraphics[width=0.67\textwidth]{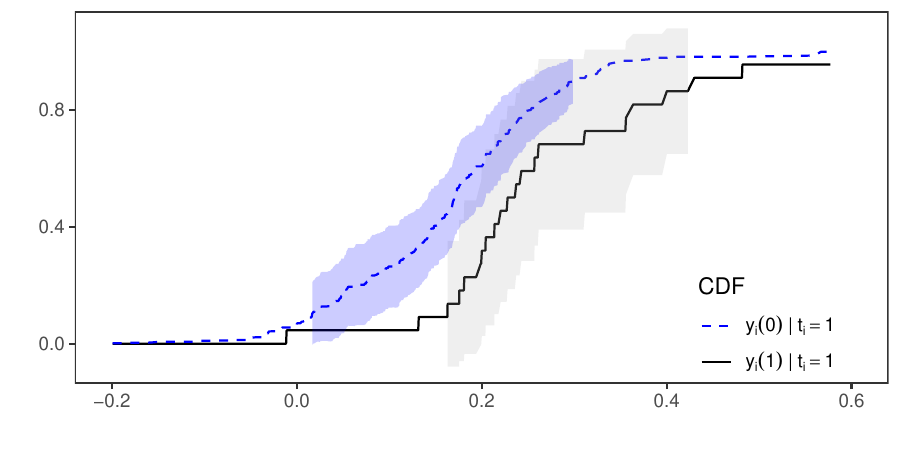}
	\label{figure:SD}
	{\footnotesize\noindent\parbox{\linewidth}{%
			\textit{Notes}: This figure plots the estimated cumulative distribution functions of potential returns for connected firms. Shaded regions are 95\% confidence bands.
	}}
\end{figure}

For each $\tau\in\mathcal{Y}$, we implement the estimation procedure as described in Section \ref{sec: estimation}. 
Varying the values of $\tau$, we obtain two estimated distribution functions for firms with connections, as shown in Figure \ref{figure:SD}. 
To better understand the estimation uncertainty, $95\%$ confidence bands for the two estimated CDFs are plotted, which are based on simulating the maximum absolute value of the corresponding (studentized) approximation processes. In each case, the value of $\tau$ is restricted to the range from the 10th to the 90th percentile of the estimated distribution.

It turns out that the estimated CDF for the treated outcome is well below that for the untreated outcome. Formally, we take all observed values of the cumulative returns as the evaluation points, and then simulate the maximum of the approximation process by taking $500$ draws of random weights $\{\omega_i\}_{i=1}^n$. In this simple example, the test statistic equals $0$, which is well below the critical value $5.81$ for a confidence level of $0.95$ obtained through simulation. Thus, SD of $\theta_1(\cdot)$ over $\theta_0(\cdot)$ cannot be rejected. This finding is consistent with the positive effects of political connections in this example being felt over the entire distribution of the stock returns of connected firms, which is a stronger conclusion than that based simply on the mean in Section \ref{sec: estimation}.
\smallskip

\FloatBarrier
\FloatBarrier

\section{Conclusion} \label{sec: conclusion}

This paper has developed a causal inference method for treatment effects models with some confounders not directly observed. Relevant information on these latent confounders is extracted from a large set of noisy proxies that admits an unknown, possibly nonlinear factor structure. Such information is then used to match comparable units in the subsequent counterfactual analysis. Large-sample properties of the proposed estimators are established. The results cover a large class of causal parameters, including average treatment effects and counterfactual distributions. The method is illustrated with an empirical application studying the effect of political connections on stock returns of financial firms.

\bibliography{Feng_2026--Bibliography}
\bibliographystyle{jasa}

\clearpage
\clearpage
\hypersetup{pageanchor=false}
\onehalfspacing

\setcounter{section}{0}
\setcounter{equation}{0}
\setcounter{figure}{0}
\setcounter{table}{0}
\setcounter{thm}{0}
\setcounter{lem}{0}
\setcounter{coro}{0}
\setcounter{assumption}{0}
\setcounter{definition}{0}

\renewcommand{\tts}{\scriptscriptstyle\mathsf{D}}
\renewcommand{\tty}{\scriptscriptstyle\mathsf{Y}}
\setenumerate[1]{label=\bf(\alph*)}

\renewcommand{\thesection}{SA-\arabic{section}}
\renewcommand{\theequation}{SA-\arabic{equation}}
\renewcommand{\thefigure}{SA-\arabic{figure}}
\renewcommand{\thetable}{SA-\arabic{table}}
\renewcommand{\thelem}{SA-\arabic{lemma}}
\renewcommand{\thethm}{SA-\arabic{theorem}}
\renewcommand{\thecoro}{SA-\arabic{coro}}
\renewcommand{\theassumption}{SA-\arabic{assumption}}
\renewcommand{\thedefinition}{SA-\arabic{definition}}

\numberwithin{equation}{section}
\numberwithin{thm}{section}
\numberwithin{lem}{section}
\numberwithin{assumption}{section}
\numberwithin{coro}{thm}

\renewcommand{\theHsection}{SA.\arabic{section}}
\renewcommand{\theHsubsection}{\theHsection.\arabic{subsection}}
\renewcommand{\theHsubsubsection}{\theHsubsection.\arabic{subsubsection}}
\renewcommand{\theHequation}{\theHsection.\arabic{equation}}
\renewcommand{\theHfigure}{SA.\arabic{figure}}
\renewcommand{\theHtable}{SA.\arabic{table}}
\renewcommand{\theHthm}{\theHsection.\arabic{thm}}
\renewcommand{\theHlem}{\theHsection.\arabic{lem}}
\renewcommand{\theHcoro}{\theHthm.\arabic{coro}}
\renewcommand{\theHassumption}{\theHsection.\arabic{assumption}}
\renewcommand{\theHdefinition}{SA.\arabic{definition}}
\renewcommand{\theHexmp}{\theHsection.\arabic{exmp}}
\expandafter\def\csname theHremark_tmp\endcsname{\theHsection.\arabic{remark_tmp}}

\makeatletter
\let\maketitle\arxivmaketitle
\let\@maketitle\arxivinternalmaketitle
\let\title\arxivtitle
\let\author\arxivauthor
\let\date\arxivdate
\let\thanks\arxivthanks
\let\and\arxivand
\makeatother

\title{\vspace{-0.25in}\Large Causal Inference in Possibly Nonlinear Factor Models\\
	\medskip
	Supplemental Appendix
\bigskip }
\author{
	Yingjie Feng\thanks{School of Economics and Management, Tsinghua University, Beijing 100084, China.}}
\date{\today}
\maketitle

\vspace{1em}

\begin{abstract}
	This supplement provides theoretical proofs of the results in the main paper, additional simulation results, as well as discussions about key assumptions used in the paper and other extensions. Section \ref{SA section: condition on x} uses a concrete example to illustrate the problems of methods that directly condition on noisy proxies. Section \ref{SA section: discuss assumptions} gives more discussions about Assumptions \ref{Assumption: Regularities}--\ref{Assumption: LPCA} imposed in the paper.  Section \ref{SA sec: general framework} sets up a multi-valued treatment effect framework, generalizing the one presented in the main paper, and describes the corresponding estimation procedure. Section \ref{SA section: theory} gives the theoretical results, and Section \ref{SA-sec: numerical} provides additional numerical results.
Section \ref{SA-sec: covariates adjustment} presents an adjusted estimation procedure 
when additional regressors are included in the nonlinear factor model for noisy proxies.
Section \ref{SA-sec: proofs} presents all proofs of results in the main paper as well as some useful technical lemmas.
\end{abstract}

\thispagestyle{empty}

\newpage
\makeatletter
\section*{\contentsname
	\@mkboth{\MakeUppercase\contentsname}{\MakeUppercase\contentsname}}
\@input{Feng_2026--Supplement-Contents.tex}
\makeatother

\thispagestyle{empty}
\clearpage

\setcounter{page}{1}
\pagestyle{plain}


\section{Conditioning on Noisy Proxies}\label{SA section: condition on x}
In the main paper we use a simple example to illustrate the problems of methods that directly condition on noisy proxies. This section provides a more complete discussion about an example that fits into our setup.

Suppose that
\begin{equation}\label{eq: Gaussian model}
	y_i(0)=\alpha_i+\epsilon_i, \quad t_i=\I(v_i\leq e(\alpha_i)),\quad x_{il}=\alpha_i+u_{il},\; l\in[p],
\end{equation}
where
\begin{align*}
	&\begin{bmatrix}
		\epsilon_i\\
		u_{i1}\\
		\vdots\\
		u_{ip}
	\end{bmatrix}
	\sim \mathsf{N}\left(\bm{0}, 
	\begin{bmatrix}
		1 &\bm{0}_{p-1}'& \sigma_{p\epsilon}\\
		\bm{0}_{p-1}& \bm{I}_{p-1} & \bm{0}_{p-1}\\
		\sigma_{p\epsilon}&\bm{0}_{p-1}'& 1 
	\end{bmatrix}
	\right),\qquad
	\alpha_i,v_i\sim \mathsf{U}[0,1],\\ 
	&(\epsilon_i, v_i, u_{i1},\cdots, u_{ip})\inde \alpha_i,\qquad
	v_i\inde (\epsilon_i,u_{i1},\cdots, u_{ip}),
\end{align*}
where $|\sigma_{p\epsilon}|\in[0,1)$, $\bm{0}_{p-1}$ denotes a $(p-1)$-vector of zeros, $\bI_{p-1}$ is an identity matrix of size $p-1$, $\mathsf{N}(\ba, \bSigma)$ denotes the multivariate normal distribution with mean $\ba$ and covariance matrix $\bSigma$, and $\mathsf{U}[0,1]$ denotes the uniform distribution over $[0,1]$.
That is, the outcome error $\epsilon_i$ is independent of  errors in all proxies but the last one. This setup, though simple, has a realistic motivation: in panel data settings the noisy proxies are usually outcome variables collected in history, and the outcome in the current period of interest is likely to be strongly correlated with those in the most recent periods.

In this case 
\begin{align*}
	(\epsilon_i,\bx_i)'|\alpha_i\sim\mathsf{N}\left(
	\begin{bmatrix}
		0\\
		\alpha_i\bm{1}_p
	\end{bmatrix}, 
	\begin{bmatrix}
		1 &\bm{0}_{p-1}'& \sigma_{p\epsilon}\\
		\bm{0}_{p-1}& \bm{I}_{p-1} & \bm{0}_{p-1}\\
		\sigma_{p\epsilon}&\bm{0}_{p-1}'& 1 
	\end{bmatrix}
	\right),
\end{align*}
where $\bm{1}_p$ denotes the $p$-vector of ones.

The conditional density of $\alpha_i$ given $\bx_i$ is proportional to
\[
\exp\Big(-\frac{p}{2}(\alpha-\bar{x}_i)^2\Big)\I(\alpha\in[0,1]),
\]
i.e., $\alpha_i$ given $\bx_i$ follows a normal distribution with mean  $\bar{x}_i=\frac{1}{p}\sum_{l=1}^px_{il}$ and variance $p^{-1}$ truncated  to $[0,1]$. Therefore,
\begin{align*}
	\E[y_i(0)|\bx_i]&=\E[\alpha_i+\E[\epsilon_i|\bx_i,\alpha_i]|\bx_i]\\
	&=\E[\alpha_i+\sigma_{p\epsilon}\biota_p'(\bx_i-\alpha_i\bm{1}_p)|\bx_i]\\
	&=\sigma_{p\epsilon}x_{ip}+(1-\sigma_{p\epsilon})\E[\alpha_i|\bx_i]\\
	&=\sigma_{p\epsilon}x_{ip}+(1-\sigma_{p\epsilon})\Big[\bar{x}_i+\frac{1}{\sqrt{p}}\frac{\phi(\sqrt{p}\bar{x}_i)-\phi(\sqrt{p}(1-\bar{x}_i))}{\Phi(\sqrt{p}(1-\bar{x}_i))-\Phi(\sqrt{p}(-\bar{x}_i))}\Big]\\
	&=\E[y_i(0)|\alpha_i]+\sigma_{p\epsilon}u_{ip}+(1-\sigma_{p\epsilon})\Big[\bar{u}_i+\frac{1}{\sqrt{p}}\frac{\phi(\sqrt{p}\bar{x}_i)-\phi(\sqrt{p}(1-\bar{x}_i))}{\Phi(\sqrt{p}(1-\bar{x}_i))-\Phi(\sqrt{p}(-\bar{x}_i))}\Big],
\end{align*} 
where $\biota_p$ denotes the $p$-th unit vector.
This formula shows that $\E[y_i(0)|\bx_i]\neq\E[y_i(0)|\alpha_i]$ in general, for any finite $p$.

In the special case of $\sigma_{p\epsilon}=0$, $\E[y_i(0)|\bx_i]-\E[y_i(0)|\alpha_i]=O_\P(p^{-1/2})$ as $p\rightarrow\infty$. However, usual high-dimensional techniques that rely on sparsity conditions still suffer from at least two issues given the calculation above.

First, ignoring the small-order term $\frac{1}{\sqrt{p}}\frac{\phi(\sqrt{p}\bar{x}_i)-\phi(\sqrt{p}(1-\bar{x}_i))}{\Phi(\sqrt{p}(1-\bar{x}_i))-\Phi(\sqrt{p}(-\bar{x}_i))}$,  we can see that $\E[y_i(0)|\bx_i]$ is linear in $x_{il}$'s, but \textit{not} sparse at all, with all coefficients close to zero when $p$ is large, which may lead to poor performance of sparsity-based techniques such as Lasso. 

Second, for any finite $p$, $\E[y_i(0)|\bx_i]$ is different from $\E[y_i(0)|\alpha_i]$, which implies that regression techniques that control for a moderate number of  proxies are still invalid, even if one employs a data-driven procedure to select a few coordinates among all $x_{il}$'s.

The issues discussed above are fundamental and do not rely on the specific structure of this simple example. Intuitively, they arise due to the existence of a non-negligible error in each proxy and the potential correlation between the measurement error and the outcome (or treatment) error. 
When $\eta_l(\cdot)$ takes some unknown nonlinear form other than the simple linear form in this example, 
the (high-dimensional) model for $\E[y_i(0)|\bx_i]$ may be nonlinear, further complicating the analysis.

For completeness, we also compute $\E[y_i(0)|\bx_i, t_i=0]$ below, which corresponds to the usual regression adjustment strategy: take the subsample of untreated, and run regression of outcomes on covariates.
\begin{align*}
	\E[y_i(0)|\bx_i, t_i=0]&=\E[\alpha_i+\E[\epsilon_i|\bx_i,\alpha_i]|\bx_i, t_i=0]\\
	&=\E[\alpha_i+\sigma_{p\epsilon}\biota_p'(\bx_i-\alpha_i\bm{1}_p)|\bx_i, t_i=0]\\
	&=\sigma_{p\epsilon}x_{ip}+(1-\sigma_{p\epsilon})\E[\alpha_i|\bx_i, t_i=0]\\
	&=\sigma_{p\epsilon}x_{ip}+(1-\sigma_{p\epsilon})
	\frac{
	\int_{0}^{1}
		\alpha \, (1 - e(\alpha))
		\exp\Big( -\frac{p}{2}(\alpha - \bar{x}_i)^{2} \Big)
		\, d\alpha
	}{
		\int_{0}^{1}
		(1 - e(\alpha))
		\exp\Big( -\frac{p}{2}(\alpha - \bar{x}_i)^{2} \Big)
		\, d\alpha
	}.
\end{align*} 
When $p$ is large, the posterior density of $\alpha$ given $\bx_i$ and $t_i=0$ ``concentrates'' at $\alpha=\bar{x}_i$. Therefore, under some mild regularity conditions, it can be shown that the second term in the last line is approximately $(1-\sigma_{p\epsilon})\bar{x}_i$. 
The main conclusions from the previous discussion also apply here.

\section{Discussion of Assumptions}\label{SA section: discuss assumptions}

In this section, we provide more discussions about our main assumptions used in the paper.

\subsection{Assumption \ref{Assumption: Regularities}}
\label{SA section: local design}
The support, moment, and boundedness restrictions in Assumption \ref{Assumption: Regularities} are already primitive. Here we only discuss its local Gram condition, also imposed in Assumption \ref{SA Assumption: Regularities} below. Recall that $\bZ_{\subi}$ collects the observed covariates in $\mathcal N_i$ and $\bP_{\subi}$ projects onto the column space of the target loading matrix $\bLambda_{\subi}$. When this matrix has full column rank,
\[
\bP_{\subi}=\bLambda_{\subi}(\bLambda_{\subi}'\bLambda_{\subi})^{-1}\bLambda_{\subi}'.
\]
The local Gram condition requires the observed covariates to retain nondegenerate variation after this projection. It is invariant to rotations and rescaling of the loadings, and is vacuous when there are no observed covariates. With normalized loadings, it ensures a nondegenerate joint design; see the proof of Lemma \ref{lem: bound on coefficients}.

We give two sufficient conditions below. In both cases, we maintain the support, boundedness, and independent sampling conditions on $(\balpha_i,\bz_i)$, and the local loading and signal bounds in Assumption \ref{SA Assumption: LPCA}(a). In addition, let $\tilde{\bz}_i=\bz_i-\E[\bz_i\mid\balpha_i]$ and suppose that for some $c>0$,
\begin{equation}\label{SA eq: conditional covariance z}
\E[\tilde{\bz}_i\tilde{\bz}_i'\mid\balpha_i]\succeq c\bI_{\ttr_z}\quad\text{almost surely}.
\end{equation}
Here $\succeq$ denotes the positive-semidefinite matrix ordering. Matrix norms in this subsection are operator norms.

\subsubsection*{Equal exponents and a matching-error rate}

Suppose Assumption \ref{Assumption: Matching} holds with $\uvarsig=\lvarsig$, $K/n=o(1)$, $\log(n\vee p)/K=o(1)$, and $a_p^{1/\lvarsig}=o(h_n)$, where $h_n=(K/n)^{1/\ttr_\alpha}$. These rate and exponent restrictions, together with \eqref{SA eq: conditional covariance z}, suffice even when $\bz_i$ and the measurement errors are dependent.

First, the support and density restrictions imply that a ball of radius proportional to $h_n$ around each $\balpha_i$ contains order $K$ observations, uniformly over $i$. Choose sufficiently small constants $0<c_1<c_2$ such that the ball of radius $c_2h_n$ contains fewer than $K$ observations and the upper distance bound for radius $c_1h_n$ is smaller than the lower distance bound for radius $c_2h_n$. Since $a_p=o(h_n^{\lvarsig})$, every unit within the smaller ball has a smaller noisy-proxy distance than every unit outside the larger ball. Thus, with
\[
\mathcal I_i=\{j\in[n]:\|\balpha_j-\balpha_i\|\leq c_1h_n\},
\qquad \mathcal I_i\subseteq\mathcal N_i,\qquad |\mathcal I_i|\asymp K,
\]
uniformly over $i$, on events of arbitrarily high probability. For any prescribed probability arbitrarily close to one, the bounds in Assumption \ref{Assumption: Matching}(b) permit fixed choices of $c_1$ and $c_2$. The fixed compact convex support with a density bounded away from zero ensures the same order of observations near its boundary.

Second, conditional on $\{\balpha_j\}_{j\in[n]}$, the sets $\mathcal I_i$ are fixed and the $\tilde{\bz}_j$'s are independent, centered, and bounded. Condition \eqref{SA eq: conditional covariance z} and Bernstein's inequality, with a union bound over neighborhoods and proxy coordinates, give
\begin{align*}
\frac1K\sum_{j\in\mathcal I_i}\tilde{\bz}_j\tilde{\bz}_j'&\succeq \big\{c|\mathcal I_i|/K-\bar{o}_\P(1)\big\}\bI_{\ttr_z},\\
\max_i\Big\|\frac1K\sum_{j\in\mathcal I_i}\tilde{\bz}_j\E[\bz_j\mid\balpha_j]'\Big\|&=O_\P\big(\sqrt{\log n/K}\big),\\
\max_{i,\,l\in\mathcal C^\ddagger}\Big\|\frac1K\sum_{j\in\mathcal I_i}\tilde{\bz}_j\eta_l(\balpha_j)\Big\|&=O_\P\big(\sqrt{\log(n\vee p)/K}\big).
\end{align*}
To handle the loadings, let $\bUpsilon_{\subi}=(\bLambda_{\subi}'\bLambda_{\subi}/K)^{1/2}$, as in Assumption \ref{SA Assumption: LPCA}. The target PCA solution satisfies $\bLambda_{\subi}=\bH_{\subi}^\ddagger\bF_{\subi}/p^\ddagger$ and $\bF_{\subi}'\bF_{\subi}/p^\ddagger=\bI_{\ttd_i}$. Hence by Cauchy-Schwarz,
\begin{align*}
\Big\|\frac1K\sum_{j\in\mathcal I_i}\tilde{\bz}_j\blambda_{j,\subi}'\bUpsilon_{\subi}^{-1}\Big\|
&\leq\frac{1}{\upsilon_{\ttd_i,\subi}}
\Big\{\frac1{p^\ddagger}\sum_{l\in\mathcal C^\ddagger}
\big\|\frac1K\sum_{j\in\mathcal I_i}\tilde{\bz}_j\eta_l(\balpha_j)\big\|^2\Big\}^{1/2}\\
&=\bar{O}_\P\big(\sqrt{\log(n\vee p)/K}/\upsilon_{\ttd_i,\subi}\big)=\bar{o}_\P(1).
\end{align*}
The last equality uses the signal bound in Assumption \ref{SA Assumption: LPCA}(a). This argument allows $\bF_{\subi}$ and $\bLambda_{\subi}$ to depend on the proxy-selected neighborhood; no independence between these matrices and $\bz_j$ is required.

Finally, for any unit vector $\ba\in\mathbb R^{\ttr_z}$, the projection of $\bZ_{\subi}\ba$ on the normalized loadings has coefficient $\bb=\bUpsilon_{\subi}^{-1}\bLambda_{\subi}'\bZ_{\subi}\ba/K$, which is uniformly bounded by compactness of the covariates and normalization of the loadings. Restrict the residual sum of squares to $\mathcal I_i$, expand $\bz_j=\E[\bz_j\mid\balpha_j]+\tilde{\bz}_j$, and drop the nonnegative square involving the conditional mean and loadings. Then
\begin{align*}
\frac1K\ba'\bZ_{\subi}'(\bI_K-\bP_{\subi})\bZ_{\subi}\ba
&\geq\frac1K\sum_{j\in\mathcal I_i}
(\ba'\bz_j-\bb'\bUpsilon_{\subi}^{-1}\blambda_{j,\subi})^2\\
&\geq\frac1K\sum_{j\in\mathcal I_i}(\ba'\tilde{\bz}_j)^2
+\frac2K\sum_{j\in\mathcal I_i}\ba'\tilde{\bz}_j
\{\ba'\E[\bz_j\mid\balpha_j]-\bb'\bUpsilon_{\subi}^{-1}\blambda_{j,\subi}\}\\
&\geq\frac{c|\mathcal I_i|}{K}+\bar{o}_\P(1)\gtrsim_\P1,
\end{align*}
uniformly over $i$ and $\|\ba\|=1$, proving the local Gram condition. No smoothness of $\E[\bz_i\mid\balpha_i]$ is needed. Thus, equal exponents and the additional restriction on $a_p$ provide one way to verify the condition, rather than restrictions imposed by the general results.

\subsubsection*{Conditional independence of covariates and measurement errors}
Alternatively, suppose
\[
\{\bz_j\}_{j\in[n]}\inde\{\bu_j\}_{j\in[n]}\mid\{\balpha_j\}_{j\in[n]}.
\]
For example, this holds if $(\balpha_i,\bz_i,\bu_i)$ are independent across units and $\bz_i\inde\bu_i\mid\balpha_i$. It is enough to impose the array independence above only for the matching block $\bu_j^\dagger$. Conditional on the latent confounders and this block, the neighborhoods and target loadings are fixed, while the $\tilde{\bz}_j$'s remain independent with mean zero and satisfy \eqref{SA eq: conditional covariance z}.

Take $\mathcal I_i=\mathcal N_i$ in the preceding calculation. By boundedness of the covariates and normalized loadings, together with $\log n/K=o(1)$, we have
\begin{align*}
&\frac1K\sum_{j\in\mathcal N_i}\tilde{\bz}_j\tilde{\bz}_j'\succeq\{c-\bar{o}_\P(1)\}\bI_{\ttr_z},\\
&\max_i\big\|\frac1K\sum_{j\in\mathcal N_i}\tilde{\bz}_j\E[\bz_j\mid\balpha_j]'\big\|
+\max_i\big\|\frac1K\sum_{j\in\mathcal N_i}\tilde{\bz}_j\blambda_{j,\subi}'\bUpsilon_{\subi}^{-1}\big\|
=O_\P\big(\sqrt{\log n/K}\big)=o_\P(1).
\end{align*}
Then, repeating the remaining argument for the first route yields the desired local Gram condition. This route does not require equal matching exponents or the additional restriction on $a_p$ in the first route.

\subsection{Assumption \ref{Assumption: indep of errors}}\label{SA section: indep of errors}

In the main paper we imposed a conditional independence restriction on $(\epsilon_i, v_i)$ and $\bu_i$, which is only for simplifying the presentation of our main results. As will be detailed in the theoretical analysis in this supplement, this condition can be relaxed. 
We introduce a key quantity $$\biaserr=\max_{i\in[n]}\;\Big(|\E[\epsilon_{i}|\mathcal{F}, \{t_i\}_{i\in[n]}]|+|\E[v_{i}|\mathcal{F}]|\Big)$$
to characterize the bias due to the potential correlation between  $(\epsilon_i, v_i)$ and $\bu_i$,
where $\mathcal{F}$ is a sigma field on which the obtained set of nearest neighbors $\mathcal{N}_i$, local factor loadings $\bLambda_{\subi}$, and confounders $\balpha_i$, $\bz_i$ are measurable. $\sigma(\{(\balpha_i, \bz_i, \bx_i)\}_{i\in[n]})$ (the sigma field generated by all proxies and confounders) is a (trivial) choice of $\mathcal{F}$, but in general, $\mathcal{F}$ can be a smaller sigma field generated only by $\balpha_i$, $\bz_i$, and the information needed to obtain $\mathcal{N}_i$ and $\bLambda_{\subi}$. See concrete examples below.

$\biaserr$ determines the magnitude of the bias resulting from conditioning on a transformation of noisy proxies. Generally, thanks to the property of the doubly robust score function, we only need $\biaserr^2=o_\P(n^{-1/2})$. In the following, we use several examples to illustrate the magnitude of this bias.

\begin{exmp}[Classical measurement error model]
	In the classical measurement error literature, it is typical to assume that conditional on the ``true'' confounders (including the observed $\bz_i$ and the unobserved $\balpha_i$), the measurement error $\bu_i$ and the error $\epsilon_i$ in the outcome equation (or $v_i$ in the  treatment equation) are independent. Since the estimated nearest neighbors and local factor loadings are completely determined by the noisy proxies $\bx_i$'s, in this case we can simply take $\mathcal{F}=\sigma(\{(\balpha_i, \bz_i, \bx_i)\}_{i\in[n]})$, and then
	\begin{align*}
		&\E[v_i|\mathcal{F}]=\E[v_i|\{(\balpha_j,\bz_j,\bu_j)\}_{j\in[n]}]=\E[v_i|\{(\balpha_j,\bz_j)\}_{j\in[n]}]=0,\\
		&\E[\epsilon_i|\mathcal{F},\{t_j\}_{j\in[n]}]=\E[\epsilon_i|\{(\balpha_j,\bz_j,\bu_j,t_j)\}_{j\in[n]}]=\E[\epsilon_i|\{(\balpha_j,\bz_j,t_j)\}_{j\in[n]}]=0,
	\end{align*}
	where the second result follows by  the unconfoundedness assumption (conditional independence between $y_i(0)$ and $t_i$).
\hfill{\LARGE$\lrcorner$}
\end{exmp}

\begin{exmp}[Martingale difference]
			In many applications, variables are collected over time. $(\epsilon_i, v_i)$ may be some errors in the current period of interest, say time $T$; $\bu_i$ is some error in proxies in the past. So we can assume that $(\epsilon_i, v_i)\in\mathcal{F}_T$, some sigma field up to time $T$, while $\bu_i\in\mathcal{F}_{T-1}$. Thus, the conditional independence between the measurement error and the outcome/treatment equation error is not necessary for $\biaserr=0$; it still holds if these errors have an MDS-style property: $\E[v_i|\{(\balpha_j, \bz_j)\}_{j\in[n]}, \mathcal{F}_{T-1}]=0$, and $\E[\epsilon_i|\{(\balpha_j, \bz_j, t_j)\}_{j\in[n]}, \mathcal{F}_{T-1}]=0$.
			\hfill{\LARGE$\lrcorner$}
\end{exmp}

However, our results in this paper are \textit{not} restricted to the case of $\biaserr=0$.
When only a small number of proxies are correlated with outcomes or treatment (conditionally on true confounders) and the proxy transformations aggregate information diffusely across many proxies, $\biaserr$ can shrink as $p$ grows.
To gain some intuition, consider the following examples.
\begin{exmp}[Local average]
	To begin with, in this example we consider local average smoothing based on the distance of averages, with the PCA step suppressed, and assume for simplicity that there is no $\bz_i$. Then, the obtained nearest neighbors are fully determined by the average of $\bx_i$'s, for $i\in[n]$.  
	Let $\mathcal{F}=\sigma(\{(\balpha_i, \bar{u}_i)\}_{i\in[n]})$ with $\bar{u}_i=p^{-1}\sum_{l=1}^pu_{il}$.
	Consider the simple Gaussian error model discussed before, where only the error in the $p$-th proxy is correlated with the outcome error $\epsilon_i$: 
	\[
	\begin{bmatrix}
		\bu_i\\
		\epsilon_i
	\end{bmatrix}
	\sim \mathsf{N}\left(\bm{0}, 
	\begin{bmatrix}
		\bI_{p-1} & 0 & 0\\
		0 & 1 & \sigma_{p\epsilon}\\
		0 & \sigma_{p\epsilon}& 1
	\end{bmatrix}
	\right),
	\]
	$\balpha_i\inde (\bu_i, \epsilon_i)$, and $v_i\inde (\bu_i, \epsilon_i, \balpha_i)$. Then, we have 
	\[
	\begin{bmatrix}
		\epsilon_i\\
		\bar{u}_i
	\end{bmatrix}
	\sim \mathsf{N}\left(\bm{0}, 
	\begin{bmatrix}
		1 & \frac{\sigma_{p\epsilon}}{p}\\
		\frac{\sigma_{p\epsilon}}{p}& \frac{1}{p}
	\end{bmatrix}
	\right),
	\]	
	and thus, for any $i\in[n]$,
	\[
	\E[\epsilon_i|\{(\balpha_j,\bar{u}_j, t_j)\}_{j=1}^n]=	\E[\epsilon_i|\bar{u}_i]=\V[\bar{u}_i]^{-1}\E[\bar{u}_i\epsilon_i]\bar{u}_i=\sigma_{p\epsilon}\bar{u}_i=O_\P(p^{-1/2}).
	\]
	By the usual Gaussian tail bound, we conclude that $\biaserr=O_\P(\sqrt{\log n/p})$. The rate restriction $\sqrt{n}\biaserr^2=o_\P(1)$ is therefore satisfied when $\sqrt{n}\log n/p=o(1)$, which has been already implied by $\sqrt{n}\sam^{-2}=o(1)$ and thus does not add extra restrictions at all.
	\hfill{\LARGE$\lrcorner$}
\end{exmp}

Next, we consider the local PCA procedure discussed in the paper.
\begin{exmp}[Local PCA]
		Maintain the Gaussian error model in the previous example and suppose that the $p$-th proxy belongs to the PCA block $\mathcal{C}^\ddagger$. Write $\epsilon_i=\sigma_{p\epsilon}u_{ip}+\tau_i$, where $\tau_i$ is independent of all other random variables in the model. To provide a more concrete example, suppose that, for $l\in\mathcal{C}^\ddagger$,
		$\eta_l(\balpha_i)=w_lg(\balpha_i)$,
		where $g(\cdot)$ may be nonlinear and the common weights $\bw=(w_l:l\in\mathcal{C}^\ddagger)$ satisfy $\max_l|w_l|\lesssim1$ and $(p^\ddagger)^{-1}\bw'\bw\gtrsim1$. The matching block is independent of $(\bU^\ddagger,\epsilon_i)$ and therefore does not affect the calculation below.

		Let $\bg=(g(\balpha_i):i\in[n])'$ and $\bW=\bU^\ddagger\bw$, whose $i$-th entry is $W_i=\bw'\bu_i^\ddagger$. We may take
		\[
		\mathcal F=\sigma\Big(\bX^\dagger,\{\balpha_i\}_{i\in[n]},
		\bW,\bU^\ddagger(\bU^\ddagger)'\Big).
		\]
		This sigma field determines the nearest-neighbor sets and all local PCA estimators because
		\[
		\bX^\ddagger(\bX^\ddagger)'
		=(\bw'\bw)\bg\bg'+\bg\bW'+\bW\bg'
		+\bU^\ddagger(\bU^\ddagger)'.
		\]

		For any orthogonal matrix $\bQ$ satisfying $\bQ\bw=\bw$, replacing $\bU^\ddagger$ by $\bU^\ddagger\bQ$ preserves both its Gaussian distribution and $\mathcal F$. It follows that $\E[\bu_i^\ddagger\mid\mathcal F]$ must be proportional to $\bw$. Since
		\[
		\bw'\E[\bu_i^\ddagger\mid\mathcal F]
		=\E[W_i\mid\mathcal F]=W_i,
		\]
		we obtain
		\[
		\E[\bu_i^\ddagger\mid\mathcal F]
		=\frac{W_i\bw}{\bw'\bw}.
		\]
		Because treatment assignment is independent of the proxy and outcome errors conditional on the latent variables,
		\[
		\E[\epsilon_i\mid\mathcal F,\{t_j\}_{j\in[n]}]
		=\frac{\sigma_{p\epsilon}w_pW_i}{\bw'\bw}.
		\]
		Conditional on the latent variables, $W_i\sim\mathsf N(0,\bw'\bw)$. Hence the last display has variance $\sigma_{p\epsilon}^2w_p^2/(\bw'\bw)\lesssim1/p^\ddagger$. The treatment-error component of $\biaserr$ is zero by the same conditional independence, and a Gaussian maximal bound gives
		\[
		\biaserr=O_\P\Big(\sqrt{\log n/p}\Big).
		\]
		Thus, as in the previous example, $\sqrt n\biaserr^2=o_\P(1)$ follows from $\sqrt n\sam^{-2}=o(1)$ and imposes no additional rate restriction. This calculation illustrates information aggregation under a common-weight nonlinear structure.
		\hfill{\LARGE$\lrcorner$}
\end{exmp}

The preceding examples illustrate how aggregation across proxies can make $\biaserr$ shrink when only a small number of measurement errors are correlated with the outcome or treatment. The same intuition can accommodate several correlated errors, multiple common-weight directions, and more general nonlinear local PCA structures, although a formal treatment must also account for the information jointly retained by the local PCA outputs and any conditional dependence across units induced by proxy transformations. We leave these extensions for future research.

\subsection{Assumption \ref{Assumption: Matching}}\label{SA section: matching}

In this section, we provide more detailed discussion about how to verify Assumption \ref{Assumption: Matching}. 
We will focus on the pseudo-max distance, our recommended choice. See \cite{Feng_2023_wp_Nonlinear} for discussion of other distances.

Let $\bx_i^\dagger$, $\bmeta_i^\dagger$ and $\bu_i^\dagger$ denote (the transpose of) the $i$-th row of $\bX^\dagger$, $\bH^\dagger$, and $\bU^\dagger$ respectively. $s_{\min}(\cdot)$ denotes the minimum singular value of a matrix, and
$\nabla\bmeta^\dagger(\balpha)=\frac{\partial\bmeta^\dagger(\balpha)}{\partial\balpha'}$ denotes the Jacobian matrix of $\bmeta^\dagger(\balpha)=(\eta_l(\balpha))_{l\in\mathcal{C}^\dagger}$.  
We define $\bQ_{\balpha}\in\mathbb{R}^{p^\dagger\times \ttr_\alpha}$ to be the matrix formed by any orthonormal basis of the column space of $\nabla\bmeta^\dagger(\balpha)$. So $\bQ_{\balpha}\bQ_{\balpha}'$ is the projection operator onto the space spanned by the tangent basis at the point $\bmeta^\dagger(\balpha)$. 

Suppose the following conditions hold:
	\begin{enumerate}[label=(\roman*)]
		\item $\balpha_i$ is i.i.d. with a density bounded and bounded away from zero on a fixed compact convex support $\mathcal{A}$ with nonempty interior; $u_{il}$ is independent over $i\in[n]$ and $l\in[p]$ conditional on $\{\balpha_i\}_{i=1}^n$, and $\E[|u_{il}|^{2+\nu}|\{\balpha_i\}_{i=1}^n]$ is bounded uniformly over $i\in[n]$ and $l\in[p]$, for some $\nu>2$.
		\item For all $l\in[p]$, $\eta_l$ is twice continuously differentiable, with first- and second-order partial derivatives bounded uniformly over $l\in[p]$ and $\balpha\in\mathcal{A}$;
	\item 	$s_{\min}\Big(\frac{1}{p^\dagger}
	\nabla\bmeta^\dagger(\balpha)'\nabla\bmeta^\dagger(\balpha)\Big)\gtrsim 1$ for all $\balpha\in\mathcal{A}$;
	\item 	For any $\varepsilon>0$, there exists $\Delta>0$ such that $\frac{1}{p^\dagger}\|\bmeta^\dagger(\balpha)-\bmeta^\dagger(\tilde{\balpha})\|^2<\Delta\Rightarrow \|\balpha-\tilde{\balpha}\|<\varepsilon$, for all $p^\dagger$ sufficiently large; 

	\item 	For some $\underline{c}>0$ and all $n, p$ sufficiently large, 
	$$\inf_{\balpha\in\mathcal{A}}\inf_{\|\bv\|=1}\sup_{\tilde{\balpha}\in\mathcal{A}}\;
	\frac{1}{p^\dagger}|\bv'\bQ_{\balpha}'\bmeta^\dagger(\tilde{\balpha})|^2\geq \underline{c}.$$
\end{enumerate} 
If $\frac{n^{\frac{4}{\nu}}\log n}{p}\lesssim 1$, then
Assumption \ref{Assumption: Matching} holds with
$$a_p=\sqrt{(\log n)/p}, \;\rho_0=0, \;\lvarsig=\uvarsig=1$$

Conditions (i)--(iii) collect support, moment, smoothness, and local rank requirements, while condition (iv) requires the proxies to distinguish latent values globally. Condition (v) directly requires the proxies to remain informative in every local tangent direction, as needed for the sharp lower bound. Although stated separately for clarity, it is \textit{not} an additional restriction under conditions (i)--(iii), as discussed below. We first verify Assumption \ref{Assumption: Matching} under these conditions and then discuss how they are justified, with illustrative examples provided.

\subsubsection*{Verification of Assumption \ref{Assumption: Matching}}
\label{SA subsection: matching verification}

We check each requirement in Assumption \ref{Assumption: Matching} and show the roles of the sufficient conditions.

\textit{Part (a)}: To verify part (a), note that for any $j\neq i$ and $k\notin\{j,i\}$,
\[
\frac{1}{p^\dagger}(\bx^{\dagger}_j-\bx^{\dagger}_i)'\bx^{\dagger}_k
= \frac{1}{p^\dagger}(\bmeta^{\dagger}_j-\bmeta^{\dagger}_i)'\bmeta^{\dagger}_k+
\frac{1}{p^\dagger}(\bmeta^{\dagger}_j-\bmeta^{\dagger}_i)'\bu^\dagger_k+
\frac{1}{p^\dagger}(\bu^\dagger_j-\bu^\dagger_i)'\bmeta^{\dagger}_k+
\frac{1}{p^\dagger}(\bu^\dagger_j-\bu^\dagger_i)'\bu^\dagger_k.
\]
By the standard maximal inequality (see Lemma \ref{SA lem: maximal inequality}), we have
$\max_{i\neq j}|\frac{1}{p^\dagger}\sum_{l\in\mathcal{C}^\dagger}u_{il}u_{jl}|\lesssim_\P \sqrt{(\log n)/p}$.
By the same argument, we can conclude that the same bound holds for the last three terms uniformly over $i,j,k$, which suffices to show $\rho_0=0$ and $a_p=\sqrt{(\log n)/p}$. For this part, the key condition we used is the independence of units and $\E[\bu_i|\balpha_i]=0$.
\bigskip

\textit{Part (b), upper bound}: For all $k\neq i,j$, by Cauchy-Schwarz inequality,
\[
\begin{split}
	\Big|\frac{1}{p^\dagger}(\bmeta^{\dagger}_j-\bmeta^{\dagger}_i)'\bmeta^{\dagger}_k\Big|
	\leq\Big(\frac{1}{p^\dagger}\sum_{l\in\mathcal{C}^\dagger}\Big(\eta_{l}(\balpha_i)-\eta_{l}(\balpha_{j})\Big)^2\Big)^{1/2}
	\Big(\frac{1}{p^\dagger}\sum_{l\in\mathcal{C}^\dagger}\eta_{l}(\balpha_k)^2\Big)^{1/2}
	\lesssim \|\balpha_j-\balpha_i\|.
\end{split}
\]
This immediately implies that $\uvarsig=1$. The key conditions used for this step are the Lipschitz continuity of $\eta_l$'s.
\bigskip

\textit{Part (b), lower bound}: We consider the following two cases.

\textit{Large deviations.} Fix a sufficiently small $\varepsilon>0$ (so that the linear term in the local expansion below dominates its second-order remainder). By condition (iv), there is a $\Delta>0$ such that, for all sufficiently large $p^\dagger$, every pair with $\|\balpha_i-\balpha_j\|\geq\varepsilon$ satisfies $\|\bmeta^\dagger(\balpha_i)-\bmeta^\dagger(\balpha_j)\|^2/p^\dagger\geq\Delta$.
By condition (i), uniformly over $i$ and $j$, we can choose $i_0,j_0\notin\{i,j\}$ such that $\max\{\|\balpha_{i_0}-\balpha_i\|,\|\balpha_{j_0}-\balpha_j\|\}=o_\P(1)$. Adding and subtracting $\bmeta_{i_0}^\dagger$ and $\bmeta_{j_0}^\dagger$ in the two inner products below, respectively, and using the bounded derivatives in condition (ii), we obtain
{\[
\begin{split}
\frac{1}{p^\dagger}\|\bmeta_i^\dagger-\bmeta_j^\dagger\|^2
&=\frac{1}{p^\dagger}(\bmeta_i^\dagger-\bmeta_j^\dagger)'\bmeta_i^\dagger-
\frac{1}{p^\dagger}(\bmeta_i^\dagger-\bmeta_j^\dagger)'\bmeta_j^\dagger\\
&\leq \frac{1}{p^\dagger}|(\bmeta_i^\dagger-\bmeta_j^\dagger)'\bmeta_{i_0}^\dagger|
+\frac{1}{p^\dagger}|(\bmeta_i^\dagger-\bmeta_j^\dagger)'\bmeta_{j_0}^\dagger|+o_\P(1)\\
&\leq 2\rho(\bmeta_i^\dagger,\bmeta_j^\dagger)+o_\P(1).
\end{split}
\]}
Consequently, in this large-deviation case, $\rho(\bmeta_i^\dagger,\bmeta_j^\dagger)$ is bounded away from zero. Since compactness of the support also bounds $\|\balpha_i-\balpha_j\|$ from above, $\rho(\bmeta_i^\dagger,\bmeta_j^\dagger)\gtrsim\|\balpha_i-\balpha_j\|$ uniformly on this region.

\textit{Small deviations.} For $0<\|\balpha_i-\balpha_j\|<\varepsilon$, write $\boldsymbol{\delta}_{ij}=\balpha_i-\balpha_j$ and define the unit vector
{\[
\bv_{ij}=\frac{\bQ_{\balpha_i}'\nabla\bmeta^\dagger(\balpha_i)\boldsymbol{\delta}_{ij}}
{\|\nabla\bmeta^\dagger(\balpha_i)\boldsymbol{\delta}_{ij}\|}.
\]}
By condition (v), for each $\balpha_i$ and $\bv_{ij}$ there is a $\tilde{\balpha}\in\mathcal{A}$ such that $|\bv_{ij}'\bQ_{\balpha_i}'\bmeta^\dagger(\tilde{\balpha})|/\sqrt{p^\dagger}\geq\sqrt{\underline{c}}$. Condition (i) ensures that one can choose an observation $\ell\notin\{i,j\}$ satisfying $\|\balpha_\ell-\tilde{\balpha}\|=o_\P(1)$, uniformly over $i$ and $j$. By condition (ii), replacing $\tilde{\balpha}$ by $\balpha_\ell$ changes this normalized inner product by $o_\P(1)$, so $|\bv_{ij}'\bQ_{\balpha_i}'\bmeta^\dagger(\balpha_\ell)|/\sqrt{p^\dagger}\gtrsim 1$. Condition (iii) and a first-order expansion with a second-order remainder then give
{\[
\begin{split}
\rho(\bmeta_i^\dagger,\bmeta_j^\dagger)
&\geq \frac{1}{p^\dagger}|(\bmeta_i^\dagger-\bmeta_j^\dagger)'\bmeta_\ell^\dagger|\\
&\geq \frac{\|\nabla\bmeta^\dagger(\balpha_i)\boldsymbol{\delta}_{ij}\|}{\sqrt{p^\dagger}}
\frac{|\bv_{ij}'\bQ_{\balpha_i}'\bmeta^\dagger(\balpha_\ell)|}{\sqrt{p^\dagger}}
-C\|\boldsymbol{\delta}_{ij}\|^2\\
&\gtrsim \|\boldsymbol{\delta}_{ij}\|-C\|\boldsymbol{\delta}_{ij}\|^2.
\end{split}
\]}
Here, $C>0$ is a finite constant since condition (ii) bounds the second derivatives of the latent functions and Assumption \ref{Assumption: Regularities} bounds their levels.
For $\|\boldsymbol{\delta}_{ij}\|$ sufficiently small, this proves the desired lower bound and hence $\lvarsig=1$.

\subsubsection*{Discussion of the Sufficient Conditions}
\label{SA subsection: matching sufficient conditions}

\textit{Conditions (i)--(iii).}
\label{SA part: matching conditions i-iii}
Conditions (i) and (ii) impose standard support, moment, independence, and smoothness restrictions used in the nearest-neighbor and maximal-inequality arguments. Condition (iii) is a local rank condition: it prevents the tangent space generated by $\bmeta^\dagger(\cdot)$ from losing any direction of $\balpha$. For instance, when $\eta_l(\balpha)=\eta(\balpha,\bm{\varpi}_l)$ and the $\bm{\varpi}_l$'s are i.i.d., a uniform law of large numbers reduces this condition to uniform positive definiteness of $\E_{\bm{\varpi}}[\nabla_{\balpha}\eta(\balpha,\bm{\varpi})\nabla_{\balpha}\eta(\balpha,\bm{\varpi})']$.
\bigskip

\noindent\textit{Condition (iv).}
\label{SA part: matching condition iv}
Condition (iv) formalizes global informativeness: proxy vectors that are close in average squared distance must correspond to close latent values. For instance, when $\eta_l(\balpha)=\eta(\balpha,\bm{\varpi}_l)$ and the $\bm{\varpi}_l$'s are i.i.d., a uniform law of large numbers gives, uniformly over $\balpha,\tilde{\balpha}\in\mathcal{A}$,
{\[
\frac{1}{p^\dagger}\|\bmeta^\dagger(\balpha)-\bmeta^\dagger(\tilde{\balpha})\|^2
=\frac{1}{p^\dagger}\sum_{l\in\mathcal{C}^\dagger}
\{\eta(\balpha,\bm{\varpi}_l)-\eta(\tilde{\balpha},\bm{\varpi}_l)\}^2
\xrightarrow{\P}\E_{\bm{\varpi}}[\{\eta(\balpha,\bm{\varpi})-\eta(\tilde{\balpha},\bm{\varpi})\}^2].
\]}
Condition (iv) can then be verified in two steps. First, require the expectation above to be zero only when $\balpha=\tilde{\balpha}$. Continuity of $\eta(\balpha,\bm{\varpi})$ in $\balpha$ and compactness of $\mathcal{A}$ then imply that, for every $\varepsilon>0$,
{\[
\inf_{\substack{\balpha,\tilde{\balpha}\in\mathcal{A}\\
\|\balpha-\tilde{\balpha}\|\geq\varepsilon}}
\E_{\bm{\varpi}}[\{\eta(\balpha,\bm{\varpi})-\eta(\tilde{\balpha},\bm{\varpi})\}^2]>0.
\]}
Second, the uniform convergence transfers this positive lower bound to the finite-$p^\dagger$ distance with probability approaching one, yielding the required $\Delta>0$. The following linear and nonlinear examples illustrate how this condition can be verified.

{
\begin{exmp}[Linear factor benchmark]
\label{SA example: matching condition iv linear}
Suppose that $\eta_l(\balpha)=\bm{\varpi}_l'\balpha$ and that the loading Gram matrix satisfies
\[
s_{\min}\left(\frac{1}{p^\dagger}\sum_{l\in\mathcal{C}^\dagger}
\bm{\varpi}_l\bm{\varpi}_l'\right)\geq c>0.
\]
Then
\[
\frac{1}{p^\dagger}\|\bmeta^\dagger(\balpha)-\bmeta^\dagger(\tilde\balpha)\|^2
=(\balpha-\tilde\balpha)'\left(\frac{1}{p^\dagger}\sum_{l\in\mathcal{C}^\dagger}
\bm{\varpi}_l\bm{\varpi}_l'\right)(\balpha-\tilde\balpha)
\geq c\|\balpha-\tilde\balpha\|^2.
\]
Thus, condition (iv) holds with $\Delta=c\varepsilon^2$. Individual loadings may be zero or arbitrarily small; the Gram-matrix restriction only requires sufficiently many proxies to be jointly informative in every latent direction. The same restriction also verifies the local rank condition (iii).
\hfill{\LARGE$\lrcorner$}
\end{exmp}
}

{
\begin{exmp}[Polynomial model]
\label{SA example: matching condition iv polynomial}
Let $\alpha,\varpi\in[0,1]$, let the $\varpi_l$'s be independently uniformly distributed, and suppose $\eta(\alpha,\varpi)=1+\varpi\alpha+\varpi^2\alpha^2$. Then
\[
\begin{split}
\int_0^1\{\eta(\alpha,\varpi)-\eta(\tilde\alpha,\varpi)\}^2d\varpi
&=(\alpha-\tilde\alpha)^2\int_0^1\{\varpi+\varpi^2(\alpha+\tilde\alpha)\}^2d\varpi\\
&\geq\frac{1}{3}|\alpha-\tilde\alpha|^2.
\end{split}
\]
Thus, condition (iv) holds. The derivative second moment is $1/3+\alpha+4\alpha^2/5\geq1/3$, so condition (iii) also holds.
\hfill{\LARGE$\lrcorner$}
\end{exmp}
}

{
\begin{exmp}[Trigonometric model]
\label{SA example: matching condition iv trigonometric}
Let $\alpha,\varpi\in[0,1]$, let the $\varpi_l$'s be independently uniformly distributed, and suppose $\eta(\alpha,\varpi)=\sin\{\pi(\varpi+\alpha)\}$. Then
\[
\int_0^1\{\eta(\alpha,\varpi)-\eta(\tilde\alpha,\varpi)\}^2d\varpi
=1-\cos\{\pi(\alpha-\tilde\alpha)\}\geq2|\alpha-\tilde\alpha|^2.
\]
Thus, condition (iv) holds. The derivative second moment is $\pi^2/2$, so condition (iii) also holds.
\hfill{\LARGE$\lrcorner$}
\end{exmp}
}

{
\begin{exmp}[Logit model]
\label{SA example: matching condition iv logit}
Let $\alpha\in[0,1]$ and $\psi(a)=1/\{1+\exp(-a)\}$. Consider the measurement-specific logit functions $\eta_l(\alpha)=\psi(\varpi_{l0}+\varpi_{l1}\alpha)$, where the intercepts are uniformly bounded and the slopes are positive, bounded, and bounded away from zero. Because the indices lie in a compact set, the first derivative $\psi^{(1)}$ of $\psi$ is uniformly bounded away from zero. The mean value theorem therefore gives $(p^\dagger)^{-1}\|\bmeta^\dagger(\alpha)-\bmeta^\dagger(\tilde\alpha)\|^2\gtrsim|\alpha-\tilde\alpha|^2$, verifying condition (iv). Writing $\partial_\alpha$ for differentiation with respect to $\alpha$, $\partial_\alpha\eta_l(\alpha)=\varpi_{l1}\psi^{(1)}(\varpi_{l0}+\varpi_{l1}\alpha)$ is uniformly bounded and bounded away from zero, so conditions (ii) and (iii) also hold. This specification can represent the conditional means of binary proxies in logit models with measurement-specific (e.g., time-specific) coefficients.
\hfill{\LARGE$\lrcorner$}
\end{exmp}
}

Condition (iv) excludes cases in which distinct latent values generate the same proxy vector. Neither smoothness nor the local rank condition (iii) alone guarantees global informativeness. For example, on $\mathcal{A}=[0,1]$, let the coordinates of $\bmeta^\dagger(\alpha)$ be repeated pairs $(\sin(2\pi\alpha),\cos(2\pi\alpha))$. Its derivative satisfies condition (iii), but $\bmeta^\dagger(0)=\bmeta^\dagger(1)$, so condition (iv) fails.
\bigskip

\noindent\textit{Condition (v).}
\label{SA part: matching condition v}
Condition (v) directly requires the proxies to remain informative in every local tangent direction, which is the property used for the sharp linear lower bound for the pseudo-max distance. We next show that it follows from conditions (i)--(iii). Fix $\balpha\in\mathcal{A}$ and a unit vector $\bv$, and define
{\[
f_{\balpha,\bv}(\tilde\balpha)
=\frac{1}{\sqrt{p^\dagger}}\bv'\bQ_{\balpha}'\bmeta^\dagger(\tilde\balpha).
\]}
Because $\bQ_{\balpha}$ spans the column space of $\nabla\bmeta^\dagger(\balpha)$, condition (iii) gives
{\[
\left\|\nabla_{\tilde\balpha}f_{\balpha,\bv}(\balpha)\right\|^2
=\frac{1}{p^\dagger}\left\|\nabla\bmeta^\dagger(\balpha)'\bQ_{\balpha}\bv\right\|^2\gtrsim1.
\]}
Condition (ii) also bounds the Hessian of $f_{\balpha,\bv}$ uniformly.

Because the fixed compact convex support $\mathcal{A}$ has nonempty interior, it contains a ball $B(\balpha_0,r)$ for some $r>0$. Let $\bg=\nabla_{\tilde\balpha}f_{\balpha,\bv}(\balpha)$, $\widetilde{\balpha}_{\pm}=\balpha_0\pm r\bg/\|\bg\|$, and $\balpha_{\pm}=(1-s)\balpha+s\widetilde{\balpha}_{\pm}$ for a sufficiently small fixed $s>0$. Both $\balpha_+$ and $\balpha_-$ belong to $\mathcal{A}$ by convexity. A second-order expansion and the uniform Hessian bound give
\[
\left|f_{\balpha,\bv}(\balpha_+)-f_{\balpha,\bv}(\balpha_-)\right|
\geq 2sr\|\bg\|-Cs^2\gtrsim1,
\]
where $C>0$ is uniform by condition (ii) and compactness of $\mathcal{A}$. Consequently,
\[
\sup_{\tilde\balpha\in\mathcal{A}}|f_{\balpha,\bv}(\tilde\balpha)|
\geq\frac{1}{2}\left|f_{\balpha,\bv}(\balpha_+)-f_{\balpha,\bv}(\balpha_-)\right|\gtrsim1,
\]
uniformly over $\balpha$ and $\bv$. Squaring this inequality yields condition (v). Thus, condition (v) is retained because it states the requirement used directly in the lower-bound calculation, rather than because it imposes an additional restriction.

The following examples calculate the projection in condition (v) directly to illustrate its meaning.

{
\begin{exmp}[Linear factor benchmark, continued]
\label{SA example: matching condition v linear}
Continue Example \ref{SA example: matching condition iv linear}. In this model, $\nabla\bmeta^\dagger(\balpha)$ is the matrix with $l$-th row $\bm{\varpi}_l'$ and does not depend on $\balpha$. We can therefore take $\bQ_{\balpha}=\nabla\bmeta^\dagger(\balpha)\{\nabla\bmeta^\dagger(\balpha)'\nabla\bmeta^\dagger(\balpha)\}^{-1/2}$. Because condition (i) implies that $\mathcal{A}$ contains a ball of some radius $r>0$, maximizing the linear form below over this ball gives, for every $\|\bv\|=1$,
\[
\begin{split}
\sup_{\tilde\balpha\in\mathcal{A}}\frac{1}{p^\dagger}
|\bv'\bQ_{\balpha}'\bmeta^\dagger(\tilde\balpha)|^2
&=\sup_{\tilde\balpha\in\mathcal{A}}\frac{1}{p^\dagger}
|\bv'\bQ_{\balpha}'\nabla\bmeta^\dagger(\balpha)\tilde\balpha|^2\\
&\geq\frac{r^2}{p^\dagger}
\|\bv'\bQ_{\balpha}'\nabla\bmeta^\dagger(\balpha)\|^2\\
&=r^2\bv'\left(\frac{1}{p^\dagger}\sum_{l\in\mathcal{C}^\dagger}
\bm{\varpi}_l\bm{\varpi}_l'\right)\bv\geq cr^2,
\end{split}
\]
where the last inequality follows from the loading Gram-matrix restriction in Example \ref{SA example: matching condition iv linear}, which also verifies condition (iii). Thus, condition (v) holds. With a scalar latent variable, the left-hand side of condition (v) equals $(p^\dagger)^{-1}\sum_{l\in\mathcal{C}^\dagger}\varpi_l^2\sup_{\tilde\alpha\in\mathcal{A}}\tilde\alpha^2$, which is bounded away from zero by the loading restriction and condition (i).
\hfill{\LARGE$\lrcorner$}
\end{exmp}
}

{
\begin{exmp}[Polynomial model, continued]
\label{SA example: matching condition v polynomial}
Continue Example \ref{SA example: matching condition iv polynomial}. Taking $\tilde\alpha=0$ gives $\int_0^1\partial_\alpha\eta(\alpha,\varpi)\eta(0,\varpi)d\varpi=1/2+2\alpha/3\geq1/2$, while the derivative second moment is at most $32/15$. Thus, after normalizing the derivative vector, the corresponding squared population inner product is at least $15/128$, and condition (v) holds.
\hfill{\LARGE$\lrcorner$}
\end{exmp}
}

{
\begin{exmp}[Trigonometric model, continued]
\label{SA example: matching condition v trigonometric}
Continue Example \ref{SA example: matching condition iv trigonometric}. Choose $|\tilde\alpha-\alpha|=1/2$, which is always possible on $[0,1]$. Then $|\int_0^1\partial_\alpha\eta(\alpha,\varpi)\eta(\tilde\alpha,\varpi)d\varpi|=\pi/2$, while the derivative second moment equals $\pi^2/2$. Thus, after normalizing the derivative vector, the corresponding squared population inner product is $1/2$, and condition (v) holds.
\hfill{\LARGE$\lrcorner$}
\end{exmp}
}

{
\begin{exmp}[Logit model, continued]
\label{SA example: matching condition v logit}
Continue Example \ref{SA example: matching condition iv logit}. Since the latent variable is scalar, $\bQ_\alpha$ is the normalized derivative vector. Uniformly over $\alpha,\tilde\alpha\in[0,1]$,
\[
\frac{1}{p^\dagger}|\bQ_\alpha'\bmeta^\dagger(\tilde\alpha)|^2
=\frac{\left\{(p^\dagger)^{-1}\sum_{l\in\mathcal{C}^\dagger}
\partial_\alpha\eta_l(\alpha)\eta_l(\tilde\alpha)\right\}^2}
{(p^\dagger)^{-1}\sum_{l\in\mathcal{C}^\dagger}\{\partial_\alpha\eta_l(\alpha)\}^2}
\gtrsim1,
\]
because both $\eta_l(\tilde\alpha)$ and $\partial_\alpha\eta_l(\alpha)$ are uniformly positive and bounded. Thus, condition (v) holds.
\hfill{\LARGE$\lrcorner$}
\end{exmp}
}

These examples illustrate how proxy vectors at suitable latent values have a nonnegligible inner product with every local tangent direction. This is precisely the feature used by the pseudo-max distance to obtain the sharp linear lower bound.

\subsection{Assumption \ref{Assumption: LPCA}}\label{SA section: lpca}

This section provides further discussion about how to understand and verify Assumption \ref{Assumption: LPCA}.

\subsubsection*{Link with Classical Nonparametric Regression}
Assumption \ref{Assumption: LPCA} is closely related to the approximation strategy used in usual nonparametric regression. 
To see this, we first introduce the standard multi-index notation:
for any $\bq = (q_1,\ldots,q_d)\in \mathbb{N}_0^d$,
let $|\bq| := \sum_{j=1}^d q_j$, $\bq! := \prod_{j=1}^d q_j!$;
for $\bx=(x_1,\ldots,x_d)\in \mathbb{R}^d$ and $\bq\in\mathbb{N}_0^d$, define
$\bx^{\bq} := \prod_{j=1}^d x_j^{q_j}$;
for a function $f:\mathbb{R}^d\to\mathbb{R}$ that is $|\bq|$-times differentiable, the
$\bq$-th partial derivative is $f^{(\bq)}(\bx):= \frac{\partial^{|\bq|} f(\bx)}{\partial x_1^{q_1}\cdots \partial x_d^{q_d}}$.
Also, let $\mathscr{D}\bmeta_l(\balpha)=(\eta^{(\bq)}_l(\balpha))_{|\bq|\leq m}$ denote a vector containing all partial derivatives of $\eta_l(\cdot)$ at $\balpha$ up to order $m$. 

Suppose the following conditions hold:
\begin{enumerate}[label=(\roman*)]
		\item $\balpha_i$ is i.i.d. and has a compact convex support with a density bounded and bounded away from zero; $u_{il}$ is independent over $i\in[n]$ and $l\in[p]$ conditional on $\{\balpha_i\}_{i=1}^n$, and $\E[|u_{il}|^{2+\nu}|\{\balpha_i\}_{i=1}^n]$ is bounded uniformly over $i\in[n]$ and $l\in[p]$, for some $\nu>2$.
	\item Assumption \ref{Assumption: Matching} holds with $\uvarsig=\lvarsig$ and 
	$a_p^{1/\lvarsig}=o(h_n)$ for $h_n=(K/n)^{1/{\ttr_\alpha}}$. 
	\item $\mu(\cdot)$, $e(\cdot)$, $\eta_l(\cdot)$, for $l\in[p]$, are $m$-times continuously differentiable with partial derivatives of order no greater than $m$ bounded by a constant, and 
	\begin{equation}\label{eq: def of nondegenerate to order kappa}
		s_{\min}\Big(\frac{1}{p^\ddagger}\sum_{l\in\mathcal{C}^\ddagger}
		(\mathscr{D}\bmeta_{l}(\balpha))(\mathscr{D}\bmeta_{l}(\balpha))'\Big)\gtrsim 1,
	\end{equation}
	uniformly over $\balpha\in\mathcal{A}$.
\end{enumerate}
If $\sam^{-1}/h_{n}^{m}=o(1)$ and $K\log n/n=o(1)$, then Assumption \ref{Assumption: LPCA} is satisfied by setting  $\ttd_i=\binom{m+\ttr_\alpha}{\ttr_\alpha}$, with 
$\rate{\eta}$, $\rate{\mu}$ and $\rate{e}$ being $\bar{o}_\P(h_n^m)$.
\begin{remark}
	Among the three conditions, \eqref{eq: def of nondegenerate to order kappa} is the key restriction for justifying the local approximation, which can be viewed as a generalization of condition (iii) discussed in the previous Section \ref{SA section: matching}. 
	Note that if we further assume $\eta_l$, $\mu$ and $e$ are $(m+1)$-times continuously differentiable but keep the same number of factors extracted $\ttd_i=\binom{m+\ttr_\alpha}{\ttr_\alpha}$, it is well known in the approximation theory that we can have a more precise characterization of the error bounds, with $\rate{\eta}, \rate{\mu}$ and $\rate{e}$ of order $\bar{O}_\P(h_n^{m+1})$.
	
	\eqref{eq: def of nondegenerate to order kappa} may be restrictive for large $m$, since it requires sufficient nonlinearities of the latent function $\bmeta(\cdot)$. However, when $m=0$ or $m=1$, \eqref{eq: def of nondegenerate to order kappa} is less stringent. 
	
	When $m=0$, the leading factor corresponds to the ``intercept'' of $\bmeta(\cdot)$, which is generally nonzero. Even if $\bmeta(\balpha_i)\approx \bm{0}$, we only miss a constant basis (for the neighborhood of unit $i$), but in practice we can always include an intercept in the local estimation of conditional means of potential outcomes or propensity scores (local constant smoothing). As long as the indirect matching on noisy proxies is valid (the key Assumption \ref{Assumption: Matching} holds), this strategy guarantees that $\rate{\mu}$ and $\rate{e}$ are $\bar{o}_\P(1)$.
	
	When $m=1$, the next $\ttr_\alpha$ factor loading corresponds to the linear basis, and
	the non-degeneracy of factors, as described before, says that the tangent plane at every point of the surface generated by $\bmeta(\cdot)$ should not collapse in any direction. Roughly speaking, it holds when each latent variable in $\balpha$ can generate sufficient variation in $\bmeta(\balpha)$. In this case, $\rate{\mu}$ and $\rate{e}$ are $\bar{o}_\P(h_n)$.
\end{remark}

Now, we explain how these sufficient conditions can be used to verify Assumption \ref{Assumption: LPCA}. 

Condition (ii) ensures that for all units $i\in[n]$, $\max_{j\in\mathcal{N}_i}\|\balpha_j-\balpha_i\|\lesssim_\P h_n$ for some ``bandwidth'' $h_n=o(1)$. So our approximation using nearest neighbors is truly \textit{local}. Specifically,  according to Lemma \ref{SA lem: lpca} below, the restrictions that $\uvarsig=\lvarsig$ and $a_p^{1/\lvarsig}=o((K/n)^{1/{\ttr_\alpha}})$ imply that
$h_n=O((K/n)^{1/{\ttr_\alpha}})$, which is the usual bandwidth one can obtain when matching directly on $\balpha_i$'s.

Then, given the smoothness of $\eta_l(\cdot)$'s, we can justify the local PCA approximation via Taylor's expansion (locally at point $\balpha_i$):
$$
\eta_l(\balpha_j)=\sum_{|\bq|\leq m}\frac{\eta^{(\bq)}_l(\balpha_i)}{\bq!}(\balpha_j-\balpha_i)^{\bq}+o(h_n^m)
$$ 
for all $l\in[p]$. In this context, $\eta^{(\bq)}_l(\balpha_i)/\bq!$ plays the role of an auxiliary local factor (varying over $l$), $(\balpha_j-\balpha_i)^{\bq}$ plays the role of an auxiliary local factor loading (varying over $j$), and the remainder is the approximation error. If $\eta_l(\cdot)$ is a polynomial of order no greater than $m$, then the approximation error is \textit{exactly} zero.
Therefore, \eqref{eq: def of nondegenerate to order kappa} can be viewed as a ``non-degenerate factors'' condition commonly imposed in the factor model literature, which is specialized to this nonlinear setting. Accordingly, the magnitude of singular values $\upsilon_{j,\subi}$'s of the noise-free structure $\bH_{\subi}^\ddagger$ is determined by  that of the loadings $(\balpha_j-\balpha_i)^{\bq}$: 
\begin{align*}
	&\upsilon_{1, \subi}\asymp_\P 1, \\
	&\upsilon_{j,\subi}\asymp_\P h_n, \text{ for } j=2, \cdots, \ttr_\alpha+1,\\
	&\qquad\vdots\\
	&\upsilon_{j,\subi}\asymp_\P h_n^m, \text{ for } j=\binom{m-1+\ttr_\alpha}{\ttr_\alpha}+1, \cdots, \binom{m+\ttr_\alpha}{\ttr_\alpha}
\end{align*}
As described in the main text, to guarantee the weakest factors are still estimable, we require $\sam^{-1}/h_{n}^{m}=o(1)$.

Note that in the main text, we define the target factors $\bF_{\subi}$ and loadings $\bLambda_{\subi}$ as the solution to the local PCA problem using the noise-free data $\bH_{\subi}^\ddagger$. By Theorem A.4 of \cite{Feng_2023_wp_Nonlinear}, $\bLambda_{\subi}$ approximates a rotation of the auxiliary monomial loadings formed by the monomial basis $(\balpha_j-\balpha_i)^{\bq}$, $|\bq|\leq m$. Therefore, the extracted factor loadings can provide approximation power for other unknown functions of the latent confounders. Therefore, Assumption \ref{Assumption: LPCA}(b) holds with $\rate{\eta}$, $\rate{\mu}$, and $\rate{e}$ all of order $\bar{o}_\P(h_n^m)$.

\subsubsection*{Approximation up to the Noise Order}
The previous discussion explicitly links the local PCA approximation with the nonparametric regression, but it sets a fixed number of approximation terms, which may not always be possible (due to degeneracy of factors or small factor loadings).
In practice, we recommend an alternative strategy that roughly allows one to achieve $\rate{\eta}=\bar{O}_\P(\sam^{-1})$, almost the best approximation rate for  $\eta_l$'s permitted by the data. 

Note that in our context, the local noise matrix $\bU_{\subi}^\ddagger$ has the largest singular value of order $\sqrt{p^\ddagger}+\sqrt{K}$. Under sufficient smoothness and nondegeneracy of the corresponding higher-order factors and loadings, we can extract all local factors and loadings associated with singular values that asymptotically dominate the noise in terms of the order. 
	To see this, consider the decomposition of $x_{jl}$:
$$
x_{jl}=\eta_l(\balpha_j)+u_{jl}=\sum_{|\bq|\leq m}\frac{\eta^{(\bq)}_l(\balpha_i)}{\bq!}(\balpha_j-\balpha_i)^{\bq}+\xi_{jl}+u_{jl}.
$$ 
Suppose that, for example, $\eta_l^{(\bq)}(\balpha_i)\approx 0$, $|\bq|=2$, for all or many $l\in[p]$, thus making us unable to ``identify'' the quadratic basis terms. However, it may be the case that the cubic term $\eta_l^{(\bq)}(\balpha_i)(\balpha_j-\balpha_i)^{\bq}$, $|\bq|=3$, is large, associated with singular values much larger than the noise order $\sqrt{p^\ddagger}+\sqrt{K}$. So we can extract the cubic terms, and reduce the approximation bias further. Finally, we stop when the remaining approximation bias is small and fails to dominate the noise.
Under the preceding conditions, any remaining Taylor term with entrywise magnitude asymptotically larger than $\sam^{-1}$ would generate a singular value that dominates the noise and hence would be extracted. Therefore, we can roughly achieve $\rate{\eta}=\bar{O}_\P(\sam^{-1})$, uniformly over all neighborhoods.

This example also shows the source of heterogeneity of $\rate{\mu}$ and $\rate{e}$ across neighborhoods. For some specific units $i$, the quadratic terms in the approximation of $\bmeta(\cdot)$ are small, and thus we miss the information about $(\balpha_j-\balpha_i)^{\bq}$, $|\bq|=2$. However, when we approximate $\mu(\cdot)$ or $e(\cdot)$, the quadratic terms may be large and dominate in bias characterization. Therefore, $\rate{\mu}$ and $\rate{e}$ are typically of order $h_n^2$. This demonstrates the importance of a key feature of our theory:
Assumption \ref{Assumption: LPCA} explicitly allows for heterogeneous approximation error, and the validity of our main results only requires the averaged approximation errors to be small.

\subsection{Comparison of Identifying Assumptions}\label{SA section: comparison of identifying assumptions}

We briefly compare the main identifying assumptions of several related approaches with those used in this paper.

\cite{miao2018identifying} uses two proxy sets: conditional on latent confounders, one is independent of treatment and the other proxy set, while the other is independent of outcome after also conditioning on treatment. A rank condition for categorical proxies or completeness for continuous proxies ensures sufficient information about latent confounders. In the continuous case, an integral equation linking the proxies to observed outcomes must also be solvable.

\cite{Deaner_2023_wp_ProxyControls} likewise uses outcome-aligned and treatment-aligned proxy sets. Conditional on latent and observed confounders, potential outcomes and outcome-aligned proxies are each independent of treatment and the treatment-aligned proxies. Both proxy sets must satisfy $L_2$-completeness, the latent confounders must have full conditional support, and at least one of the two conditional moment equations must admit a solution.

\cite{singh2020kernel} uses a negative-control treatment, which does not affect the outcome, and a negative-control outcome, which is affected by neither treatment nor the negative-control treatment. Identification also requires latent exchangeability and overlap, existence of a confounding bridge linking the outcome regression to the controls, and a completeness condition. The bridge-existence condition captures the need for both negative controls to be informative about the latent confounders.

\cite{Altonji-Mansfield_2018_AER} uses group averages of observed characteristics to absorb group-level variation in unobserved characteristics, relying on a group-choice model linking sorting on observables and unobservables.

\cite{Nagasawa_2021_wp} constructs a control function from the conditional distribution of a noisy proxy given treatment and an excluded variable. Conditional on latent confounders, potential outcomes and the noisy proxy are each independent of treatment and the excluded variable. The proxy must satisfy completeness and hence be sufficiently informative, while the excluded variable must generate enough variation to ensure overlap at comparable control-function values. Restrictions on the outcome model can relax this overlap requirement.

Our method assigns neither excluded-variable nor negative-control roles to individual proxies. Instead, many proxies must collectively satisfy Assumptions \ref{Assumption: Matching} and \ref{Assumption: LPCA}: proxy matching must select units close in latent confounders as neighbors, and proxy-based local bases must approximate the outcome and treatment functions. These approaches therefore rely on different data structures, and none is uniformly weaker or stronger.

\section{General Setup}\label{SA sec: general framework}
This section presents a theoretical framework that generalizes the one in the main text in the following aspects:
(i) a multi-valued treatment is allowed; (ii) a local quasi-maximum likelihood method, allowing for general likelihood (and link) functions, can be used to estimate the conditional means of potential outcomes and generalized propensity scores; and (iii) the conditional independence between the measurement error and the outcome/treatment is relaxed.

Suppose that a random sample 
$\{(y_{i}, t_{i}, \bx_{i}, \bz_i)\}_{i\in[n]}$ 
is available, where $y_{i}\in\mathbb{R}$ is the outcome of interest, $t_i\in\mathcal{J}=\{0,\cdots, J\}$ denotes the multi-valued treatment status, 
$\bx_i$ is a vector of noisy proxies of some unobserved confounders $\balpha_i\in\mathbb{R}^{\ttr_\alpha}$, and
$\bz_i\in\mathbb{R}^{\ttr_z}$ is a vector of observed confounders. 
The asymptotic theory is developed assuming $n$ and $p$ simultaneously increase to infinity whereas $\ttr_\alpha+\ttr_z$ is fixed.

Let $y_i(t)$ be the potential outcome of unit $i$ at treatment level $t\in\mathcal{J}$. The observed outcome is $y_{i}=\sum_{t=0}^{J}d_{i}(t)y_{i}(t)$ 
where $d_i(t)=\I(t_{i}=t)$ is an indicator for each treatment level $t\in\mathcal{J}$.
For each treatment level $t\in\mathcal{J}$, the outcome of interest is characterized by a possibly nonlinear, reduced-form model:
\begin{equation}\label{SA eq: outcome}
	y_{i}(t)=\zeta_{i,t}+\epsilon_{i,t}, \qquad \zeta_{i,t}=\psi_{\tty}(\mu_t(\balpha_i)+\bz_i'\bbeta_t), \qquad \E[\epsilon_{i,t}|\bz_i, \balpha_i]=0,
\end{equation}
where $\zeta_{i,t}$ is the conditional expectation of the potential outcome at treatment level $t$ given the observed confounders $\bz_i$ and unobserved confounders $\balpha_i$, and $\psi_{\tty}^{-1}:\mathbb{R}\to\mathbb{R}$ is a (known) link function associated with the outcome equation. Let $\bepsilon_i=(\epsilon_{i,0},\cdots,\epsilon_{i,J})$.

On the other hand, given a (known) link function $\bpsi^{-1}_{\tts}:(0,1)^{J+1}\to\mathbb{R}^{J}$ associated with the treatment equation,
setting $t=0$ as the base level, the assignment mechanism is described by
\begin{equation}\label{SA eq: treatment}
	\bm{d}_{i}=\bvrho_{i}+\bv_{i},\qquad
	\bvrho_{i}=\bpsi_{\tts}(\be(\balpha_i)+\bGamma\bz_i),\qquad 
	\E[\bv_{i}|\bz_i,\balpha_i]=0, 
\end{equation}
where we write $\bm{d}_i=(d_i(0),\cdots, d_i(J))'$, $\bvrho_i=(\varrho_{i,0},\cdots, \varrho_{i,J})'$, $\bv_i=(v_{i,0},\cdots, v_{i,J})'$, $\be(\cdot)=(e_1(\cdot),\cdots,e_J(\cdot))'$, and $\bGamma=(\bgamma_1, \cdots, \bgamma_J)'$ with $\bgamma_t\in\mathbb{R}^{\ttr_z}$ for each $t=1, \cdots, J$. Also, let $\bpsi_{\tts}=(\bpsi_{\tts,0}, \cdots, \bpsi_{\tts, J})'$.

The observed noisy covariates $\bx_i=(x_{i1},\cdots, x_{ip})'$ are linked with the latent $\balpha_i$ through a possibly nonlinear factor model:
\begin{equation}\label{SA eq: HD covariate}
	\bx_{i}=\bmeta(\balpha_i)+\bu_{i}, \quad \E[\bu_{i}|\mathcal{F}_0]=0.
\end{equation}
$\bmeta: \mathbb{R}^{\ttr_\alpha}\to\mathbb{R}^p$ is a vector of latent functions, $\bu_{i}=(u_{i1}, \cdots, u_{ip})'$ is a vector of idiosyncratic errors, and $\mathcal{F}_0=\sigma(\{\balpha_i\}_{i\in[n]})$ is the sigma field generated by $\{\balpha_i\}_{i\in[n]}$.
For simplicity, in this paper we view the latent function $\bmeta$ as fixed (conditioned upon).
Define $n\times p$ matrices $\bX=(\bx_1,\cdots, \bx_n)'$, $\bH=(\bmeta(\balpha_1),\cdots, \bmeta(\balpha_n))'$ and $\bU=(\bu_1,\cdots, \bu_n)'$.
Then, \eqref{SA eq: HD covariate} can be written in matrix form:
$\bX=\bH+\bU$.

\subsection{Estimation Procedure}\label{SA sec: estimation}

In this section, we adapt the estimation procedure described in the paper to this more general multi-valued treatment effects setup.

\begin{itemize}
	\item \textit{Step 1:  Latent-Confounder Information Extraction}. The procedure presented in Section \ref{sec: estimation}  in the paper remains the same.
	Recall that indirect matching and PCA use two subsamples obtained by proxy-wise splitting. Accordingly, the symbols $\dagger$ and $\ddagger$ denote that a quantity is based on the subsample for matching and for PCA, respectively.
	We use $\mathcal{N}_i$ to denote the set of the $K$ nearest neighbors for a unit $i$, and $\bX_{\subi}^\ddagger$, $\bH_{\subi}^\ddagger$ and $\bU_{\subi}^\ddagger$ denote the submatrices of $\bX$, $\bH$ and $\bU$ formed by units in $\mathcal{N}_i$ and proxies in the PCA block.
	
	We define
	\begin{equation*}
		(\bF_{\subi}, \bLambda_{\subi})=
		\underset{\scalebox{0.7}{ $\tilde\bF_{\subi}\in\mathbb{R}^{p^\ddagger\times \ttd_i},  \tilde\bLambda_{\subi}\in\mathbb{R}^{K\times \ttd_i}$}}
		{\argmin}\tr\Big[\Big(\bH_{\subi}^\ddagger-\tilde\bLambda_{\subi}\tilde\bF_{\subi}'\Big)\Big(\bH_{\subi}^\ddagger-\tilde\bLambda_{\subi}\tilde\bF_{\subi}')'\Big]
	\end{equation*}
	such that $\frac{1}{p^\ddagger}\tilde\bF_{\subi}'\tilde\bF_{\subi}=\bI_{\ttd_i}$ and $\frac{1}{K}\tilde\bLambda_{\subi}'\tilde\bLambda_{\subi}$ is diagonal. 
	Then, we can write
	\begin{equation*}
		\bX_{\subi}^\ddagger=\bLambda_{\subi}\bF_{\subi}'+\bXi_{\subi}+\bU_{\subi}^\ddagger,
	\end{equation*} 
	where $\bXi_{\subi}$ is the approximation error in the first step.
	
	\item \textit{Step 2: Local QMLE}. We implement more general local quasi-maximum likelihood estimation. For the outcome equation \eqref{SA eq: outcome}, consider a quasi-log-likelihood function $\mathcal{L}_{\tty}(\zeta, y)$ such that
	$\frac{\partial}{\partial \zeta}
	\mathcal{L}_{\tty}(\zeta,y)=\frac{y-\zeta}{V_{\tty}(\zeta)}$.
	Then, the estimator of $\zeta_{i,t}$ is given by
	\begin{equation}\label{SA eq: local reg of outcome}
		\begin{split}
			\widehat{\zeta}_{i,t}&=
			\psi_{\tty}
			\Big(\widehat{\mu}_t(\balpha_i)+\bz_{i}'\widehat{\bbeta}_{t,\subi}\Big),\quad \widehat{\mu}_t(\balpha_i)=
			\widehat{\blambda}_{i,\subi}'\widehat{\bb}_{t,\subi},\quad \text{where}\\[1em]
			\Big(\widehat{\bb}_{t,\subi},\;\widehat{\bbeta}_{t,\subi}\Big)&=
			\argmax_{\bb\in\mathbb{R}^{\ttd_i},\; \bbeta\in\mathbb{R}^{\ttr_z}}\;
			\sum_{j\in\mathcal{N}_i}
			d_j(t)\mathcal{L}_{\tty}\Big(\psi_{\tty}(\widehat{\blambda}_{j,\subi}'\bb+\bz_j'\bbeta),\; y_j\Big).
		\end{split}
	\end{equation}
	Similarly, for the treatment equation \eqref{SA eq: treatment}, given a quasi-likelihood $\mathcal{L}_{\tts}(\cdot,\cdot)$ associated with some functions $\{V_{\tts,t}(\cdot)\}_{t\in[J]}$ such that  $\frac{\partial}{\partial\kappa_{t}}\mathcal{L}_{\tts}(\bpsi_{\tts}(\bm{\kappa}),\bm{d})=
	\frac{d(t)-\psi_{\tts,t}(\bm{\kappa})}{V_{\tts,t}(\bm{\kappa})}$ for $\bkappa=(\kappa_1, \cdots, \kappa_J)'$, 
	the estimated conditional treatment probability is given by
	\begin{align*}
	\widehat{\bvrho}_{i}&=\bpsi_{\tts}(\widehat{\be}(\balpha_i)+\widehat{\bGamma}_{\subi}\bz_{i}),\quad
	\widehat{\be}(\balpha_i)=
	\widehat{\bB}_{\subi}\widehat{\blambda}_{i,\subi},\quad \text{where}\\[1em]
	\Big(\widehat{\bB}_{\subi},\;\widehat{\bGamma}_{\subi}\Big)&=
	\argmax_{\bB\in\mathbb{R}^{J\times\ttd_i},\,\bGamma\in\mathbb{R}^{J\times\ttr_z}}\;
	\sum_{j\in\mathcal{N}_i}
	\mathcal{L}_{\tts}\Big(\bpsi_{\tts}(\bB\widehat{\blambda}_{j,\subi}+\bGamma\bz_j),\; \bm{d}_j\Big).
	\end{align*}
	
	\item \textit{Step 3: Counterfactual Analysis}. We estimate the counterfactual means of potential outcomes, which form the basis of estimators for other causal parameters. Specifically, define $\theta_{t,s}:=\E[y_{i}(t)|t_{i}=s]$. (For $\theta_{t,t}$, we can simply define an estimator $\widehat{\theta}_{t,t}=\sum_{i=1}^nd_i(t)y_i/\sum_{i=1}^nd_i(t)$.) For $t\neq s$, note that unconfoundedness (see Assumption \ref{SA Assumption: ID-counterfactual}) implies that
	\[
	\theta_{t,s}=\E\left[\frac{d_{i}(s)\zeta_{i,t}}{\varrho_{s}}+\frac{\varrho_{i,s}}{\varrho_{s}}\frac{d_{i}(t)(y_{i}-\zeta_{i,t})}{\varrho_{i,t}}\right],
	\]
	where $\varrho_t:=\P(t_{i}=t)$ for any $t\in\mathcal{J}$.
	Then, an estimator of $\theta_{t,s}$ is given by
	\begin{equation}\label{SA eq: final estimator}
		\widehat{\theta}_{t,s}:=
		\frac{1}{n}\sum_{i=1}^{n}\left[\frac{d_{i}(s)\widehat{\zeta}_{i,t}}{\widehat{\varrho}_{s}}+\frac{\widehat{\varrho}_{i,s}}{\widehat{\varrho}_{s}}\frac{d_{i}(t)(y_{i}-\widehat{\zeta}_{i,t})}{\widehat{\varrho}_{i,t}}\right],
	\end{equation}
	where $\widehat{\zeta}_{i,t}$, $\widehat{\varrho}_{i,t}$ and $\widehat{\varrho}_{i,s}$ are obtained in the second step, and $\widehat{\varrho}_\ell=\frac{1}{n}\sum_{i=1}^{n}d_i(\ell)$ for $\ell\in\mathcal{J}$.
	For the purpose of inference, a simple plug-in variance estimator for $\widehat{\theta}_{t,s}$ is  
	\begin{equation}\label{SA eq: standard error}
		\widehat{\sigma}_{t,s}^2:=
		\frac{1}{n}\sum_{i=1}^{n}\bigg[\frac{d_i(s)(\widehat{\zeta}_{i,t}-\widehat{\theta}_{t,s})^2}{\widehat{\varrho}_{s}^2}\bigg]+
		\frac{1}{n}\sum_{i=1}^{n}\bigg[\frac{\widehat{\varrho}_{i,s}^2d_i(t)(y_i-\widehat{\zeta}_{i,t})^2}{\widehat{\varrho}_{s}^2\widehat{\varrho}^2_{i,t}}\bigg]		
	\end{equation}	
\end{itemize}

\subsection{Notation}
\textbf{Matrices.}
For a vector $\bv\in\mathbb{R}^\mathsf{d}$, $\|\bv\|=\sqrt{\bv'\bv}$ is the Euclidean norm of $\bv$, and for an $m\times n$ matrix $\bA$, $\|\bA\|=s_{\max}(\bA)$ is the operator norm and $\|\bA\|_{\max}=\max_{1\leq i\leq m, 1\leq j\leq n}|a_{ij}|$ is the entrywise sup-norm of $\bA$. $s_{\max}(\bA)$ and $s_{\min}(\bA)$ denote the largest and smallest singular values of $\bA$ respectively.
Moreover,  $\bA_{i\cdot}$ and $\bA_{\cdot j}$ denote the $i$th row and the $j$th column of $\bA$ respectively,
and $\bm{1}_{\ttd}$ denotes a $\ttd$-vector of ones.

\textbf{Asymptotics.} For sequences of numbers or random variables, we use $a_n\lesssim b_n$ or $a_n=O(b_n)$ to denote $\limsup_n|a_n/b_n|$ is finite, and $a_n\lesssim_\P b_n$ or $a_n=O_\P(b_n)$ to denote $\limsup_{\varepsilon\rightarrow\infty}\limsup_n\P[|a_n/b_n|\geq\varepsilon]=0$. $a_n=o(b_n)$ implies $a_n/b_n\rightarrow 0$, and $a_n=o_\P(b_n)$ implies that $a_n/b_n\rightarrow_\P 0$, where $\rightarrow_\P$ denotes convergence in probability. $a_n\asymp b_n$ implies that $a_n\lesssim b_n$ and $b_n\lesssim a_n$. $\rightsquigarrow$ denotes convergence in distribution.
$a_n\asymp_\P b_n$ means that $a_n\lesssim_\P b_n$ and $b_n\lesssim_\P a_n$.
Moreover, for a possibly random sequence $\{a_{i,n}\}_{i\in[n]}$ and
a strictly positive sequence $\{r_{i,n}\}_{i\in[n]}$, 
we write $a_{i,n}=\bar{O}_\P(r_{i,n})$ if 
$\max_{i\in[n]}|a_{i,n}/r_{i,n}|\lesssim_\P 1$, and 
$a_{i,n}=\bar{o}_\P(r_{i,n})$ if 
$\max_{i\in[n]}|a_{i,n}/r_{i,n}|\rightarrow_\P 0$.
``w.p.a.  1'' means ``with probability approaching one''.

\textbf{Others.}
For two numbers $a$ and $b$, $a\vee b=\max\{a,b\}$ and $a\wedge b=\min\{a,b\}$. For a collection $\mathcal{R}$ of random variables, $\sigma(\mathcal{R})$ denotes the sigma field it generates; for sigma fields $\mathcal{G}_1$ and $\mathcal{G}_2$, $\mathcal{G}_1\vee\mathcal{G}_2$ denotes the smallest sigma field containing both. We write $\sam=(K^{1/2}\wedge p^{1/2})/\sqrt{\log (n\vee p)}$. For a finite set $\mathcal{S}$, $|\mathcal{S}|$ denotes its cardinality.  
For an integer $m>0$, $[m]=\{1, 2, \cdots, m\}$.
For a sequence $\{a_{i,n}\}_{i\in[n]}$, $\E_n[a_{i,n}]=\frac{1}{n}\sum_{i=1}^{n}a_{i,n}$.

\subsubsection*{Summary of Important Quantities and Rate Conditions}
We summarize important quantities and rates below:
\begin{itemize}
	\item $\sam=(K^{1/2}\wedge p^{1/2})/\sqrt{\log(n\vee p)}$; $\sam^{-1}$ is the stochastic error scale.
	\item $h_n=(K/n)^{1/\ttr_\alpha}$ is the latent-neighborhood radius in the smooth examples of Section \ref{SA section: lpca} (when the matching exponents agree).
	\item For each neighborhood $\mathcal N_i$,
	\[
	\begin{aligned}
	\rate{\eta}&=\|\bXi_{\subi}\|_{\max},\\
	\rate{\mu}&=\max_{t\in\mathcal J}\max_{j\in\mathcal N_i}
	\big|\mu_t(\balpha_j)-\bfrb_{t,\subi}'\blambda_{j,\subi}\big|,\\
	\rate{e}&=\max_{t\in[J]}\max_{j\in\mathcal N_i}
	\big|e_t(\balpha_j)-\bfrc_{t,\subi}'\blambda_{j,\subi}\big|,
	\end{aligned}
	\]
	the local approximation errors for the proxy signal, outcome, and treatment functions, respectively.
	\item $\biaserr$ measures the dependence-induced bias and equals zero under the independence condition in the paper. Bars over $\biaserr$ and $\rate{\mu}$ denote their uniform-in-$\tau$ counterparts.
\end{itemize}
Assumption \ref{SA Assumption: LPCA} requires $\max_i\sam^{-1}/\upsilon_{\ttd_i,\subi}=o_\P(1)$, $\rate{\eta}=\bar{o}_\P(\upsilon_{\ttd_i,\subi})$, and $\rate{\mu}\vee\rate{e}=\bar{o}_\P(1)$. All three main theorems require $(np)^{2/\nu}\sam^{-2}\lesssim1$, and Theorem \ref{SA thm: QMLE} also requires $\biaserr=o_\P(1)$. The additional inference conditions are
\[
\begin{aligned}
\text{pointwise:}\quad
&\sqrt{n\log n}\left\{\sam^{-2}+\biaserr^2+
\E_n\big[\rate{\eta}^2+\rate{\mu}^2+\rate{e}^2\big]\right\}=o_\P(1),\\
\text{uniform:}\quad
&\sqrt{n\log n}\left\{\sam^{-2}+\barbiaserr^2+
\E_n\big[\rate{\eta}^2+\barrate{\mu}^2+\rate{e}^2\big]\right\}=o_\P(1).
\end{aligned}
\]

\subsubsection*{Matching Notation (Paper \& SA)}
In the main paper, we present a binary framework, and thus some notation is simplified relative to the general setup above. For clarity, we provide a mapping between quantities in the paper and in the SA:
\begin{itemize}
	\item $t_i$ (paper) $\leftrightarrow$ $d_i(1)$ (SA), and $1-t_i$ (paper) $\leftrightarrow$ $d_i(0)$ (SA);
	\item $\theta$ (paper) $\leftrightarrow$ $\theta_{0,1}$ (SA);
	\item $\theta(\tau)$ (paper) $\leftrightarrow$ $\theta_{0,1}(\tau)$ (SA);
	\item $\zeta_i$ (paper) $\leftrightarrow$ $\zeta_{i,0}$ (SA);
	\item $\zeta_{i,\tau}$ (paper) $\leftrightarrow$ $\zeta_{i,0,\tau}$ (SA);
	\item $\varrho_i$ (paper) $\leftrightarrow$ $\varrho_{i,1}$ (SA);
	\item $\varrho$ (paper) $\leftrightarrow$ $\varrho_1$ (SA);
	\item $\mu(\cdot)$ (paper) $\leftrightarrow$ $\mu_0(\cdot)$ (SA);
	\item $\mu_\tau(\cdot)$ (paper) $\leftrightarrow$ $\mu_{0,\tau}(\cdot)$ (SA);
	\item $e(\cdot)$ (paper) $\leftrightarrow$ $e_1(\cdot)$ (SA);
\end{itemize}
The quantities $\sam$, $h_n$, and $\rate{\eta}$ use the same notation in both documents. The paper's $\rate{\mu}$, $\rate{e}$, and $\barrate{\mu}$ are the corresponding binary-treatment components of the supplemental quantities, which take maxima over treatment levels.

\subsection{Assumptions}

For completeness, we collect all assumptions needed for our theoretical analysis in this section. Some of them are extensions of the analogous assumptions in the paper, which accommodate the more general framework and the estimation procedure in this supplement.

First, we impose the unconfoundedness and overlap conditions, which generalize Assumption \ref{Assumption: ID-counterfactual} in the paper.

\begin{assumption}[Unconfoundedness and Overlap] 
	\label{SA Assumption: ID-counterfactual}
	$(\{y_i(t)\}_{t\in\mathcal{J}},t_i,\balpha_i, \bz_i)$ is i.i.d. over $i\in[n]$ and satisfies that (a) $y_i(t)\inde d_i(s)\,|\, \balpha_i, \bz_i$, $\forall t,s\in\mathcal{J}$;
	and (b) for all $t\in\mathcal{J}$, $\P(t_i=t|\balpha_i, \bz_i)\geq c_{\min}>0$ for some constant $c_{\min}$ almost surely.
\end{assumption}

Second, we impose some regularity conditions on covariates, errors and latent functions, analogously to Assumption \ref{Assumption: Regularities} in the paper. 
Let $\mathcal{F}_1$ be some sigma field on which $\{(\mathcal{N}_i, \widehat{\bLambda}_{\subi})\}_{i\in[n]}$ are measurable, and
define $\mathcal{F}=\mathcal{F}_0\vee\mathcal{F}_1\vee\sigma(\{\bz_i\}_{i\in[n]})$.
Let $\bZ_{\subi}$ have rows $\bz_j'$, $j\in\mathcal{N}_i$, in the same order as $\bLambda_{\subi}$, and let $\bP_{\subi}$ be the orthogonal projection onto the column space of $\bLambda_{\subi}$.

\begin{assumption}[Regularities] 
	\label{SA Assumption: Regularities}
	$\balpha_i$ has a compact convex support $\mathcal{A}$ with a density bounded and bounded away from zero; $\bz_i$ has a compact support, and $K^{-1}\bZ_{\subi}'(\bI_K-\bP_{\subi})\bZ_{\subi}$ has eigenvalues uniformly bounded away from zero in probability over $i\in[n]$;
	the errors satisfy $\max_{i\in [n], l\in [p]}
	\E[|u_{il}|^{2+\nu}|\mathcal{F}_0]\lesssim_\P 1$, and $\max_{i\in [n]}
	\E[\|\bepsilon_{i}\|^{2+\nu}|\mathcal{F}]\lesssim_\P 1$, for some $\nu>2$;
	for all $l\in[p]$, $\eta_l(\cdot)$ is bounded by a universal constant.
\end{assumption}
The local Gram condition is vacuous when $\bz_i$ is absent; Section \ref{SA section: local design} gives sufficient conditions for it.

Next, we impose an assumption restricting the errors in our model, analogously to Assumption \ref{Assumption: indep of errors} in the paper.
\begin{assumption}[Independence of Errors]\label{SA Assumption: indep of errors}
	$u_{il}$ is independent across $i\in[n]$ and over $l\in[p]$, conditional on $\mathcal{F}_0$;
	$(\bepsilon_i, \bv_i)$ is independent over $i$ conditional on $\mathcal{F}$.
\end{assumption}

\begin{remark}
	Compared to Assumptions \ref{Assumption: Regularities} and \ref{Assumption: indep of errors} in the paper, Assumptions \ref{SA Assumption: Regularities} and \ref{SA Assumption: indep of errors} do \textit{not} assume the measurement error $\bu_i$ is (conditionally) independent of outcome and treatment errors $(\bepsilon_i, \bv_i)$. Instead, later we will introduce a quantity $\biaserr$ to characterize the correlation of $(\bepsilon_i, \bv_i)$ with $\bu_i$, reflected through the correlation with the extracted information $(\mathcal{N}_i, \widehat{\bLambda}_{\subi})$. 
	Our results are valid if $\biaserr^2=o_\P(n^{-1/2})$. Also, the moment restriction on $\bepsilon_i$ now is imposed conditionally on confounders $(\balpha_i, \bz_i)$ \textit{and}  $(\mathcal{N}_i, \widehat{\bLambda}_{\subi})$.
	
	In the main paper, since we assume $\bu_i$ is (conditionally) independent of $(\bepsilon_i, \bv_i)$, $\biaserr=0$ exactly, and the moment condition
	$\E[\|\bepsilon_{i}\|^{2+\nu}|\mathcal{F}]$ is exactly the same as $\E[\|\bepsilon_{i}\|^{2+\nu}|\{(\balpha_i, \bz_i)\}_{i\in[n]}]$. So Assumptions \ref{SA Assumption: Regularities} and \ref{SA Assumption: indep of errors} are immediately implied.
\end{remark}

Next, we impose a high-level condition on the indirect matching step, which is the same as Assumption \ref{Assumption: Matching} in the paper.
\begin{assumption}[Indirect Matching] 
	\label{SA Assumption: Matching}\leavevmode
	For some fixed constants $\rho_0\in\mathbb{R}$, $\lvarsig>0$, $\uvarsig>0$ and some positive sequence $a_p=o(1)$, 
	the following conditions hold:
	\begin{enumerate}[label=(\alph*)]		
		\item $\underset{i,j\in[n]:\,i\neq j}{\max}\;|\rho(\bX_{i\cdot}^\dagger, \bX_{j\cdot}^\dagger)-
		\rho(\bH_{i\cdot}^\dagger, \bH_{j\cdot}^\dagger)
		-\rho_0|\lesssim_\P a_p$;
		
		\item 
		$\underset{i,j\in[n]:\,\balpha_i\neq \balpha_j}{\max}\;\frac{\rho(\bH_{i\cdot}^\dagger,\, \bH_{j\cdot}^\dagger)}{\|\balpha_i-\balpha_j\|^{\uvarsig}}
		\lesssim_\P 1$ and 
		$\underset{i,j\in[n]:\,\balpha_i\neq \balpha_j}{\min}\;\frac{\rho(\bH_{i\cdot}^\dagger,\, \bH_{j\cdot}^\dagger)}{\|\balpha_i-\balpha_j\|^{\lvarsig}}\gtrsim_\P 1$.
	\end{enumerate}
\end{assumption}

Next, we impose a high-level condition on the local approximation of the nonlinear factor model, corresponding to Assumption \ref{Assumption: LPCA} in the paper.
Let $\bUpsilon_{\subi}:=(K^{-1}\bLambda_{\subi}'\bLambda_{\subi})^{1/2}=\diag\{\upsilon_{1,\subi},\cdots,\upsilon_{\ttd_i,\subi}\}$, where $\upsilon_{1,\subi}\geq\cdots\geq\upsilon_{\ttd_i,\subi}>0$. Thus, $K^{-1}(\bLambda_{\subi}\bUpsilon_{\subi}^{-1})'(\bLambda_{\subi}\bUpsilon_{\subi}^{-1})=\bI_{\ttd_i}$.
\begin{assumption}[Local Approximation] \label{SA Assumption: LPCA}
	For each $i\in[n]$, the following conditions hold:
	\begin{enumerate}[label=(\alph*)]
		\item
		$\max_{i\in[n]}\|\bF_{\subi}\|_{\max}\lesssim_\P1$ and $\underset{i\in[n]}{\max}\|\bLambda_{\subi}\bUpsilon_{\subi}^{-1}\|_{\max}\lesssim_\P 1$. 
		Also, either $\upsilon_{j,\subi}/\upsilon_{j+1,\subi}\lesssim_\P 1$ or $\upsilon_{j+1,\subi}/\upsilon_{j,\subi}=o_\P(1)$ for $j\in[\ttd_i-1]$, and
		$\max_{i\in[n]}\sam^{-1}/\upsilon_{\ttd_i,\subi}=o_\P(1)$.
		\item 
		$\rate{\eta}:=\|\bXi_{\subi}\|_{\max}=\bar{o}_\P(\upsilon_{\ttd_i,\subi})$, and  
		there exist some $\bfrb_{t,\subi}, \bfrc_{t,\subi}\in\mathbb{R}^{\ttd_i}$, uniformly bounded in probability over $i$ and $t$, such that
		$\rate{\mu}:=\max_{t\in\mathcal{J}}\max_{j\in\mathcal{N}_i}|\mu_t(\balpha_j)-\bfrb_{t,\subi}'\blambda_{j,\subi}|=\bar{o}_\P(1)$ and 
		$\rate{e}:=\max_{t\in[J]}\max_{j\in\mathcal{N}_i}|e_t(\balpha_j)-\bfrc_{t,\subi}'\blambda_{j,\subi}|=\bar{o}_\P(1)$.
	\end{enumerate}
\end{assumption}

Next, we impose an assumption on the regularities of the local QMLE.
Let $\overline{\mathcal{Y}}$ denote the support of $y_i$, and 
$\psi_{\tty}^{(1)}$ and $\psi_{\tts,t}^{(1)}$ denote their first derivatives. In addition, define  $\mathcal{L}_{\ell,\tty}(\kappa, y)=\frac{\partial^\ell}{\partial \kappa^\ell}\mathcal{L}_{\tty}(\psi_{\tty}(\kappa), y)$ for $\ell=1,2$,   $\bm{\mathcal{L}}_{1,\tts}(\bm{\kappa},\bm{d})=\frac{\partial}{\partial \bm{\kappa}}\mathcal{L}_{\tts}(\bpsi_{\tts}(\bm{\kappa}), \bm{d})$, and 
$\bm{\mathcal{L}}_{2,\tts}(\bm{\kappa},\bm{d})=\frac{\partial^2}{\partial \bm{\kappa}\partial\bm{\kappa}'}\mathcal{L}_{\tts}(\bpsi_{\tts}(\bm{\kappa}), \bm{d})$.
\begin{assumption}[Local QMLE] \label{SA Assumption: QMLE}
	\leavevmode
	\begin{enumerate}[label=(\alph*), ref={\theassumption(\alph*)}]
		\item \label{assump qmle a} For some fixed constant $\Delta>0$,
		\[
		\E\bigg[\max_{1\leq i\leq n}
		\frac{1}{K}\sum_{j\in\mathcal{N}_i}\sup_{0<|\tilde{\kappa}-\kappa|\leq \Delta}\frac{|\mathcal{L}_{2,\tty}(\tilde{\kappa},y_{j})-\mathcal{L}_{2,\tty}(\kappa,y_{j})|}{|\tilde{\kappa}-\kappa|}\Big|\mathcal{F}
		\bigg]
		\lesssim_\P 1,
		\]
		and the same condition also holds for every element of  $\bm{\mathcal{L}}_{2,\tts}(\bm{\kappa},\bm{d})$;
		
		\item $\mathcal{L}_{2,\tty}(\kappa,y)<0$ for $\kappa\in\mathbb{R}$ and $y\in\overline{\mathcal{Y}}$, and $\bm{\mathcal{L}}_{2,\tts}(\bm{\kappa},\bm{d})$ is negative definite for $\bm{\kappa}\in\mathbb{R}^{J}$ and $\bm{d}\in\{0,1\}^{J+1}$;
		
		\item For all $t\in\mathcal{J}$, $\psi_{\tty}(\cdot)$, $\psi_{\tts,t}(\cdot)$, $V_{\tty}(\cdot)$ and $V_{\tts,t}(\cdot)$ are twice continuously differentiable, 
		$\psi_{\tty}^{(1)}(\mu_t(\balpha)+\bz'\bbeta_t)$ and 
		$\psi_{\tts,t}^{(1)}(\be(\balpha)+\bGamma\bz)$ are nonzero, 
		$V_{\tty}(\psi_{\tty}(\mu_t(\balpha)+\bz'\bbeta_t))>0$, and 
		$V_{\tts,t}(\bm{e}(\balpha)+\bGamma\bz)>0$ 
		over the support of $\balpha$ and $\bz$.
		
	\end{enumerate}
\end{assumption}
\begin{remark}
	Recall that in the main paper, the outcome model is estimated using the Gaussian likelihood  (local least squares), and the treatment model is estimated using the Logit likelihood (local logit). Assumption \ref{SA Assumption: QMLE} immediately follows in that special case, and therefore, we do not need extra conditions in the paper.
\end{remark}

Finally, when uniform inference on counterfactual distributions is of interest, we impose the following assumption that strengthens some conditions imposed before. Let $\mathcal{Y}\subset\mathbb{R}$ be a compact set of thresholds for outcome values.
\begin{assumption}[Uniform Inference] \label{SA Assumption: QMLE, uniform}
	\leavevmode
	\begin{enumerate}[label=(\alph*), ref={\theassumption(\alph*)}]
		\item $\mu_{t,\tau}(\balpha)$ is Lipschitz in $\tau$ uniformly over $t\in\mathcal{J}$ and $\balpha\in\mathcal{A}$, and $\mu_{t,\tau}(\balpha)+\bz'\bbeta_{t,\tau}$ lies in a common compact set on which $|\psi_{\tty}^{(1)}(\cdot)|$ and $V_{\tty}(\psi_{\tty}(\cdot))$ are bounded away from zero.
		
		\item 
		There exists some $\bfrb_{t,\tau,\subi}\in\mathbb{R}^{\ttd_i}$, 
		uniformly bounded in probability over $i$, $t$, and $\tau$, such that
		$\barrate{\mu}:=\max_{t\in\mathcal{J}}\sup_{\tau\in\mathcal{Y}}\max_{j\in\mathcal{N}_i}|\mu_{t,\tau}(\balpha_j)-\bfrb_{t,\tau,\subi}'\blambda_{j,\subi}|=\bar{o}_\P(1)$.
			
		\item For some fixed constant $\Delta>0$,
		\[
		\E\bigg[\sup_{\tau\in\mathcal{Y}}\max_{1\leq i\leq n}
		\frac{1}{K}\sum_{j\in\mathcal{N}_i}\sup_{0<|\tilde{\kappa}-\kappa|\leq \Delta}\frac{|\mathcal{L}_{2,\tty}(\tilde{\kappa},y_{j,\tau})-\mathcal{L}_{2,\tty}(\kappa,y_{j,\tau})|}{|\tilde{\kappa}-\kappa|}\bigg|
		\mathcal{F}\bigg]
		\lesssim_\P 1;
		\]
	\end{enumerate}
\end{assumption}

\section{Theoretical Results}\label{SA section: theory}

Define 
$$\biaserr:=\max_{i\in[n]}\;(\|\E[\bepsilon_{i}|\mathcal{F}, \{\bm d_i\}_{i\in[n]}]\|+\|\E[\bv_{i}|\mathcal{F}]\|),$$
which will be a key quantity that determines the bias due to the potential correlation between the measurement error and the outcomes (or treatment assignment). Under Assumption \ref{Assumption: indep of errors} in the main paper, $\biaserr=0$.

The following is our first main result, which shows that the conditional means of potential outcomes and the propensity scores can be consistently estimated by the proposed local quasi-maximum likelihood method.

\begin{thm}[Local QMLE]\label{SA thm: QMLE}
	Suppose that Assumptions \ref{SA Assumption: ID-counterfactual}--\ref{SA Assumption: QMLE} hold. In addition, 
	$(np)^{\frac{2}{\nu}}\sam^{-2}\lesssim 1$ and $\biaserr=o_\P(1)$.
	Then, for each $t\in\mathcal{J}$,  
	$$
	|\widehat{\zeta}_{i,t}-\zeta_{i,t}|=
	\bar{O}_\P (\sam^{-1}+\rate{\eta}+\rate{\mu}+\biaserr)\quad \text{and}\quad
	|\widehat{\varrho}_{i,t}-\varrho_{i,t}|=
	\bar{O}_\P(\sam^{-1}+\rate{\eta}+\rate{e}+\biaserr). 
	$$
	A detailed asymptotic expansion is given by Equation \eqref{eq: asymptotic expansion} in the proof.
\end{thm}

\begin{thm}[Causal Inference] 
	\label{SA thm: pointwise inference}
	Suppose that Assumptions \ref{SA Assumption: ID-counterfactual}--\ref{SA Assumption: QMLE} hold. If 
	$(np)^{\frac{2}{\nu}}\sam^{-2}\lesssim 1$ and 
	$\sqrt{n\log n}(\sam^{-2}+\biaserr^2+
	\E_n[\rate{\mu}^2+
	\rate{e}^2+\rate{\eta}^2]
	)=o_\P(1)$, then
	\begin{enumerate}
		\item 
		$\sqrt{n}(\widehat{\theta}_{t,s}-\theta_{t,s})=
		\frac{1}{\sqrt{n}}\sum_{i=1}^{n}\varphi_{i,t,s}
		+o_\P(1)$
		where $\varphi_{i,t,s}=
		\frac{d_i(s)(\zeta_{i,t}-\theta_{t,s})}{\varrho_{s}}+
		\frac{\varrho_{i,s}}{\varrho_{s}}\frac{d_i(t)(y_{i}-\zeta_{i,t})}{\varrho_{i,t}}$;
		
		\item 	If $\sigma_{t,s}^2:=\V[\varphi_{i,t,s}]>0$, then $\sqrt{n}(\widehat{\theta}_{t,s}-\theta_{t,s})/\widehat{\sigma}_{t,s}\rightsquigarrow\mathsf{N}(0,1)$.
		
	\end{enumerate}	
\end{thm}

Next, we discuss uniform inference on counterfactual distributions, the analysis of which relies on a
particular function class $\mathcal{G}=\{y\mapsto \I(y\leq \tau):\tau\in\mathcal{Y}\}$. Each $g(\cdot)\in\mathcal{G}$ corresponds to a particular value $\tau\in\mathcal{Y}$. Therefore, we write  $y_{i,\tau}(t)=\I(y_i(t)\leq \tau)$ and $y_{i,\tau}=\I(y_i\leq \tau)$. Accordingly, Equation \eqref{SA eq: outcome} becomes $$y_{i,\tau}(t)=\zeta_{i,t,\tau}+\epsilon_{i,t,\tau},\quad\zeta_{i,t,\tau}=\psi_{\tty}\big(\mu_{t,\tau}(\balpha_i)+\bz_i'\bbeta_{t,\tau}\big),$$
where $\zeta_{i,t,\tau}=\P(y_i(t)\leq \tau|\balpha_i, \bz_i)$. 
Let $\bepsilon_{i,\tau}=(\epsilon_{i,0,\tau},\cdots,\epsilon_{i,J,\tau})'$.
For each $\tau$, the second step of the estimation procedure in Section \ref{SA sec: estimation} is implemented to obtain an estimator $\widehat{\zeta}_{i,t,\tau}$ of $\zeta_{i,t,\tau}$. The parameter of interest is 
$\theta_{t,s}(\tau)=\E[\I(y_i(t)\leq \tau)|t_i=s]$, the counterfactual distribution function of the potential outcome $y_i(t)$ for the group with treatment status $s$. 
From the perspective of uniform inference, $\theta_{t,s}(\cdot)$ is a parameter in $\ell^\infty(\mathcal{Y})$, a function space of bounded functions on $\mathcal{Y}$ equipped with sup-norm.

The following theorem shows that the (scaled) counterfactual distribution process weakly converges to a limiting Gaussian process indexed by $\tau\in\mathcal{Y}$, which forms the basis of uniform inference. See \cite{vandeVaart_book_1996} for underlying technical details. Define
$$\barbiaserr:=\max_{i\in[n]}\;(\sup_{\tau\in\mathcal{Y}}\|\E[\bepsilon_{i,\tau}|\mathcal{F}, \{\bm d_i\}_{i\in[n]}]\|+\|\E[\bv_{i}|\mathcal{F}]\|),$$

\begin{thm}[Uniform Inference]\label{SA thm: uniform inference}
	Under Assumptions \ref{SA Assumption: ID-counterfactual}--\ref{SA Assumption: QMLE, uniform}, if  $(np)^{\frac{2}{\nu}}\sam^{-2}\lesssim 1$ and
	$\sqrt{n\log n}(\sam^{-2}+\barbiaserr^2+
	\E_n[\barrate{\mu}^2+
	\rate{e}^2+\rate{\eta}^2])=o_\P(1)$, 
	then
	\[
	\sqrt{n}\Big(\widehat{\theta}_{t,s}(\cdot)-\theta_{t,s}(\cdot)\Big)=
	\frac{1}{\sqrt{n}}\sum_{i=1}^n\varphi_{i,t,s}(\cdot)+o_\P(1)\rightsquigarrow \mathsf{Z}_{t,s}(\cdot) \quad \text{in } \; \ell^\infty(\mathcal{Y}),
	\]
	where $\varphi_{i,t,s}(\cdot)=
	\frac{d_i(s)(\zeta_{i,t,\cdot}-\theta_{t,s}(\cdot))}{\varrho_{s}}+
	\frac{\varrho_{i,s}}{\varrho_{s}}\frac{d_i(t)(y_{i,\cdot}-\zeta_{i,t,\cdot})}{\varrho_{i,t}}$ and $\mathsf{Z}_{t,s}(\cdot)$ is a zero-mean Gaussian process with covariance kernel $\E[\varphi_{i,t,s}(\tau_1)\varphi_{i,t,s}(\tau_2)]$
	for $\tau_1,\tau_2\in\mathcal{Y}$.
\end{thm}

Under proper regularity conditions, the weak convergence above can be applied to construct inference procedures for other quantities such as quantile treatment effects by the functional delta method, as shown in the following corollary.

\begin{coro}[Functional Delta Method]\label{SA coro: functional delta method}
	Let the conditions in Theorem \ref{SA thm: uniform inference} hold. Consider the parameter $\tilde\theta$ as an element of a
	parameter space $D_{\tilde\theta}\subseteq \ell^\infty(\mathcal{Y})$, a function space of bounded functions on $\mathcal{Y}$, with $D_{\tilde\theta}$ containing the true value $\theta_{t,s}$. Let a functional $\Psi(\tilde\theta)$ mapping $D_{\tilde\theta}$ to $\ell^\infty(\mathcal{Q})$ for some index set $\mathcal Q$ be Hadamard differentiable in $\tilde\theta$ at $\theta_{t,s}$ with derivative
	$\dot{\Psi}_\theta$. Then,
	$\sup_{q\in\mathcal Q}|\sqrt{n}\{\Psi(\widehat{\theta}_{t,s})(q)-\Psi(\theta_{t,s})(q)\}-
	n^{-1/2}\sum_{i=1}^{n}\dot{\Psi}_{\theta}(\varphi_{i,t,s})(q)|=o_\P(1)$ and $\sqrt{n}(\Psi(\widehat{\theta}_{t,s})(\cdot)-\Psi(\theta_{t,s})(\cdot))\rightsquigarrow\dot{\Psi}_\theta(\mathsf{Z}_{t,s})(\cdot)$ that is a Gaussian process  in $\ell^\infty(\mathcal{Q})$ with mean zero and covariance kernel defined by the limit of the second
	moment of $\dot{\Psi}_\theta(\varphi_{i,t,s})$.
\end{coro}

The limiting Gaussian process can be approximated based on a practically feasible multiplier bootstrap procedure. Take an i.i.d. sequence of random variables $\{\omega_i\}_{i\in[n]}$ independent of the data with mean zero and variance one. Define a uniformly consistent estimator of $\varphi_{i,t,s}(\cdot)$:
$$\widehat{\varphi}_{i,t,s}(\cdot)=
\frac{d_i(s)(\widehat{\zeta}_{i,t,\cdot}-\widehat{\theta}_{t,s}(\cdot))}{\widehat{\varrho}_{s}}+
\frac{\widehat{\varrho}_{i,s}}{\widehat{\varrho}_{s}}\frac{d_i(t)(y_{i,\cdot}-\widehat{\zeta}_{i,t,\cdot})}{\widehat{\varrho}_{i,t}}.$$ 
The following corollary shows that conditional on the data, $\frac{1}{\sqrt{n}}\sum_{i=1}^n\omega_i\widehat{\varphi}_{i,t,s}(\cdot)$ weakly converges to the same limiting process $\mathsf{Z}_{t,s}(\cdot)$ as in Theorem \ref{SA thm: uniform inference}. In practice, one only needs to simulate this feasible approximation process
by taking random draws of $\{\omega_i\}_{i\in [n]}$.

\begin{coro}[Multiplier Bootstrap] \label{SA coro: multiplier bootstrap}
	Let the conditions of Theorem \ref{SA thm: uniform inference} hold. Then, conditional on the data, $n^{-1/2}\sum_{i=1}^{n}\omega_i\widehat{\varphi}_{i,t,s}(\cdot)\rightsquigarrow\mathsf{Z}_{t,s}(\cdot)$ that is the Gaussian process defined in Theorem \ref{SA thm: uniform inference}, with probability approaching one. 
\end{coro}

\section{Additional Numerical Results}\label{SA-sec: numerical}

\subsection{Additional Simulation Results}
We consider six specifications of $\eta_l(\cdot)$'s:
\begin{itemize}
	\item Model 1: $\eta_l(\alpha_{i1})=-4.5\,(\alpha_{i1} \varpi_{l1}) + 18.4\,(\alpha_{i1} \varpi_{l1})^2 - 16.2\,(\alpha_{i1} \varpi_{l1})^3 $.
	\item Model 2: $\eta_l(\alpha_{i1})=2\exp\!\left(-10(\alpha_{i1} - \varpi_{l1})^2\right)$.
	\item Model 3: $\eta_l(\alpha_{i1})=2\exp\!\left(-10|\alpha_{i1} - \varpi_{l1}|\right)$.
	\item Model 4: $\eta_l(\balpha_i)=2s_{il} + s_{il}^2 + \tfrac{1}{2}s_{il}^3$, $s_{il}
	= \tilde{\alpha}_{i1}\tilde{\varpi}_{l1}+\tilde{\alpha}_{i2}\tilde{\varpi}_{l2}$, $\tilde{\alpha}_{ij}=2(\alpha_{ij}-0.5), \tilde{\varpi}_{lj}=2(\varpi_{lj}-0.5)$ for $j=1,2$.
	\item Model 5: $\eta_l(\balpha_i)=2\exp\!\left(-10[(\alpha_{i1}-\varpi_{l1})^2 + (\alpha_{i2}-\varpi_{l2})^2]\right)$.
	\item Model 6: $\eta_l(\balpha_i)=2\exp\!\left(-2[|\alpha_{i1}-\varpi_{l1}| + |\alpha_{i2}-\varpi_{l2}|]\right)$.
\end{itemize}
The first three have one latent confounder $\alpha_{i1}$, and the last three have two latent confounders $\alpha_{i1}, \alpha_{i2}$. 
All the latent variables $\alpha_{i1}, \alpha_{i2}, \varpi_{l1}, \varpi_{l2}$ follow $\mathsf{U}[0,1]$ and are independent of each other.

For Models 1--3, we generate potential outcomes and binary treatment as follows:
\begin{align*}
	&y_i(0)=\alpha_{i1}+\alpha_{i1}^2+\epsilon_{i,0},\quad y_i(1)=2\alpha_{i1}+\alpha_{i1}^2+1+\epsilon_{i,1}, \\
	&t_i=\I(v_i\leq \varrho_i), \;\;
	\varrho_i=\psi((\alpha_{i1}-0.5)+(\alpha_{i1}-0.5)^2),
\end{align*}
where $\psi$ is the standard logistic function, $v_i\sim\mathsf{U}[0,1]$, $\epsilon_{i,0}$, $\epsilon_{i,1}\sim\mathsf{N}(0,1)$, and they are independent of each other.
For Models 4--6, we use the following specification:
\begin{align*}
	&y_i(0)=2 + 10\sin\!\bigl(\pi \alpha_{i1} \alpha_{i2}\bigr) + 2(\alpha_{i2} - 0.5)^2 + 2\alpha_{i1}+\epsilon_{i,0}, \\
	&y_i(1)=\mu_0(\alpha_{i1},\alpha_{i2})+1 + 0.6(\alpha_{i1} - 0.5) + 0.4\sin(2\pi \alpha_{i2})
	+ 0.3(\alpha_{i1} - 0.5)(\alpha_{i2} - 0.5)+\epsilon_{i,1},\\
	&\varrho_i=\psi(-0.4 + 1.2(\alpha_{i1}-0.5) + 1.2(\alpha_{i2}-0.5)
	+ 0.8\sin(2\pi\alpha_{i1}) - 0.8(\alpha_{i2}-0.5)^2).
\end{align*}

We check the robustness of our methods with respect to measurement error features: 
\begin{itemize}
	\item i.i.d. error:  $u_{il}\sim\text{ i.i.d. }\mathsf{N}(0,1)$ over $i\in[n], l\in[p]$; 
	\item heteroskedastic error: each $u_{il}=\exp(0.1((\alpha_{i1}-0.5)+(\varpi_{l1}-0.5)))\times\zeta_{il}$ for $\zeta_{il}\sim\text{ i.i.d. }\mathsf{N}(0,1)$;
	\item AR(1) error: for each $i\in[n]$, the sequence $(u_{i1}, u_{i2}, \cdots, u_{ip})$ follows a zero-mean, unit-variance AR(1) process with the coefficient equal to $0.3$, and then we let $\epsilon_{i,t}=0.3u_{ip}+\zeta_{i,t}$ for $\zeta_{i,t}\sim\mathsf{N}(0, 1-0.3^2)$, $t=0,1$, to generate correlation between $u_{ip}$ and the outcome errors.
	
\end{itemize}

We consider 2\,000 simulated datasets for each specification. For each simulated dataset,
a point estimate of $ATT= \E[y_i(1)-y_i(0)|t_i = 1]$ is obtained. We report root mean squared error (RMSE), coverage rate (CR)
of the nominal 95\% confidence interval and its average length (AL). 
``Local PCA'' refers to results based on our proposed method, and ``local constant'' refers to the approach that estimates conditional means of potential outcomes and propensity scores by local averaging. Results are reported in Tables \ref{table:simul 1}--\ref{table:simul loc const} at the end of this supplement.

\subsection{Additional Diagnostics for the Empirical Application}
\label{SA subsection: application diagnostics}

In this section, we provide additional diagnostics for our empirical application.

\subsubsection{Sensitivity analyses}
We compare the ATT and its standard error across alternative local-dimension thresholds and contiguous proxy-wise splits, and after dropping control firms with the largest normalized matching discrepancies. In the poor-match trimming checks, all treated firms remain in the sample, and matching and local estimation are repeated. Table \ref{SA table: application sensitivity} reports the results at the benchmark $K=99$.

\begin{table}[!htbp]
\centering
\small
\caption{Sensitivity to Dimension Selection, Proxy Splitting, and Trimming}
\label{SA table: application sensitivity}
\begin{tabular}{lcccc}
\hline\hline
& \multicolumn{2}{c}{No covariates} & \multicolumn{2}{c}{With covariates} \\
\cline{2-3}\cline{4-5}
Specification & Full & Base & Full & Base \\
\hline
Benchmark 125/125 & 0.105 (0.028) & 0.087 (0.026) & 0.049 (0.027) & 0.074 (0.032) \\
$0.75\times$ dimension threshold & 0.131 (0.036) & 0.091 (0.033) & 0.047 (0.027) & 0.069 (0.033) \\
$1.25\times$ dimension threshold & 0.106 (0.028) & 0.087 (0.026) & 0.048 (0.027) & 0.066 (0.035) \\
Swapped halves & 0.109 (0.028) & 0.099 (0.027) & 0.061 (0.026) & 0.079 (0.026) \\
Matching 100, PCA 150 & 0.106 (0.028) & 0.093 (0.027) & 0.075 (0.026) & 0.081 (0.027) \\
Matching 150, PCA 100 & 0.107 (0.028) & 0.086 (0.026) & 0.051 (0.028) & 0.074 (0.029) \\
Trim top 10\% & 0.101 (0.028) & 0.101 (0.026) & 0.088 (0.027) & 0.091 (0.026) \\
Trim top 5\% & 0.105 (0.028) & 0.101 (0.026) & 0.081 (0.026) & 0.089 (0.026) \\
Propensity trim $>0.9$ & -- & -- & 0.048 (0.028) & 0.061 (0.032) \\
\hline
\end{tabular}
\par\smallskip\noindent\parbox{\textwidth}{\footnotesize\textit{Notes:} Entries are ATT estimates with standard errors in parentheses. The benchmark uses the first 125 return dates for matching and the last 125 for PCA. The two dimension-threshold rows multiply the benchmark selection threshold by the indicated factor, holding other choices fixed. ``Swapped halves'' interchanges the roles of the two proxy blocks; the 100/150 and 150/100 rows use the first 100 or 150 dates for matching and the remaining dates for PCA. Dates remain chronological within each block. ``Trim top 10\%'' and ``Trim top 5\%'' exclude controls at or above the 90th and 95th percentiles of normalized maximum matching discrepancy, respectively: 57 and 52 controls in the full and base samples under the 10\% rule, and 29 and 26 under the 5\% rule. No treated firm is dropped. All rows use the pseudo-max distance; rows other than the two dimension-threshold checks use the benchmark local-dimension rule. ``Propensity trim $>0.9$'' drops one treated firm selected by the covariate-adjusted benchmark score in each sample, then repeats matching and local estimation; it estimates the ATT for the remaining treated firms. Without covariates, no treated firm exceeds 0.9, so an analogous trim would drop no firm and is not reported.}
\end{table}

The ATT remains positive in these checks and comparable to the benchmark results reported in the paper, although its magnitude is more sensitive when observed covariates are included. 
For example, trimming the top 10\% of poorly matched controls raises the full-sample covariate-adjusted estimate from 0.049 to 0.088.
The dimension-threshold results provide some reassurance about the point estimates: the $0.75\times$ rule selects substantially more local components (mean selected dimensions are 2.707 and 2.941 in the full and base samples, respectively), while the $1.25\times$ rule selects one component in every neighborhood, and all corresponding ATT estimates remain positive. However, we note that the covariate-adjusted local-logit fits under the $0.75\times$ rule exhibit more frequent numerical nonconvergence. The separate checks varying the distance function also yield positive estimates, while Table \ref{table:empirical illustration} in the main paper reports the existing broader sensitivity to $K$.

We also compare the nearest-neighbor sets and local loading spaces across proxy-wise splits. Neighbor overlap is the number of firms shared by the benchmark and alternative neighborhoods divided by the number of firms in the benchmark neighborhood (the center firm of the neighborhood is excluded). For the loading-space comparison, we hold the benchmark neighborhood fixed and repeat PCA using the alternative proxy block. The distance is the Frobenius norm of the difference between orthogonal projections onto the two selected loading spaces, divided by the square root of the sum of their dimensions. It is zero for identical spaces and one for orthogonal spaces, and does not depend on factor rotations. Table \ref{SA table: application split stability} reports the medians of both measures.

\begin{table}[!htbp]
\centering
\small
\caption{Stability of Neighbor Sets and Local Loading Spaces Across Proxy Splits}
\label{SA table: application split stability}
\begin{tabular}{llcccc}
\hline\hline
& & \multicolumn{2}{c}{Neighbor overlap} & \multicolumn{2}{c}{Loading-space distance} \\
\cline{3-4}\cline{5-6}
Sample & Split & Treated & All & Treated & All \\
\hline
Full & Swapped halves & 0.398 & 0.398 & 0.260 & 0.367 \\
Full & Matching 100, PCA 150 & 0.689 & 0.714 & 0.033 & 0.032 \\
Full & Matching 150, PCA 100 & 0.847 & 0.827 & 0.018 & 0.020 \\
Base & Swapped halves & 0.429 & 0.408 & 0.289 & 0.569 \\
Base & Matching 100, PCA 150 & 0.694 & 0.714 & 0.040 & 0.036 \\
Base & Matching 150, PCA 100 & 0.852 & 0.827 & 0.023 & 0.023 \\
\hline
\end{tabular}
\par\smallskip\noindent\parbox{\textwidth}{\footnotesize\textit{Notes:} Entries are medians across treated-centered neighborhoods (22 in the full sample and 12 in the base sample) or all neighborhood centers (583 in the full sample and 526 in the base sample). The neighbor-overlap comparison excludes each center. Loading spaces are compared on the same benchmark neighborhood so that the alternative PCA block, rather than a change in the matched firms, determines this distance. All splits retain chronological order within each proxy block and use $K=99$.}
\end{table}

Swapping the two proxy halves yields relatively lower neighbor overlap and larger loading-space distances, as expected because it changes the matching and PCA blocks most substantially. The other splits, which only move the boundary between the two blocks, show much greater stability.

\subsubsection{Prediction improvement}
To examine whether proxy-based local factors add predictive information beyond the observed covariates, we conduct five-fold out-of-sample prediction of returns in a held-out proxy block for treated firms. The matching, PCA, and prediction blocks are $[-280,-181]$, $[-180,-81]$, and $[-80,-31]$, respectively; they are disjoint, contiguous, and use all 250 proxy dates. We also predict the observed untreated event outcome CAR[0,1] for held-out control firms. Table \ref{SA table: application prediction} compares the local-PCA approach with linear and nearest-neighbor predictors using only observed covariates; the test-fold firms are excluded when constructing the local factors and fitting the prediction models.

\begin{table}[!htbp]
\centering
\small
\caption{Held-Out Prediction Beyond Observed Covariates}
\label{SA table: application prediction}
\begin{tabular}{llrcc}
\hline\hline
Sample & Prediction target & Firms & Linear gain (\%) & $K$-NN gain (\%) \\
\hline
Full & Treated daily proxies & 22 & 24.15 & 10.26 \\
Base & Treated daily proxies & 12 & 23.75 & 3.08 \\
Full & Untreated CAR[0,1] & 561 & 13.61 & 13.42 \\
Base & Untreated CAR[0,1] & 514 & 16.98 & 16.03 \\
\hline
\end{tabular}
\par\smallskip\noindent\parbox{\textwidth}{\footnotesize\textit{Notes:} Each gain is $100\{1-\mathrm{MSE}(\text{local PCA with covariates})/\mathrm{MSE}(\text{indicated benchmark})\}\%$, using out-of-fold mean squared errors. The indicated benchmark is a linear or $K$-NN predictor using observed covariates only; no significance claim is made. Daily proxy errors are averaged over the 50 prediction dates $[-80,-31]$. The event-outcome check uses observed outcomes of control firms, not unobserved counterfactuals for treated firms. All comparisons use $K=99$.}
\end{table}

For treated firms, held-out proxy prediction gains are positive against both covariate-only benchmarks, especially linear prediction. The prediction gains for untreated event outcomes among held-out control firms are also positive in both samples. Thus, these comparisons illustrate useful predictive information in the reported settings.

\subsubsection{Placebo tests}
We apply the same causal estimator to 30 daily returns and four cumulative returns in the pre-event period $[-30,-1]$, which was excluded from the proxy sample. Table \ref{SA table: application placebos} reports how many effects exclude zero after simultaneous adjustment across all 34 outcomes in each specification.

\begin{table}[!htbp]
\centering
\small
\caption{Pre-Event Outcome Checks}
\label{SA table: application placebos}
\begin{tabular}{lccc}
\hline\hline
Sample & Covariates & Daily rejections (30) & Cumulative rejections (4) \\
\hline
Full & No & 4 & 0 \\
Full & Yes & 1 & 0 \\
Base & No & 1 & 0 \\
Base & Yes & 0 & 0 \\
\hline
\end{tabular}
\par\smallskip\noindent\parbox{\textwidth}{\footnotesize\textit{Notes:} Rejections use 5\% simultaneous intervals from 1,999 multiplier draws based on the estimated influence functions. The same $K=99$ specification is used for each pre-event outcome.}
\end{table}
\FloatBarrier

Overall, the results are reassuring: consistent with the event design, only a small number of daily pre-event effects are significant after simultaneous adjustment, and no cumulative effect is significant. In the full sample with covariates, the remaining significant daily effect is for day $-27$ ($-0.047$, standard error $0.011$, family-adjusted $p$-value $=0.001$). Pre-event differences may reflect anticipation, earlier news, or other differences between firms; these checks do not establish an explanation or automatically reject the event-outcome specification.

\subsubsection{Other summary statistics}
\begin{figure}[!htbp]
\centering
\includegraphics[width=0.86\textwidth]{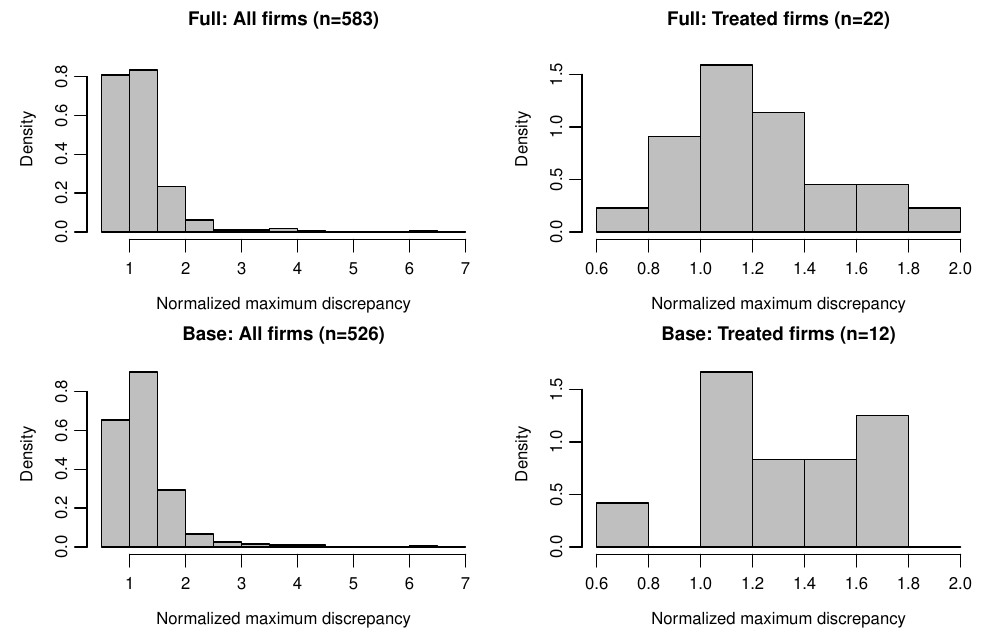}
\caption{Distributions of Normalized Maximum Matching Discrepancy}
\label{SA figure: application matching discrepancies}
\par\smallskip\noindent\parbox{0.86\textwidth}{\footnotesize\textit{Notes:} The four histograms use all 583 full-sample and 526 base-sample firms and the corresponding 22 and 12 treated firms. Normalized maximum discrepancy is defined as in Table \ref{table: summary KNN}; matching uses the benchmark pseudo-max distance with $K=99$. All-firm panels include treated firms. Full and base histograms share bins and axis scales within each column; horizontal scales differ between columns.}
\end{figure}

Table \ref{table: summary KNN} in the main paper already summarizes maximum matching discrepancies in the full sample. Figure \ref{SA figure: application matching discrepancies} shows their distributions for all firms and for treated firms in the full and base samples. The median number of treated firms in a treated-centered neighborhood is 10 in the full sample and 9 in the base sample. Each neighborhood contains $K=99$ firms in total.

Table \ref{SA table: application signals} complements Figure \ref{figure:eigenvalue} of the main paper by reporting the frequencies of the selected local dimensions and quartiles of the first two local eigenvalues relative to the implemented selection threshold in both the full and base samples. For implementation convenience, our benchmark selection rule in this application is a conservative choice and may yield fewer neighborhoods with $\ttd_i=2$. Recall that Table \ref{SA table: application sensitivity} reports sensitivity to this choice and shows that lowering the threshold increases the selected values of $\ttd_i$.

\begin{table}[!htbp]
\centering
\small
\caption{Selected Local Dimensions and Signals Across Neighborhoods}
\label{SA table: application signals}
\begin{tabular}{lcccc}
\hline\hline
Sample & $\ttd_i=1$ & $\ttd_i=2$ & First eigenvalue ratio & Second eigenvalue ratio \\
\hline
Full & 560 & 23 & 1.46 / 4.59 / 6.85 & 0.62 / 0.77 / 0.88 \\
Base & 504 & 22 & 1.43 / 2.22 / 5.97 & 0.68 / 0.81 / 0.89 \\
\hline
\end{tabular}
\par\smallskip\noindent\parbox{\textwidth}{\footnotesize\textit{Notes:} Eigenvalue-ratio entries are the first quartile, median, and third quartile across all 583 full-sample and 526 base-sample neighborhood centers. The denominator is the squared local-data-scale threshold used by the selection rule.}
\end{table}

\begin{figure}[!htbp]
\centering
\includegraphics[width=0.86\textwidth]{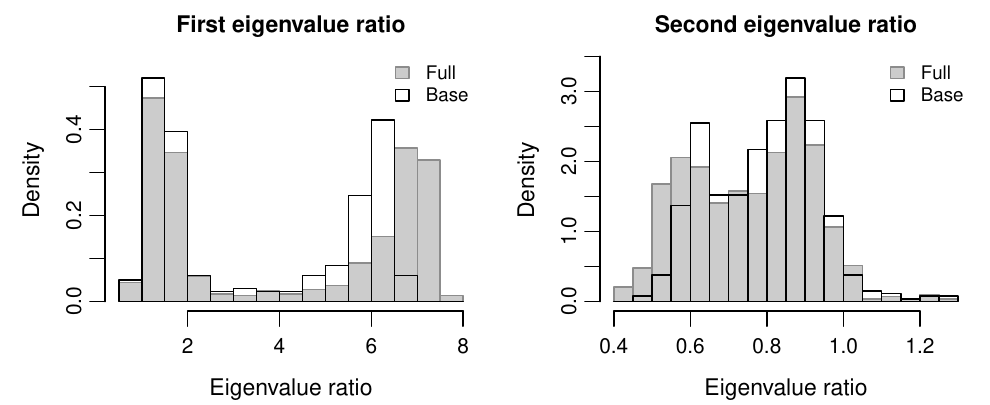}
\caption{Distributions of the First Two Local Eigenvalue Ratios}
\label{SA figure: application eigen ratios}
\par\smallskip\noindent\parbox{0.86\textwidth}{\footnotesize\textit{Notes:} Each histogram uses all 583 full-sample and 526 base-sample neighborhood centers. The ratios are squared local singular values divided by the squared selection threshold, as in Table \ref{SA table: application signals}. Gray bars show the full sample; black outlines show the base sample. Horizontal scales differ across panels.}
\end{figure}
\FloatBarrier

Table \ref{SA table: application covariate balance} compares observed-covariate balance before and after reweighting.

\begin{table}[!htbp]
\centering
\small
\caption{Observed-Covariate Balance Before and After Reweighting}
\label{SA table: application covariate balance}
\begin{tabular}{llccc}
\hline\hline
Sample & Control weights & Log assets & ROE & Leverage \\
\hline
Full & Unweighted & 1.328 & 0.588 & 0.096 \\
Full & Reweighted & 0.195 & 0.111 & 0.303 \\
Base & Unweighted & 1.056 & 0.704 & 0.492 \\
Base & Reweighted & 0.393 & 0.088 & 0.064 \\
\hline
\end{tabular}
\par\smallskip\noindent\parbox{\textwidth}{\footnotesize\textit{Notes:} Entries are absolute treated-control mean differences divided by the treated-group standard deviation, which is unchanged across rows. Reweighted controls use propensity odds from local logits based on the estimated principal components and all three observed covariates jointly.}
\end{table}

Reweighting substantially improves balance for most covariates, although the improvement is not uniform across covariates and samples.

\begin{table}[!htbp]
\centering
\small
\caption{Propensity-Score Diagnostics for the ATT}
\label{SA table: application propensity overlap}
\begin{tabular}{lcccc}
\hline\hline
& \multicolumn{2}{c}{Propensity: mean (min, max)} & \multicolumn{2}{c}{ATT diagnostics} \\
\cline{2-3}\cline{4-5}
Sample & Treated & Controls & Beyond range & Control ESS \\
\hline
Full & 0.399 (0.028, 0.967) & 0.021 ($<0.001$, 0.892) & 1/22 & 6.9 \\
Base & 0.425 (0.014, 0.901) & 0.015 ($<0.001$, 0.649) & 4/12 & 17.4 \\
Trim 5\%, full & 0.401 (0.017, $>0.999$) & 0.021 ($<0.001$, 0.871) & 2/22 & 7.9 \\
Trim 5\%, base & 0.379 (0.015, 0.978) & 0.015 ($<0.001$, 0.801) & 2/12 & 7.9 \\
\hline
\end{tabular}
\par\smallskip\noindent\parbox{\textwidth}{\footnotesize\textit{Notes:} All rows retain the three observed covariates and use the pseudo-max distance with $K=99$. ``Trim 5\%'' excludes 29 full-sample and 26 base-sample controls with the largest normalized matching discrepancies, retains all treated firms, and repeats matching and local estimation. ``Beyond range'' counts treated scores above the maximum observed control score.}
\end{table}

Control effective sample size (ESS) summarizes how evenly the propensity-odds weights $\widehat{\varrho}_i/(1-\widehat{\varrho}_i)$ are distributed across controls in the residual correction. It is the squared sum of these weights divided by the sum of their squares. Equal weights give ESS equal to the number of controls, regardless of their size; uneven weights reduce it. This measure does not describe the precision of the full doubly robust ATT estimator.

The diagnostics point to limited overlap for some treated firms: 1 of 22 treated firms in the full sample and 4 of 12 in the base sample have scores above the observed control-score range, and the corresponding control ESS values are 6.9 and 17.4. These findings indicate potential overlap limitations but do not by themselves establish a failure of overlap.

For the ATT, propensity scores close to one are the relevant overlap concern because they indicate limited availability of comparable controls and make the propensity-odds weights large; scores close to zero do not create the analogous problem. In each untrimmed benchmark fit with observed covariates, only one treated firm has an estimated propensity above 0.9; no control firm exceeds this threshold. The propensity-score trimming check above assesses sensitivity to them.

\FloatBarrier

\section{Extension: Covariate-Adjusted Factor Models}
\label{SA-sec: covariates adjustment}

In the main paper, we focus on a nonlinear factor model for the noisy proxies which has no additional covariates, 
but our results can be extended to the case where $\bx_i$ takes a  covariate-adjusted nonlinear factor structure:
\begin{equation*}
	x_{il}=\sum_{\ell=1}^{\ttr_w}
	\vartheta_\ell w_{il,\ell}+\eta_{l}(\balpha_i)+u_{il}, \quad \E[u_{il}|\{(\balpha_i, \bw_i)\}_{i\in[n]}]=0,
\end{equation*}
Write
$\bw_{i,\ell}=(w_{i1, \ell}, \cdots, w_{ip,\ell})'$,
$\ell\in[\ttr_w]$, $\bW_{i}=(\bw_{i,1},\cdots \bw_{i,\ttr_w})$, and $\bvth=(\vartheta_1,\cdots, \vartheta_{\ttr_w})'$.
Recall $\bx_{i}=(x_{i1}, \cdots, x_{ip})'$, 
$\bmeta_i=(\eta_1(\balpha_i),\cdots, \eta_p(\balpha_i))'$, and 
$\bu_{i}=(u_{i1},\cdots,u_{ip})'$. 
Then, we can rewrite the model in vector-form:
\[
\bx_i=\bW_{i}\bvth+\bmeta_i+\bu_{i}, \quad i=1, \cdots, n.
\]

The main analysis of this paper can still be applied once we have some consistent estimators of $\bvth$. They can be obtained
using a strategy similar to that for estimating the partially linear regression model. Specifically, we refer to Step 1 in Section \ref{subsec: step 1, latent variable extraction} as a general local principal subspace approximation procedure, which will be applied to other sequences in addition to $\{\bx_i\}_{i\in[n]}$. A slightly revised estimation procedure can be used to obtain the nearest neighbors and local factor loadings that provide information on the latent variables $\balpha_i$'s:
\begin{enumerate}[label=(\alph*)]\setlength\itemsep{.01em}
	\item Conduct a proxy-wise split of the index set $\mathcal{C}=[p]$ into three (non-overlapping) portions: $\mathcal{C}=\mathcal{C}^\dagger\cup\mathcal{C}^\ddagger\cup\mathcal{C}^\wr$.

	\item On $\mathcal{C}^\dagger\cup\mathcal{C}^\ddagger$, for each $\ell=1, \cdots, \ttr_w$, apply local principal subspace approximation to $\{\bw_{i,\ell}\}_{i\in[n]}$. Obtain the residual vectors on $\mathcal{C}^\ddagger$, $\widehat{\be}_{i,\ell}:=\bw_{i,\ell}^\ddagger-\widehat{\bw}_{i,\ell}^\ddagger$. Here $\widehat{\bw}_{i,\ell}^\ddagger$ is the transposed row corresponding to unit $i$ of the local PCA reconstruction for this covariate, using the neighborhood centered at $i$. Use $\mathcal{C}^\dagger$ for $K$-NN matching and $\mathcal{C}^\ddagger$ for local PCA.

	\item  On $\mathcal{C}^\dagger\cup\mathcal{C}^\ddagger$, apply the same procedure to $\{\bx_i\}_{i\in[n]}$.
	Let the residual vector on $\mathcal{C}^\ddagger$ be $\widehat{\bu}^{\natural}_{i}=\bx_{i}^\ddagger-\widehat{\bx}_{i}^\ddagger$, where $\widehat{\bx}_{i}^\ddagger$ is defined analogously as the transposed reconstructed row for unit $i$ in its own neighborhood.
	
	\item Let $\widehat{\be}_{i}=(\widehat{\be}_{i,1},\cdots, \widehat{\be}_{i,\ttr_w})'$. Estimate $\bvth=(\vartheta_1, \cdots, \vartheta_{\ttr_w})'$ by
	\[
	\widehat{\bvth}=\Big(\frac{1}{n}\sum_{i=1}^{n}\widehat{\be}_{i}\widehat{\be}_{i}'\Big)^{-1}\Big(\frac{1}{n}\sum_{i=1}^{n}\widehat{\be}_{i}\widehat{\bu}^{\natural}_{i}\Big).
	\]
	
	\item On $\mathcal{C}^\ddagger\cup\mathcal{C}^\wr$, apply the local principal subspace approximation to the covariate-adjusted $\bx_{i}$'s, i.e., $\{\bx_{i}-\sum_{\ell=1}^{\ttr_w}\bw_{i,\ell}\widehat{\vartheta}_\ell\}_{i\in[n]}$.
	Use $\mathcal{C}^\ddagger$ for $K$-NN matching and $\mathcal{C}^\wr$ for local PCA. The index sets for nearest neighbors and factor loadings obtained from this step are denoted by  $\{\mathcal{N}_i\}_{i\in[n]}$ and $\{\widehat{\bLambda}_{\subi}\}_{i\in[n]}$ respectively, which are the only quantities carried to counterfactual analysis.
\end{enumerate}
Under additional regularity conditions on $\bw_{i,\ell}$'s, it can be shown that $\widehat{\bvth}$ converges to $\bvth$ sufficiently fast and the main results established in the paper still hold. Formal analysis is available in \cite{Feng_2023_wp_Nonlinear}.

\section{Proofs}
\label{SA-sec: proofs}

\subsection{Additional notation and useful facts}
\label{SA subsection: preliminary facts}

For ease of exposition, we first prepare some additional notation and give several facts that will be useful for the proofs of our main results.

For a univariate function $f(x)$, write $f^{(1)}(x)=\frac{d}{dx}f(x)$. 
For each $j\in\mathcal{N}_i$, write $\mu_{t}(\balpha_j)=
\blambda_{j,\subi}'\bfrb_{t,\subi}+r_{t,\subi}(\balpha_j)$. 
By Assumption \ref{SA Assumption: LPCA},
$\max_{j\in\mathcal{N}_i}|r_{t,\subi}(\balpha_j)|
\lesssim \rate{\mu}$. 
Although we can only identify $\bLambda_{\subi}$ up to some rotation matrix $\bR_{\subi}$ (see   
Lemma \ref{SA lem: lpca} below), we are free to insert $\bR_{\subi}$ and its inverse in the leading approximation terms $\blambda_{j,\subi}'\bfrb_{t,\subi}$ without changing the approximation error. 
Thus, throughout the proofs, we absorb the corresponding rotations into $\bF_{\subi}$, $\bLambda_{\subi}$, and $\bfrb_{t,\subi}$ without changing their notation. We normalize the rotated loading matrix by its own Gram-matrix square root, still denoted by $\bUpsilon_{\subi}$. This matrix need not be diagonal, but its smallest eigenvalue is comparable to $\upsilon_{\ttd_i,\subi}$, and its normalized loading matrix differs from the original normalized matrix only by an orthogonal rotation. The bounds on the normalized loadings and approximation coefficients are therefore preserved.

Let $\kappa_{j}=\bz_{j}'\bbeta_{t}+\blambda_{j,\subi}'\bfrb_{t,\subi}$ and
$\widehat{\kappa}_{j}=\bz_{j}'\bbeta_{t}+\widehat{\blambda}_{j,\subi}'\bfrb_{t,\subi}$. The dependence of $\kappa_j$ and $\widehat{\kappa}_j$ on $t$ and $\subi$ is suppressed whenever there is no confusion.
For any $(\check{\bbeta}_{t,\subi}', \check{\bb}_{t,\subi}')'\in\mathbb{R}^{\ttr_z+\ttd_i}$, define 
\[
\begin{split}
	&\bb^\circ=\Big((\check{\bbeta}_{t,\subi}-\bbeta_{t})',\;
	(\bUpsilon_{\subi}(\check{\bb}_{t,\subi}-\bfrb_{t,\subi}))'\Big)',\quad
	\widehat{\bb}^\circ=((\widehat{\bbeta}_{t,\subi}-\bbeta_{t})', (\bUpsilon_{\subi}(\widehat{\bb}_{t,\subi}-\bfrb_{t,\subi}))')'\\
	&\widehat{\bpi}_{j}=(\bz_{j}',\, \widehat{\blambda}_{j,\subi}'\bUpsilon_{\subi}^{-1})',\quad
	\bpi_{j}=(\bz_{j}',\, \blambda_{j,\subi}'\bUpsilon_{\subi}^{-1})',\quad
	\bG_{j}=\bpi_{j}\bpi_{j}', \quad
	\widehat{\bG}_{j}=\widehat{\bpi}_{j}\widehat{\bpi}_{j}'.
\end{split}
\]
The dependence of $\bb^\circ$ on $t$ and $\subi$ and the dependence of $\bpi_j$ and $\widehat{\bpi}_j$ on $\subi$ are  suppressed.
Assumption \ref{SA Assumption: LPCA}  and Lemma \ref{SA lem: lpca} below imply the following two useful bounds:
{
\begin{align}
\max_{j\in\mathcal{N}_i}\|\widehat{\bpi}_j-\bpi_j\|
&=\bar{O}_\P\!\left(\frac{\sam^{-1}+\rate{\eta}}{\upsilon_{\ttd_i,\subi}}\right)=\bar{o}_\P(1),\label{SA eq: normalized loading error}\\
\max_{j\in\mathcal{N}_i}|\widehat{\kappa}_j-\kappa_j|
&=\max_{j\in\mathcal{N}_i}|(\widehat{\blambda}_{j,\subi}-\blambda_{j,\subi})'\bfrb_{t,\subi}|
=\bar{O}_\P(\sam^{-1}+\rate{\eta}).\label{SA eq: projected loading error}
\end{align}
}
The first bound gives uniform boundedness in probability of $\widehat{\bpi}_j$; the second controls $\widehat{\kappa}_j-\kappa_j$ using boundedness of the approximation coefficients.

{
Let $\mathcal{H}=\mathcal{F}\vee\sigma(\{\bm d_i\}_{i\in[n]})$. By overlap and $\biaserr=o_\P(1)$,
\[
\min_{i,t}\P(t_i=t\mid\mathcal F)
=\min_{i,t}\{\varrho_{i,t}+\E[v_{i,t}\mid\mathcal F]\}
\geq c_{\min}-\biaserr\geq c_{\min}/2
\]
with probability approaching one. Assumption \ref{SA Assumption: indep of errors} also implies independence of $\bepsilon_i$ over $i$ conditional on $\mathcal H$. On the same event,
\begin{align*}
\max_i\E[\|\bepsilon_i\|^{2+\nu}\mid\mathcal H]
&\leq\max_{i,t}\frac{\E[\|\bepsilon_i\|^{2+\nu}d_i(t)\mid\mathcal F]}{\P(t_i=t\mid\mathcal F)}\\
&\leq\frac{2}{c_{\min}}\max_i\E[\|\bepsilon_i\|^{2+\nu}\mid\mathcal F]
=O_\P(1).
\end{align*}
}
In addition, for each $t\in\mathcal{J}$,  define $\mathcal{N}_{i,t}=\{j: j\in\mathcal{N}_i, t_j=t\}$.
So we can rewrite the objective function of the local QMLE for the outcome equation as
\[
l_n(\bb^\circ):=\frac{1}{K}\sum_{j\in\mathcal{N}_{i,t}}\Big(
\mathcal{L}_{\tty}(\psi_{\tty}(\widehat{\kappa}_{j}+\widehat{\bpi}_{j}'\bb^\circ), y_{j})-
\mathcal{L}_{\tty}(\psi_{\tty}(\widehat{\kappa}_{j}), y_{j})\Big),
\]
and by optimization $l_n(\widehat{\bb}^\circ)\geq 0$.

Applying the scalar Taylor expansion to $s\mapsto l_n(s\bb^\circ)$ at $s=0$ gives
\[
\begin{split}
	l_n(\bb^\circ)=\frac{1}{K}\sum_{j\in\mathcal{N}_{i,t}}
	\mathcal{L}_{1,\tty}(\widehat{\kappa}_{j},y_{j})\widehat{\bpi}_{j}'\bb^\circ+
	\frac{1}{2K}\sum_{j\in\mathcal{N}_{i,t}}
	\mathcal{L}_{2,\tty}(\bar{\kappa}_{j},y_{j})(\bb^\circ)'\widehat{\bG}_{j}\bb^\circ=:\bA_n'\bb^\circ+(\bb^\circ)'\bQ_n\bb^\circ/2,
\end{split}
\]	
where $\bar\kappa_j=\widehat\kappa_j+s\widehat{\bpi}_j'\bb^\circ$ for one common $s\in(0,1)$ across all $j$. Thus a uniform Hessian bound over a ball around zero also applies to $\bQ_n$.

\subsection{Technical Lemmas}

We first give the following lemma, which shows that the nearest neighbors and extracted local factor loadings are informative about the unobserved confounders.

\begin{lem}\label{SA lem: lpca}
	Suppose that Assumptions \ref{SA Assumption: ID-counterfactual}, \ref{SA Assumption: Regularities}, \ref{SA Assumption: indep of errors},
	and \ref{SA Assumption: Matching} hold.
	If $\frac{\log (n/K)}{K}=o(1)$ and $\frac{K\log n}{n}=o(1)$, then, 
	\[
	\max_{i\in[n]}\max_{k\in[K]}
	\|\balpha_i-\balpha_{j_k(i)}\|
	\lesssim_\P (K/n)^{\uvarsig/(\lvarsig \ttr_\alpha)}+a_p^{1/\lvarsig}.
	\]
	If, in addition, Assumption \ref{SA Assumption: LPCA} holds
	and $(np)^{\frac{2}{\nu}}\sam^{-2}\lesssim 1$, then there exist $\{\bR_{\subi}\}_{i\in[n]}$ such that, uniformly over $i\in[n]$ and $\ell\in[\ttd_i]$,
	\[
	\|\widehat{\bLambda}_{\cdot \ell,\subi}-\bLambda_{\subi}\bR_{\cdot \ell,\subi}\|_{\max}=\bar{O}_\P
	(\sam^{-1}+\rate{\eta}),
	\]
	where $1\lesssim_\P\min_{i\in[n]}s_{\min}(\bR_{\subi})\leq \max_{i\in[n]}s_{\max}(\bR_{\subi})\lesssim_\P 1$.
\end{lem}
\begin{remark}
	To be precise, since the first step estimation only concerns the nonlinear factor model, the conditions on $\bz_i$ and the moment conditions on $\bepsilon_i$ in Assumption \ref{SA Assumption: Regularities}, the condition on the outcome and treatment errors in Assumption \ref{SA Assumption: indep of errors}, and the outcome and treatment approximation and coefficient bounds are unnecessary for this lemma. Thus, $\biaserr$ plays no role in either conclusion.
\end{remark}
\begin{proof}
	For the matching conclusion, Assumptions \ref{SA Assumption: ID-counterfactual} and \ref{SA Assumption: Regularities} provide the i.i.d. sampling and support conditions for $\balpha_i$, 
	while Assumption \ref{SA Assumption: Matching} is the indirect-matching condition in \cite{Feng_2023_wp_Nonlinear}. Thus, Theorem 4.1 therein gives the stated rate.

	For the loading conclusion, the relevant parts of Assumptions \ref{SA Assumption: Regularities}, \ref{SA Assumption: indep of errors}, and \ref{SA Assumption: LPCA} supply the boundedness, measurement-error, signal-strength, and approximation conditions of Theorem 4.2 of \cite{Feng_2023_wp_Nonlinear}.
	Therefore, the loading conclusion and the uniform bounds on the rotation matrices follow from that theorem.
\end{proof}

\begin{lem}[Maximal Inequality]\label{SA lem: maximal inequality}
	For each $i\in[n]$, let $\{Z_{j,i,n}\}_{j\in[m]}$ be a sequence of independent but not necessarily identically distributed random variables with zero mean and uniformly bounded $(2+\nu)$-th moment ($\nu>0$). Assume $m\rightarrow\infty$ as $n\rightarrow\infty$.
	If $\frac{n^{2/\nu}\log n}{m}=o(1)$, then $\max_{i\in[n]}|\frac{1}{m}\sum_{j=1}^{m}Z_{j,i,n}|\lesssim_\P 
	\Big(\frac{\log n}{m}\Big)^{1/2}$.
\end{lem}
\begin{proof}
	Write $Z_{j,i,n}=Z_{j,i,n}^- + Z_{j,i,n}^+$ where $Z_{j,i,n}^-:=Z_{j,i,n}\I(|Z_{j,i,n}|\leq \tau_n)
	-\E[Z_{j,i,n}\I(|Z_{j,i,n}|\leq \tau_n)]$ and $Z^+_{j,i,n}:=Z_{j,i,n}\I(|Z_{j,i,n}|>\tau_n)
	-\E[Z_{j,i,n}\I(|Z_{j,i,n}|>\tau_n)]$ for $\tau_n=\sqrt{m/\log n}$.
	Then, for any $\delta>0$, by Bernstein's inequality,
	\begin{align*}
		\P\Big(\Big|\frac{1}{m}\sum_{j\in[m]}Z_{j,i,n}^-\Big|>\delta\Big)\leq
		2\exp\bigg(-\frac{\delta^2/2}
		{\frac{1}{m^2}\sum_{j\in[m]}\E[(Z_{j,i,n}^-)^2]
			+\frac{1}{3m}\tau_n\delta}\bigg).
	\end{align*}
	On the other hand,
	\[
	\P\Big(\max_{1\leq i\leq n}
	\Big|\frac{1}{m}\sum_{j\in[m]}Z_{j,i,n}^+\Big|>\delta\Big)
	\leq n\max_{1\leq i\leq n}
	\frac{\sum_{j\in[m]}\E[(Z_{j,i,n}^+)^2]}
	{m^2\delta^2}
	\lesssim \frac{n}{m\tau_n^\nu\delta^2}.
	\]
	Then, by the imposed rate restriction, setting $\delta\asymp\sqrt{\log n/m}$, 
	we have
	$\max_{i\in[n]}
	|\frac{1}{m}\sum_{j\in[m]}Z_{j,i,n}|\\
	\lesssim_\P \sqrt{\frac{\log n}{m}}$.
\end{proof}

\begin{lem}\label{lem: An expansion}
	Under the conditions of Theorem \ref{SA thm: QMLE}, 
	\begin{align*}
	\bA_n&=\frac{1}{K}\sum_{j\in\mathcal{N}_{i,t}}\mathcal{L}_{1,\tty}(\kappa_j, y_j)\bpi_j
	-\frac{1}{K}\sum_{j\in\mathcal{N}_{i,t}}
	\frac{\psi_{\tty}^{(1)}(\kappa_{j})(\widehat{\kappa}_j-\kappa_j)\bpi_{j}}{\bar{V}_{\tty}(\kappa_{j})}
	+\bar{o}_\P(\sam^{-1}+\rate{\eta}+\rate{\mu}+\biaserr)\\
	&=\bar{O}_\P(\sam^{-1}+\rate{\eta}+\rate{\mu}+\biaserr).
	\end{align*}
\end{lem}
\begin{proof}
Rewrite
\[
\begin{split}
	\bA_n=&\frac{1}{K}\sum_{j\in\mathcal{N}_{i,t}}\mathcal{L}_{1,\tty}(\kappa_{j}, y_{j})\bpi_{j}+
	\frac{1}{K}\sum_{j\in\mathcal{N}_{i,t}}\mathcal{L}_{1,\tty}(\kappa_{j}, y_{j})(\widehat{\bpi}_{j}-\bpi_{j})+\\
	&\frac{1}{K}\sum_{j\in\mathcal{N}_{i,t}}(\mathcal{L}_{1,\tty}(\widehat{\kappa}_{j}, y_{j})-\mathcal{L}_{1,\tty}(\kappa_{j}, y_{j}))\widehat{\bpi}_{j}
	=:J_1+J_2+J_3.
\end{split}
\]

Recall that 
$$
\mathcal{L}_{1,\tty}(\kappa_{j}, y_{j}(t))=\frac{y_{j}(t)-\psi_{\tty}(\kappa_j)}{\bar{V}_{\tty}(\kappa_{j})}=
\frac{\epsilon_{j,t}+\bar{r}_{j,t,\subi}}{\bar{V}_{\tty}(\kappa_{j})},
$$ 
where $\bar{V}_{\tty}(\kappa_{j})=V_{\tty}(\psi_{\tty}(\kappa_{j}))/\psi_{\tty}^{(1)}(\kappa_{j})$ and
$\bar{r}_{j,t,\subi}=\psi_{\tty}(\bz_j'\bbeta_{t}+\mu_t(\balpha_j))-\psi_{\tty}(\kappa_j)=\bar{O}_\P(\rate{\mu})$. 
Conditional on $\mathcal{H}=\mathcal{F}\vee\sigma(\{\bm{d}_j\}_{j\in[n]})$, $\{\epsilon_{j,t}\}_{j\in[n]}$ is independent over $j$ with
$|\E[\epsilon_{j,t}|\mathcal{H}]|\leq\biaserr$.  Then, by Lemma \ref{SA lem: maximal inequality}, we conclude that $J_1=\bar{O}_\P(\sqrt{\log n}K^{-1/2}+\rate{\mu}+\biaserr)$.

Next, by \eqref{SA eq: normalized loading error},
\[
J_2=
\underbrace{\frac{1}{K}\sum_{j\in\mathcal{N}_{i,t}}
\bar{V}_{\tty}(\kappa_{j})^{-1}\epsilon_{j,t}
\Big(\bm{0}_{\ttr_z}',\;(\widehat{\blambda}_{j,\subi}-\blambda_{j,\subi})'\bUpsilon_{\subi}^{-1}\Big)'}_{J_{2,1}}+\bar{O}_\P\!\left(\rate{\mu}\frac{\sam^{-1}+\rate{\eta}}{\upsilon_{\ttd_i,\subi}}\right)
\]
where $\bm{0}_{\ttr_z}$ denotes a $\ttr_z$-vector of zeros. $J_{2,1}$ is a sum of conditionally independent random variables given $\mathcal{H}$, then by Lemma \ref{SA lem: maximal inequality}, 
$J_{2,1}=\bar{O}_\P((\sqrt{\log n}K^{-1/2}+\biaserr)(\sam^{-1}+\rate{\eta})/\upsilon_{\ttd_i,\subi})$. Since $(\sam^{-1}+\rate{\eta})/\upsilon_{\ttd_i,\subi}=\bar{o}_\P(1)$, it follows that $J_2=\bar{o}_\P(\sam^{-1}+\rate{\eta}+\rate{\mu}+\biaserr)$.

$J_3$ can be written as
\[
\begin{split}
	&-\frac{1}{K}\sum_{j\in\mathcal{N}_{i,t}}
	\Big(\frac{\psi_{\tty}(\widehat{\kappa}_{j})-\psi_{\tty}(\kappa_{j})}{\bar{V}_{\tty}(\kappa_{j})}\Big)\bpi_{j}+
	\frac{1}{K}\sum_{j\in\mathcal{N}_{i,t}}
	\Big(\bar{V}_{\tty}(\widehat{\kappa}_{j})^{-1}-
	\bar{V}_{\tty}(\kappa_{j})^{-1}\Big)(\epsilon_{j,t}+\zeta_{j,t}-\psi_{\tty}(\widehat{\kappa}_{j}))\bpi_{j}\\
	&+\frac{1}{K}\sum_{j\in\mathcal{N}_{i,t}}
	\Big(\frac{y_j-\psi_{\tty}(\widehat{\kappa}_{j})}{\bar{V}_{\tty}(\widehat{\kappa}_{j})}-\frac{y_j-\psi_{\tty}(\kappa_{j})}{\bar{V}_{\tty}(\kappa_{j})}\Big)(\widehat{\bpi}_{j}-\bpi_{j})
	=: -J_{3,1}+J_{3,2}+J_{3,3}.
\end{split}
\]
Now, by Taylor expansion, \eqref{SA eq: projected loading error} and Assumption \ref{SA Assumption: QMLE},
\[
J_{3,1}=\frac{1}{K}\sum_{j\in\mathcal{N}_{i,t}}
\frac{\bpi_{j}\psi_{\tty}^{(1)}(\tilde{\kappa}_{j})(\widehat{\kappa}_j-\kappa_j)}{\bar{V}_{\tty}(\kappa_{j})}=\bar{O}_\P(\sam^{-1}+\rate{\eta}),
\]
where $\tilde{\kappa}_j$ is between $\kappa_j$ and $\widehat{\kappa}_j$.
For $J_{3,2}$, center $\epsilon_{j,t}$ conditional on $\mathcal{H}$ and apply Lemma \ref{SA lem: maximal inequality}; its bound is $\bar{O}_\P((\sam^{-1}+\rate{\eta})(\sam^{-1}+\rate{\eta}+\rate{\mu}+\biaserr))$. For $J_{3,3}$, also use \eqref{SA eq: normalized loading error} to bound $\widehat{\bpi}_j-\bpi_j$. Consequently,
{
\begin{align*}
J_{3,2}+J_{3,3}
&=\bar{O}_\P\!\left((\sam^{-1}+\rate{\eta})
\left[\sam^{-1}+\rate{\eta}+\rate{\mu}+\biaserr+
\frac{\sam^{-1}+\rate{\eta}}{\upsilon_{\ttd_i,\subi}}\right]\right)\\
&=\bar{o}_\P(\sam^{-1}+\rate{\eta}+\rate{\mu}+\biaserr).
\end{align*}
}
\end{proof}

\begin{lem}\label{lem: bound on coefficients}
	Under the conditions of Theorem \ref{SA thm: QMLE}, the following conclusions hold:
	\begin{enumerate}[label=(\roman*)]
		\item the minimum eigenvalue of $-\bQ_n$ is bounded away from zero in probability, uniformly over neighborhoods $\subi$ and $\bb^\circ$ in a sufficiently small neighborhood of zero;
		\item $\widehat\bb^\circ=\bar{O}_\P(\sam^{-1}+\rate{\eta}+\rate{\mu}+\biaserr)$.
	\end{enumerate}
\end{lem}
\begin{proof}
Notice that 
$$\mathcal{L}_{2,\tty}(\bar{\kappa}_j, y_j(t))=-\frac{(y_j(t)-\psi_{\tty}(\bar{\kappa}_j))\bar{V}_{\tty}^{(1)}(\bar\kappa_j)}{\bar{V}^2_{\tty}(\bar{\kappa}_j)}
-\frac{(\psi_{\tty}^{(1)}(\bar{\kappa}_j))^2}{V_{\tty}(\psi_{\tty}(\bar{\kappa}_j))}=
-\frac{\epsilon_{j,t}\bar{V}_{\tty}^{(1)}(\bar\kappa_j)}{\bar{V}^2_{\tty}(\bar{\kappa}_j)}-\tilde{\mathcal{L}}(\bar{\kappa}_j,\zeta_{j,t}),$$ 
where $\tilde{\mathcal{L}}(\bar{\kappa}_j,\zeta_{j,t})=
\frac{(\zeta_{j,t}-\psi_{\tty}(\bar{\kappa}_j))\bar{V}_{\tty}^{(1)}(\bar\kappa_j)}{\bar{V}^2_{\tty}(\bar{\kappa}_j)}+\frac{(\psi_{\tty}^{(1)}(\bar{\kappa}_j))^2}{V_{\tty}(\psi_{\tty}(\bar{\kappa}_j))}$. Then,
\[
-\bQ_n=\frac{1}{K}\sum_{j\in\mathcal{N}_{i,t}}\frac{\epsilon_{j,t}\bar{V}_{\tty}^{(1)}(\bar\kappa_j)}{\bar{V}^2_{\tty}(\bar{\kappa}_j)}\widehat{\bpi}_j\widehat{\bpi}_j'+\frac{1}{K}\sum_{j\in\mathcal{N}_{i,t}}\tilde{\mathcal{L}}(\bar{\kappa}_j,\zeta_{j,t})\widehat{\bpi}_j\widehat{\bpi}_j'.
\]

For the second term, first establish the joint local Gram bound. For $\ba\in\mathbb R^{\ttr_z}$ and $\bb\in\mathbb R^{\ttd_i}$, normalization of the loadings gives
\begin{align*}
\frac1K\sum_{j\in\mathcal N_i}
(\ba'\bz_j+\bb'\bUpsilon_{\subi}^{-1}\blambda_{j,\subi})^2
&=\frac1K\ba'\bZ_{\subi}'(\bI_K-\bP_{\subi})\bZ_{\subi}\ba
+\big\|\bb+\frac1K\bUpsilon_{\subi}^{-1}\bLambda_{\subi}'\bZ_{\subi}\ba\big\|^2.
\end{align*}
Compactness of the covariates bounds $\|\bUpsilon_{\subi}^{-1}\bLambda_{\subi}'\bZ_{\subi}\ba/K\|$ by a constant times $\|\ba\|$. The residualized Gram condition in Assumption \ref{SA Assumption: Regularities} therefore implies
\[
\min_i s_{\min}\big(\frac1K\sum_{j\in\mathcal N_i}\bpi_j\bpi_j'\big)\gtrsim_\P1.
\]

Next, conditional on $\mathcal F$, the neighborhoods and regressors are fixed, treatments are independent, and $\min_{j,t}\P(t_j=t\mid\mathcal F)\geq c_{\min}/2$ with probability approaching one, as shown in Section \ref{SA subsection: preliminary facts}. The theorem's rate condition implies $\log n/K=o(1)$. Bernstein's inequality and a union bound yield
\begin{align*}
\max_{i,t}\big|\frac{|\mathcal N_{i,t}|}{K}
-\frac1K\sum_{j\in\mathcal N_i}\P(t_j=t\mid\mathcal F)\big|
&=O_\P\big(\sqrt{\log n/K}\big)=o_\P(1),\\
\max_{i,t}\big\|\frac1K\sum_{j\in\mathcal N_i}
\{d_j(t)-\P(t_j=t\mid\mathcal F)\}\bpi_j\bpi_j'\big\|_{\max}
&=O_\P\big(\sqrt{\log n/K}\big)=o_\P(1),
\end{align*}
where the second bound also uses uniform boundedness of the normalized regressors. Consequently, $\min_{i,t}|\mathcal N_{i,t}|/K\geq c_{\min}/4$ with probability approaching one, and
\begin{align*}
\frac1K\sum_{j\in\mathcal N_{i,t}}\bpi_j\bpi_j'
&=\frac1K\sum_{j\in\mathcal N_i}\P(t_j=t\mid\mathcal F)\bpi_j\bpi_j'+\bar{o}_\P(1)\\
&\succeq\frac{c_{\min}}{2K}\sum_{j\in\mathcal N_i}\bpi_j\bpi_j'+\bar{o}_\P(1).
\end{align*}
The uniform normalized consistency in \eqref{SA eq: normalized loading error} further gives
\[
\max_{i,t}\big\|\frac1K\sum_{j\in\mathcal N_{i,t}}
(\widehat{\bpi}_j\widehat{\bpi}_j'-\bpi_j\bpi_j')\big\|_{\max}=o_\P(1),
\qquad
\min_{i,t}s_{\min}\big(\frac1K\sum_{j\in\mathcal N_{i,t}}\widehat{\bpi}_j\widehat{\bpi}_j'\big)\gtrsim_\P1.
\]
Assumption \ref{SA Assumption: QMLE} bounds $\tilde{\mathcal L}(\bz_j'\bbeta_t+\mu_t(\balpha_j),\zeta_{j,t})$ away from zero. Thus the second term has a uniformly positive minimum eigenvalue at the true indices. By continuity and the uniform consistency of $\widehat\kappa_j$ for these indices, the same conclusion holds with $\tilde{\mathcal L}(\widehat\kappa_j+\widehat{\bpi}_j'\bb^\circ,\zeta_{j,t})$ for $\bb^\circ$ in a sufficiently small closed ball $\mathcal B$ centered at zero.

For the first term, note that $\{\epsilon_{j,t}\}$ is independent over $j$ conditional on $\mathcal{H}=\mathcal{F}\vee\sigma(\{\bm{d}_i\}_{i\in[n]})$ with $|\E[\epsilon_{j,t}|\mathcal{H}]|\leq \biaserr$.
We apply the truncation argument to show uniform convergence of the first term. Define $\tilde{\epsilon}_{j,t}=\epsilon_{j,t}-\E[\epsilon_{j,t}|\mathcal{H}]$, 
$\epsilon_{j,t}^-=\tilde\epsilon_{j,t}\I(|\tilde\epsilon_{j,t}|\leq \tau_n)$ and 
$\epsilon_{j,t}^+=\tilde\epsilon_{j,t}-\epsilon_{j,t}^-$, 
and take $\tau_n=n^{1/\nu}$. The theorem's rate condition implies $\tau_n\sqrt{\log n/K}=O(p^{-1/\nu})=o(1)$. The contribution of the conditional means $\E[\epsilon_{j,t}\mid\mathcal H]$ is uniformly $O_\P(\biaserr)=o_\P(1)$, so it remains to bound the centered errors. For the truncated part, conditional on the data, define
\[
\begin{split}
		\mathcal{T}=\Big\{\bt=(a_1,\cdots, a_n)\in\mathbb{R}^n:\; &a_j=\frac{d_{j}(t)\I(j\in\mathcal{N}_i)\bar{V}_{\tty}^{(1)}(\widehat{\kappa}_j+\widehat{\bpi}_{j}'\bb^\circ)\widehat{\pi}_{k,j}\widehat{\pi}_{l,j}}{\bar{V}^2_{\tty}(\widehat{\kappa}_j+\widehat{\bpi}_j'\bb^\circ)}\Big(\epsilon_{j,t}^--\E[\epsilon_{j,t}^-|\mathcal{H}]\Big),\\
	&\bb^\circ\in\mathcal{B}, 1\leq i\leq n, 1\leq k, l\leq \ttr_z+\ttd_i
	\Big\}.
\end{split}
\]
Define the norm $\|\cdot\|_{n,2}$ on $\mathbb{R}^n$ by $\|\ba\|^2_{n,2}=\E_n[a_j^2]$. For any $\varepsilon>0$, the covering number $N(\mathcal{T},\|\cdot\|_{n,2},\varepsilon)$ is the infimum of the cardinality of $\varepsilon$-nets of $\mathcal{T}$.
Let $\{\omega_i\}_{i\in[n]}$ be a sequence of independent Rademacher random variables that are independent of the data, and $\E_{\bomega}[\cdot]$ denotes the expectation with respect to the distribution of $\bomega=(\omega_1,\cdots,\omega_n)$.
By Symmetrization Inequality and Dudley's inequality \citep{Dudley_1967_JFA},
{\small
	\[
	\E\bigg[\max_{\substack{1\leq i\leq n,\\1\leq k,l\leq \ttr_z+\ttd_i}}\sup_{\bb^\circ\in\mathcal{B}}
	\Big|\frac{1}{\sqrt{n}}
	\sum_{j=1}^{n}a_j\Big|\bigg|\mathcal{H}\bigg]
	\leq 2\E\bigg[\E_{\bomega}
	\bigg[\sup_{\bm{a}\in\mathcal{T}}\Big|\frac{1}{\sqrt{n}}
	\sum_{j=1}^{n}a_j\omega_j\Big|\bigg]\bigg|\mathcal{H}\bigg]
	\lesssim
	\int_0^{\Theta}\sqrt{\log N(\mathcal{T},\|\cdot\|_{n,2},\varepsilon)}d\varepsilon,
	\]
}

\noindent where the diameter $\Theta:=2\sup_{\ba\in\mathcal{T}}\|\ba\|_{n,2}\lesssim_\P\tau_n\sqrt{K/n}$ by truncation.
Since $\mathcal B$ is compact and the weights are uniformly Lipschitz in $\bb^\circ$, the covering number is bounded by $Cn\{1+C\tau_n\sqrt{K/n}/\varepsilon\}^{\ttr_z+\max_i\ttd_i}$ on events of arbitrarily high probability, for a constant $C>0$. Integrating this bound gives $\sup_{\bb^\circ\in\mathcal B}\max_{i,k,l}|\frac{n}{K}\E_n[a_j]|=O_\P(\tau_n\sqrt{\log n/K})=o_\P(1)$.
For the tails, the conditional moment bound gives
\[
\P\big(\max_{j,t}|\tilde\epsilon_{j,t}|>\tau_n\mid\mathcal H\big)
\leq\frac{\sum_{j,t}\E[|\tilde\epsilon_{j,t}|^{2+\nu}\mid\mathcal H]}{\tau_n^{2+\nu}}
=O_\P(n^{-2/\nu})=o_\P(1).
\]
On the complementary event, all $\epsilon_{j,t}^+$ are zero. Retaining the maximum over all neighborhoods, the centered tail contribution is then bounded by
\begin{align*}
&\max_{i,k,l}\sup_{\bb^\circ\in\mathcal B}
\Big|\frac1K\sum_{j\in\mathcal N_{i,t}}
\frac{\bar V_{\tty}^{(1)}(\widehat\kappa_j+\widehat{\bpi}_j'\bb^\circ)\widehat\pi_{k,j}\widehat\pi_{l,j}}
{\bar V_{\tty}^2(\widehat\kappa_j+\widehat{\bpi}_j'\bb^\circ)}
\{\epsilon_{j,t}^+-\E[\epsilon_{j,t}^+\mid\mathcal H]\}\Big|\\
&\qquad\lesssim_\P\max_{j,t}\E[|\tilde\epsilon_{j,t}|\I(|\tilde\epsilon_{j,t}|>\tau_n)\mid\mathcal H]\\
&\qquad\leq\tau_n^{-(1+\nu)}\max_{j,t}\E[|\tilde\epsilon_{j,t}|^{2+\nu}\mid\mathcal H]
=O_\P(\tau_n^{-(1+\nu)})=o_\P(1).
\end{align*}
Together with the truncated and conditional-mean terms, this proves that the first term in the decomposition of $-\bQ_n$ is $o_\P(1)$ uniformly over $i$ and $\bb^\circ\in\mathcal B$.

Therefore, if the maximizer $\widehat{\bb}^\circ$ is searched within $\mathcal{B}$, by optimality, the above results suffice to show that  $\widehat{\bb}^\circ=\bar{O}_\P(\sam^{-1}+\rate{\eta}+\rate{\mu}+\biaserr)$. But note that by concavity of $\mathcal{L}_{\tty}(\cdot, \cdot)$ assumed in Assumption \ref{SA Assumption: QMLE}, it is also the maximizer over the whole parameter space.
\end{proof}

\subsection{Proof of Theorem \ref{SA thm: QMLE}}
We will only study the outcome equation; the treatment equation can be treated exactly the same way. 
To derive the asymptotic expansion, note that the first-order condition leads to
\[
\begin{split}
	\bm{0}&=\frac{1}{K}\sum_{j\in\mathcal{N}_{i,t}}\mathcal{L}_{1,\tty}(\widehat{\kappa}_j+\widehat{\bpi}'_{j}\widehat{\bb}^\circ,y_j)\widehat{\bpi}_j\\
	&=\frac{1}{K}\sum_{j\in\mathcal{N}_{i,t}}\mathcal{L}_{1,\tty}(\widehat{\kappa}_j,y_j)\widehat{\bpi}_j+
	\frac{1}{K}\sum_{j\in\mathcal{N}_{i,t}}\mathcal{L}_{2,\tty}(\tilde{\kappa}_j,y_j)\widehat{\bpi}_j\widehat{\bpi}_j'\widehat{\bb}^\circ
	=\bA_n+\bV_n\widehat{\bb}^\circ,
\end{split}
\]
where $\tilde{\kappa}_j$ is between $\widehat{\kappa}_j$ and $\widehat{\kappa}_j+\widehat{\bpi}_j'\widehat{\bb}^\circ$.
For $\bV_n$, consider the following decomposition:
{
\begin{align*}
\bV_n={}&\frac{1}{K}\sum_{j\in\mathcal{N}_{i,t}}
\mathcal{L}_{2,\tty}(\psi_{\tty}^{-1}(\zeta_{j,t}),y_j)\widehat{\bG}_j\\
&+\frac{1}{K}\sum_{j\in\mathcal{N}_{i,t}}
\big\{\mathcal{L}_{2,\tty}(\tilde\kappa_j,y_j)
-\mathcal{L}_{2,\tty}(\psi_{\tty}^{-1}(\zeta_{j,t}),y_j)\big\}\widehat{\bG}_j.
\end{align*}
}
Define the comparison Hessian conditional on $\mathcal F$ by
{
\[
\bV_{\subi}:=\frac{1}{K}\sum_{j\in\mathcal{N}_i}
\E\!\left[d_j(t)\mathcal{L}_{2,\tty}(\psi_{\tty}^{-1}(\zeta_{j,t}),y_j)
\,\middle|\,\mathcal{F}\right]\widehat{\bG}_j.
\]
}
The treatment indicator is included inside the expectation, so $\bV_{\subi}$ is measurable on $\mathcal F$. Conditional on $\mathcal F$, the neighborhoods and $\widehat{\bG}_j$ are fixed and the outcome--treatment pairs are independent across units. For the first term of $\bV_n$, applying Lemma \ref{SA lem: maximal inequality} gives
{
\begin{align*}
&\frac{1}{K}\sum_{j\in\mathcal{N}_{i,t}}
\mathcal{L}_{2,\tty}(\psi_{\tty}^{-1}(\zeta_{j,t}),y_j)\widehat{\bG}_j-\bV_{\subi}\\
&\quad=\frac{1}{K}\sum_{j\in\mathcal{N}_i}
\Big\{d_j(t)\mathcal{L}_{2,\tty}(\psi_{\tty}^{-1}(\zeta_{j,t}),y_j)
-\E\!\left[d_j(t)\mathcal{L}_{2,\tty}(\psi_{\tty}^{-1}(\zeta_{j,t}),y_j)\,\middle|\,\mathcal F\right]\Big\}\widehat{\bG}_j\\
&\quad=\bar{O}_\P(\sam^{-1}).
\end{align*}
}
For the second term, \eqref{SA eq: projected loading error} gives
{
\[
\max_{j\in\mathcal{N}_{i,t}}|\tilde\kappa_j-\psi_{\tty}^{-1}(\zeta_{j,t})|
=\bar{O}_\P(\|\widehat{\bb}^\circ\|+\sam^{-1}+\rate{\eta}+\rate{\mu}).
\]
}
Thus, Assumption \ref{SA Assumption: QMLE}(a) and Lemma \ref{lem: bound on coefficients} imply $\bV_n-\bV_{\subi}=\bar{O}_\P(\sam^{-1}+\rate{\eta}+\rate{\mu}+\biaserr)$. The uniform local Gram bound used in Lemma \ref{lem: bound on coefficients} implies that $\bV_n^{-1}$ is uniformly bounded. Since $\bV_n-\bV_{\subi}=\bar{o}_\P(1)$, $\bV_{\subi}^{-1}$ is also uniformly bounded. The matrix-inverse identity then gives
{
\[
\bV_n^{-1}-\bV_{\subi}^{-1}
=\bV_n^{-1}(\bV_{\subi}-\bV_n)\bV_{\subi}^{-1}
=\bar{O}_\P(\sam^{-1}+\rate{\eta}+\rate{\mu}+\biaserr).
\]
}

Now, further decompose the following difference: 
\begin{align*}
	&\Big(\bz_{i}'\widehat{\bbeta}_{t,\subi}+\widehat{\blambda}_{i,\subi}'\widehat{\bb}_{t,\subi}\Big)-
	\Big(\bz_{i}'\bbeta_t+\mu_t(\balpha_i)\Big)\\
	=\;&\Big(\bz_{i}'\widehat{\bbeta}_{t,\subi}+\widehat{\blambda}_{i,\subi}'\widehat{\bb}_{t,\subi}-
	\bz_{i}'\bbeta_t-\widehat{\blambda}_{i,\subi}'\bfrb_{t,\subi}\Big)+
	\Big(\widehat{\blambda}_{i,\subi}'\bfrb_{t,\subi}-\blambda_{i,\subi}'\bfrb_{t,\subi}\Big)+
	\Big(\blambda_{i,\subi}'\bfrb_{t,\subi}-\mu_t(\balpha_i)\Big).
\end{align*}
By Lemma \ref{lem: bound on coefficients} and Assumption \ref{SA Assumption: Regularities}, the first term is $\bar{O}_\P(\sam^{-1}+\rate{\eta}+\rate{\mu}+\biaserr)$.
The second term, by \eqref{SA eq: projected loading error}, is $\bar{O}_\P(\sam^{-1}+\rate{\eta})$.
Finally, the last term is simply the approximation error, which is $\bar{O}_\P(\rate{\mu})$. 
Therefore, we conclude the above difference is  $\bar{O}_\P(\sam^{-1}+\rate{\eta}+\rate{\mu}+\biaserr)$. Since $\psi_{\tty}^{(1)}(\cdot)$ is assumed to be continuous, the first result of the theorem follows.

Finally, consider the asymptotic expansion for $\widehat{\zeta}_{i,t}$. Expanding $\mathcal{L}_{1,\tty}(\widehat\kappa_j,y_j)$ at $\psi_{\tty}^{-1}(\zeta_{j,t})$ gives
{
\begin{align*}
\bA_n={}&\frac{1}{K}\sum_{j\in\mathcal{N}_{i,t}}
\left\{\mathcal{L}_{1,\tty}(\psi_{\tty}^{-1}(\zeta_{j,t}),y_j)
-\frac{\psi_{\tty}^{(1)}(\psi_{\tty}^{-1}(\zeta_{j,t}))}{\bar V_{\tty}(\psi_{\tty}^{-1}(\zeta_{j,t}))}
\big(\widehat\kappa_j-\psi_{\tty}^{-1}(\zeta_{j,t})\big)\right\}\widehat{\bpi}_j\\
&+\bar{O}_\P\big((\sam^{-1}+\rate{\eta}+\rate{\mu}+\biaserr)^2\big).
\end{align*}
}
To obtain the remainder, note that $\widehat\kappa_j-\psi_{\tty}^{-1}(\zeta_{j,t})=(\widehat\blambda_{j,\subi}-\blambda_{j,\subi})'\bfrb_{t,\subi}-r_{t,\subi}(\balpha_j)=\bar{O}_\P(\sam^{-1}+\rate{\eta}+\rate{\mu})$. The Taylor remainder is quadratic in this difference by Assumption \ref{SA Assumption: QMLE}(a). The remaining term linear in $\epsilon_{j,t}$ has order $\bar{O}_\P((\sam^{-1}+\rate{\eta}+\rate{\mu})(\sam^{-1}+\biaserr))$ by conditional centering and Lemma \ref{SA lem: maximal inequality}. Combining this expansion with $\widehat{\bb}^\circ=-\bV_n^{-1}\bA_n$ and the inverse-Hessian bound above gives
{
\begin{equation}\label{eq: asymptotic expansion}
	\begin{split}
		\widehat{\zeta}_{i,t}-\zeta_{i,t}=&\;
		\psi_{\tty}^{(1)}\Big(\psi_{\tty}^{-1}(\zeta_{i,t})\Big)\times\\
		&\bigg\{\widehat{\bpi}_i'\bV_{\subi}^{-1}
		\Big[\frac{1}{K}\sum_{j\in\mathcal{N}_{i,t}}
		\Big(-\mathcal{L}_{1,\tty}(\psi_{\tty}^{-1}(\zeta_{j,t}), y_{j})
		+\frac{\psi_{\tty}^{(1)}(\psi_{\tty}^{-1}(\zeta_{j,t}))}{\bar{V}_{\tty}(\psi_{\tty}^{-1}(\zeta_{j,t}))}(\widehat{\blambda}_{j,\subi}-\blambda_{j,\subi})'\bfrb_{t,\subi}\Big)\widehat{\bpi}_j\Big]\\
		&+(\widehat{\blambda}_{i,\subi}-\blambda_{i,\subi})'\bfrb_{t,\subi}
		-r_{t,\subi}(\balpha_i)
		-\widehat{\bpi}_i'\bV_{\subi}^{-1}\Big[\frac{1}{K}\sum_{j\in\mathcal{N}_{i,t}}\frac{\psi_{\tty}^{(1)}(\psi_{\tty}^{-1}(\zeta_{j,t}))r_{t,\subi}(\balpha_j)}{\bar{V}_{\tty}(\psi_{\tty}^{-1}(\zeta_{j,t}))}\widehat{\bpi}_j\Big]\bigg\}\\
		&+\bar{O}_\P\big((\sam^{-1}+\rate{\eta}+\rate{\mu}+\biaserr)^2\big).
	\end{split}
\end{equation}
}

\subsection{Proof of Theorem \ref{SA thm: pointwise inference}}
	Let $\tilde{\varphi}_{i,t,s}:=\frac{d_i(s)\zeta_{i,t}}{\varrho_{s}}+\frac{\varrho_{i,s}d_i(t)(y_i-\zeta_{i,t})}{\varrho_{s}\varrho_{i,t}}-\theta_{t,s}$.
	By some algebra, 
	\begin{align*}
		\sqrt{n}(\widehat{\theta}_{t,s}-
		\theta_{t,s})
		=&\frac{1}{\sqrt{n}}\sum_{i=1}^n\tilde{\varphi}_{i,t,s}+\frac{1}{\sqrt{n}}\sum_{i=1}^n\mathscr{R}_{i}+\Big(\frac{\varrho_{s}}{\widehat{\varrho}_{s}}-\frac{\varrho_{s}}{\varrho_{s}}\Big)\cdot\frac{1}{\sqrt{n}}\sum_{i=1}^n\mathscr{R}_{i}\\
		&+\sqrt{n}\frac{\varrho_{s}-\widehat{\varrho}_{s}}{\varrho_{s}}\theta_{t,s}+
		\sqrt{n}(\varrho_{s}-\widehat{\varrho}_{s})\theta_{t,s}\Big(\frac{1}{\widehat{\varrho}_{s}}-\frac{1}{\varrho_{s}}\Big)\\
		&+\sqrt{n}\frac{\varrho_{s}-\widehat{\varrho}_{s}}{\widehat{\varrho}_{s}}\Big(\E_n\left[
		\frac{d_i(s)\zeta_{i,t}}{\varrho_{s}}+
		\frac{\varrho_{i,s}}{\varrho_{s}}\frac{d_i(t)(y_{i}-\zeta_{i,t})}{\varrho_{i,t}}\right]-\theta_{t,s}\Big)
	\end{align*}
	where $\frac{1}{\sqrt{n}}\sum_{i=1}^n\mathscr{R}_{i}$ is given by
	\begin{align*}
		&\frac{1}{\sqrt{n}}\sum_{i=1}^n\bigg[\frac{\varrho_{i,s}}{\varrho_{s}}d_i(t)(y_{i}-\zeta_{i,t})\Big(
		\frac{1}{\widehat{\varrho}_{i,t}}-\frac{1}{\varrho_{i,t}}\Big)\bigg]+
		\frac{1}{\sqrt{n}}\sum_{i=1}^n\bigg[\Big(\frac{\widehat{\varrho}_{i,s}-\varrho_{i,s}}{\varrho_{s}}\Big)\frac{d_i(t)(y_{i}-\zeta_{i,t})}{\widehat{\varrho}_{i,t}}\bigg]\\
		&+\frac{1}{\sqrt{n}}\sum_{i=1}^n\bigg[\Big(\frac{d_i(s)}{\varrho_{s}}-
		\frac{\widehat{\varrho}_{i,s}d_i(t)}{\varrho_{s}\widehat{\varrho}_{i,t}}\Big)(\widehat{\zeta}_{i,t}-\zeta_{i,t})\bigg]
		=:I_1+I_2+I_3.
	\end{align*}
	
	Note that $\frac{1}{\sqrt{n}}\sum_{i=1}^{n}\tilde{\varphi}_{i,t,s}+\sqrt{n}\frac{\varrho_{s}-\widehat{\varrho}_{s}}{\varrho_{s}}\theta_{t,s}=\frac{1}{\sqrt{n}}\sum_{i=1}^{n}\varphi_{i,t,s}$. 
	Thus, it remains to show that all other terms in the expansion are $o_\P(1)$. By Assumption \ref{SA Assumption: ID-counterfactual} and the fact that $\sqrt{n}(\widehat{\varrho}_{s}-\varrho_{s})=O_\P(1)$, 
	the last two terms on the right-hand side are $o_\P(1)$, and so is the third term if we can show $\frac{1}{\sqrt{n}}\sum_{i=1}^{n}\mathscr{R}_i=o_\P(1)$.
	
	In the following we focus on the expansion of $\frac{1}{\sqrt{n}}\sum_{i=1}^{n}\mathscr{R}_i$. 
	{
	For $I_1$, condition on $\mathcal H$. The propensity estimates and treatment indicators are fixed, and the outcome errors are independent across units with conditional means bounded by $\biaserr$. Uniform consistency and overlap bound the estimated inverse propensities. Therefore,
	\begin{align*}
	|\E[I_1\mid\mathcal H]|
	&\lesssim_\P\sqrt n\,\biaserr
	\big(\E_n[(\widehat\varrho_{i,t}-\varrho_{i,t})^2]\big)^{1/2}=o_\P(1),\\
	\V[I_1\mid\mathcal H]
	&\lesssim_\P\E_n[(\widehat\varrho_{i,t}-\varrho_{i,t})^2]=o_\P(1).
	\end{align*}
	Hence $I_1=o_\P(1)$. The same argument, using the propensity error for $s$, gives $I_2=o_\P(1)$.
	}

	Next, rewrite $\varrho_{s}I_3$ as
	\begin{align*}
	&\frac{1}{\sqrt{n}}\sum_{i=1}^n\Big[(\widehat{\zeta}_{i,t}-\zeta_{i,t})\Big(d_i(s)-\frac{\varrho_{i,s}d_i(t)}{\varrho_{i,t}}\Big)\Big]+
	\frac{1}{\sqrt{n}}\sum_{i=1}^n\Big[(\widehat{\zeta}_{i,t}-\zeta_{i,t})\Big(\frac{\varrho_{i,s}d_i(t)}{\varrho_{i,t}}-\frac{\widehat{\varrho}_{i,s}d_i(t)}{\widehat{\varrho}_{i,t}}\Big)\Big]\\
	=&:I_{3,1}+I_{3,2}.
	\end{align*}
	{
	For $I_{3,2}$, Cauchy--Schwarz, overlap, Theorem \ref{SA thm: QMLE}, and the stated rate condition give
	\[
	|I_{3,2}|\lesssim_\P\sqrt n
	\big(\E_n[(\widehat\zeta_{i,t}-\zeta_{i,t})^2]\big)^{1/2}
	\big(\E_n[(\widehat\varrho_{i,t}-\varrho_{i,t})^2+
	(\widehat\varrho_{i,s}-\varrho_{i,s})^2]\big)^{1/2}
	=o_\P(1).
	\]
	}
	For $I_{3,1}$, recall that $v_{i,t}=d_{i}(t)-\varrho_{i,t}$, and then we further write
	$$
	I_{3,1}=\frac{1}{\sqrt{n}}\sum_{i=1}^n[(\widehat{\zeta}_{i,t}-\zeta_{i,t})v_{i,s}]-
	\frac{1}{\sqrt{n}}\sum_{i=1}^n\Big[(\widehat{\zeta}_{i,t}-\zeta_{i,t})v_{i,t}\frac{\varrho_{i,s}}{\varrho_{i,t}}\Big]=:I_{3,1,1}-I_{3,1,2}.
	$$
	{
	We focus on $I_{3,1,1}$; the additional factor $\varrho_{i,s}/\varrho_{i,t}$ in $I_{3,1,2}$ is bounded and measurable on $\mathcal F$, and thus $I_{3,1,2}$ can be treated similarly. Recall that $\bar r_{i,t,\subi}=\zeta_{i,t}-\psi_{\tty}(\kappa_i)$. We first expand $I_{3,1,1}$ as follows:
	\begin{align*}
	I_{3,1,1}
	&=\frac{1}{\sqrt n}\sum_{i=1}^n v_{i,s}
	\Big\{\psi_{\tty}(\bz_i'\widehat{\bbeta}_{t,\subi}+\widehat{\blambda}_{i,\subi}'\widehat{\bb}_{t,\subi})
	-\psi_{\tty}(\kappa_i)-\bar r_{i,t,\subi}\Big\}\\
	&=\frac{1}{\sqrt n}\sum_{i=1}^n v_{i,s}
	\Big\{\psi_{\tty}(\bz_i'\widehat{\bbeta}_{t,\subi}+\widehat{\blambda}_{i,\subi}'\widehat{\bb}_{t,\subi})
	-\psi_{\tty}(\kappa_i)\Big\}+o_\P(1)\\
	&=\frac{1}{\sqrt n}\sum_{i=1}^n v_{i,s}
	\Big\{\psi_{\tty}(\bz_i'\widehat{\bbeta}_{t,\subi}+\widehat{\blambda}_{i,\subi}'\widehat{\bb}_{t,\subi})
	-\psi_{\tty}(\widehat\kappa_i)\Big\}+o_\P(1)\\
	&=\frac{1}{\sqrt n}\sum_{i=1}^n v_{i,s}
	\psi_{\tty}^{(1)}(\psi_{\tty}^{-1}(\zeta_{i,t}))
	\Big\{\bz_i'(\widehat{\bbeta}_{t,\subi}-\bbeta_t)
	+\widehat{\blambda}_{i,\subi}'(\widehat{\bb}_{t,\subi}-\bfrb_{t,\subi})\Big\}+o_\P(1)\\
	&=\frac{1}{\sqrt n}\sum_{i=1}^n v_{i,s}
	\psi_{\tty}^{(1)}(\psi_{\tty}^{-1}(\zeta_{i,t}))
	\widehat{\bpi}_i'\widehat{\bb}^{\circ}+o_\P(1).
	\end{align*}
	For the second and third equalities, the discarded terms are measurable on $\mathcal F$ apart from $v_{i,s}$. Conditional variance and the bound on $\E[v_{i,s}\mid\mathcal F]$ give
	\begin{align*}
	\frac{1}{\sqrt n}\sum_{i=1}^n v_{i,s}\bar r_{i,t,\subi}
	&=O_\P\!\left(\big(\E_n[\rate{\mu}^2]\big)^{1/2}
	(1+\sqrt n\,\biaserr)\right)=o_\P(1),\\
	\frac{1}{\sqrt n}\sum_{i=1}^n v_{i,s}\big\{\psi_{\tty}(\widehat\kappa_i)-\psi_{\tty}(\kappa_i)\big\}
	&=O_\P\!\left(\left\{\sam^{-1}+\big(\E_n[\rate{\eta}^2]\big)^{1/2}\right\}
	(1+\sqrt n\,\biaserr)\right)=o_\P(1).
	\end{align*}
	The fourth equality follows by Taylor expansion at $\psi_{\tty}^{-1}(\zeta_{i,t})$, and the last follows by the definitions of $\widehat{\bpi}_i$ and $\widehat{\bb}^{\circ}$. The Taylor remainder, multiplied by the bounded treatment residual and averaged, is
	\[
	O_\P\!\left(\sqrt n\,\E_n[
	(\sam^{-1}+\rate{\eta}+\rate{\mu}+\biaserr)^2]\right)=o_\P(1).
	\]
	Finally, expanding $\widehat{\bb}^{\circ}=-\bV_n^{-1}\bA_n$ as above gives
	\[
	\begin{split}
		I_{3,1,1}={}&-\frac{1}{K\sqrt n}\sum_{i=1}^{n}v_{i,s}\psi_{\tty}^{(1)}(\psi_{\tty}^{-1}(\zeta_{i,t}))\widehat{\bpi}_{i,\subi}'\bV_{\subi}^{-1}\times
		\bigg[\sum_{j\in\mathcal{N}_{i,t}}
		\bigg(\mathcal{L}_{1,\tty}(\psi_{\tty}^{-1}(\zeta_{j,t}),y_j)\\
		&\qquad
		+\frac{\psi_{\tty}^{(1)}(\psi_{\tty}^{-1}(\zeta_{j,t}))r_{t,\subi}(\balpha_j)}{\bar{V}_{\tty}(\psi_{\tty}^{-1}(\zeta_{j,t}))}
		-\frac{\psi_{\tty}^{(1)}(\psi_{\tty}^{-1}(\zeta_{j,t}))}{\bar{V}_{\tty}(\psi_{\tty}^{-1}(\zeta_{j,t}))}(\widehat{\blambda}_{j,\subi}-\blambda_{j,\subi})'\bfrb_{t,\subi}\bigg)\widehat{\bpi}_{j,\subi}\bigg]
		+o_\P(1).
	\end{split}
	\]
	}
    Note that for clarity of expression, the subscript $\subi$ is added back to $\widehat{\bpi}_j$.
	Expanding this expression yields three terms.  Specifically,
	the first term we need to consider is
	{\small
		\[
		\begin{split}
			&\sqrt{n}\cdot\frac{1}{Kn}\sum_{i=1}^{n}v_{i,s}\psi_{\tty}^{(1)}(\psi_{\tty}^{-1}(\zeta_{i,t}))
			\widehat{\bpi}_{i,\subi}'\bV_{\subi}^{-1}\sum_{j=1}^{n}\widehat{\bpi}_{j,\subi}d_j(t)\I(j\in\mathcal{N}_i)\epsilon_{j,t}/\bar{V}_{\tty}(\psi_{\tty}^{-1}(\zeta_{j,t}))\\
			=&\sqrt{n}\cdot\frac{1}{n}
			\sum_{j=1}^{n}\bigg\{
			\frac{\epsilon_{j,t}d_j(t)}{\bar{V}_{\tty}(\psi_{\tty}^{-1}(\zeta_{j,t}))}\cdot
			\Big(\frac{1}{K}\sum_{i=1}^{n}v_{i,s}
			\psi_{\tty}^{(1)}(\psi_{\tty}^{-1}(\zeta_{i,t}))
			\widehat{\bpi}_{i,\subi}'\bV_{\subi}^{-1}\widehat{\bpi}_{j,\subi}
			\I(j\in\mathcal{N}_i)\Big)\bigg\}\\
			=&:\sqrt{n}\cdot\frac{1}{n}
			\sum_{j=1}^{n}\frac{(\epsilon_{j,t}-\E[\epsilon_{j,t}|\mathcal{H}]+\E[\epsilon_{j,t}|\mathcal{H}])d_j(t)}{\bar{V}_{\tty}(\psi_{\tty}^{-1}(\zeta_{j,t}))}\cdot a_j,
		\end{split}
		\] 
	}
	
	\vspace{-1em}
	{
	\noindent Conditional on $\mathcal H$, the centered outcome errors in this expression are independent and the weights $a_j$ are fixed. To bound these weights, write $D_j=\sum_{i=1}^n\I(j\in\mathcal N_i)$; then $\sum_jD_j=nK$ and $D_j\leq n$. All coefficients multiplying $v_{i,s}$ in $a_j$ are measurable on $\mathcal F$ and uniformly bounded in probability. Consequently,
	\[
	\V[a_j\mid\mathcal F]\lesssim_\P\frac{D_j}{K^2},
	\qquad
	|\E[a_j\mid\mathcal F]|\lesssim_\P
	\frac{D_j}{K}\max_i|\E[v_{i,s}\mid\mathcal F]|.
	\]
	Using $\sum_jD_j^2\leq n^2K$, conditional Markov's inequality, and Jensen's inequality,
	\[
	\E_n[a_j^2]=O_\P\!\left(K^{-1}+\frac{n}{K}\biaserr^2\right)=o_\P(1),
	\qquad
	\E_n[|a_j|]=O_\P(K^{-1/2}+\biaserr).
	\]
	Consequently,
	\begin{align*}
	&\V\!\left[\frac{1}{\sqrt n}\sum_{j=1}^n
	\frac{(\epsilon_{j,t}-\E[\epsilon_{j,t}\mid\mathcal H])d_j(t)a_j}
	{\bar V_{\tty}(\psi_{\tty}^{-1}(\zeta_{j,t}))}
	\,\middle|\,\mathcal H\right]\\
	&\qquad=\frac{1}{n}\sum_{j=1}^n
	\frac{d_j(t)a_j^2\V[\epsilon_{j,t}\mid\mathcal H]}
	{\bar V_{\tty}^2(\psi_{\tty}^{-1}(\zeta_{j,t}))}
	\lesssim_\P\E_n[a_j^2]=o_\P(1).
	\end{align*}
	Moreover,
	\[
	\begin{split}
	\left|\frac{1}{\sqrt n}\sum_{j=1}^n
	\frac{\E[\epsilon_{j,t}\mid\mathcal H]d_j(t)a_j}
	{\bar V_{\tty}(\psi_{\tty}^{-1}(\zeta_{j,t}))}\right|
	&\leq\frac{\biaserr}{\sqrt n}\sum_{j=1}^n
	\left|\frac{d_j(t)a_j}{\bar V_{\tty}(\psi_{\tty}^{-1}(\zeta_{j,t}))}\right|
	\\&=O_\P\!\left(\sqrt n\,\biaserr(K^{-1/2}+\biaserr)\right)=o_\P(1).
	\end{split}
	\]
	Here the stated rate condition implies $\sqrt n/K=o(1)$ and $\sqrt n\,\biaserr^2=o_\P(1)$.
	}

	 The second term we need to consider is
	 \begin{align*}
	 &\hspace{1.5em}\sqrt{n}\cdot\frac{1}{Kn}\sum_{i=1}^{n}v_{i,s}\psi_{\tty}^{(1)}(\psi_{\tty}^{-1}(\zeta_{i,t}))\widehat{\bpi}_{i,\subi}'\bV_{\subi}^{-1}
	 \sum_{j=1}^n\frac{\psi_{\tty}^{(1)}(\psi_{\tty}^{-1}(\zeta_{j,t}))r_{t,\subi}(\balpha_j)}{\bar{V}_{\tty}(\psi_{\tty}^{-1}(\zeta_{j,t}))}\widehat{\bpi}_{j,\subi}d_j(t)\I(j\in\mathcal{N}_i)\\
	 &=\sqrt{n}\cdot\frac{1}{Kn}
	 \sum_{j=1}^n\frac{d_j(t)\psi_{\tty}^{(1)}(\psi_{\tty}^{-1}(\zeta_{j,t}))}{\bar{V}_{\tty}(\psi_{\tty}^{-1}(\zeta_{j,t}))}
	 \sum_{i=1}^{n}v_{i,s}\psi_{\tty}^{(1)}(\psi_{\tty}^{-1}(\zeta_{i,t}))\widehat{\bpi}_{i,\subi}'\bV_{\subi}^{-1}
	 \widehat{\bpi}_{j,\subi}r_{t,\subi}(\balpha_j)\I(j\in\mathcal{N}_i)\\
	 &=:\sqrt{n}\cdot\frac{1}{Kn}
	 \sum_{j=1}^n\frac{d_j(t)\psi_{\tty}^{(1)}(\psi_{\tty}^{-1}(\zeta_{j,t}))}{\bar{V}_{\tty}(\psi_{\tty}^{-1}(\zeta_{j,t}))}
	 \sum_{i=1}^nv_{i,s}c_{i,j}.
	 \end{align*}
	{
	The coefficients $c_{i,j}$ are measurable on $\mathcal F$, with
	$|c_{i,j}|\lesssim_\P\rate{\mu}\I(j\in\mathcal N_i)$.
	Let $\tilde v_{i,s}=v_{i,s}-\E[v_{i,s}\mid\mathcal F]$.
	Conditional independence and Cauchy--Schwarz give
	\begin{align*}
	\E\!\left[\left|\sum_{i=1}^n\tilde v_{i,s}c_{i,j}\right|
	\,\middle|\,\mathcal F\right]
	&\lesssim_\P\left(\sum_{i=1}^n\I(j\in\mathcal N_i)\rate{\mu}^2\right)^{1/2},\\
	\frac{1}{K\sqrt n}\sum_{j=1}^n
	\E\!\left[\left|\sum_{i=1}^n\tilde v_{i,s}c_{i,j}\right|
	\,\middle|\,\mathcal F\right]
	&\lesssim_\P\frac{1}{K\sqrt n}\sum_{j=1}^n
	\left(\sum_{i=1}^n\I(j\in\mathcal N_i)\rate{\mu}^2\right)^{1/2}\\
	&\leq\frac{\sqrt n}{K}
	\left(\frac{1}{n}\sum_{i=1}^n\rate{\mu}^2\sum_{j=1}^n\I(j\in\mathcal N_i)\right)^{1/2}\\
	&=\sqrt{\frac nK}\big(\E_n[\rate{\mu}^2]\big)^{1/2}.
	\end{align*}
	The last two steps use Jensen's inequality and $\sum_j\I(j\in\mathcal N_i)=K$.

	For the second term in the decomposition of $I_{3,1,1}$, we obtain
	\[
	\begin{split}
	&\left|\frac{1}{K\sqrt n}\sum_{j=1}^n
	\frac{d_j(t)\psi_{\tty}^{(1)}(\psi_{\tty}^{-1}(\zeta_{j,t}))}
	{\bar V_{\tty}(\psi_{\tty}^{-1}(\zeta_{j,t}))}
	\sum_{i=1}^n v_{i,s}c_{i,j}\right|\\
	&\quad\lesssim_\P\frac{1}{K\sqrt n}\sum_{j=1}^n
	\left|\sum_{i=1}^n\tilde v_{i,s}c_{i,j}\right|
	+\frac{\biaserr}{K\sqrt n}\sum_{j=1}^n\sum_{i=1}^n\I(j\in\mathcal N_i)\rate{\mu}\\
	&\quad=\frac{1}{K\sqrt n}\sum_{j=1}^n
	\left|\sum_{i=1}^n\tilde v_{i,s}c_{i,j}\right|
	+\sqrt n\,\biaserr\E_n[\rate{\mu}]\\
	&\quad=O_\P\!\left(\sqrt{\frac nK}\big(\E_n[\rate{\mu}^2]\big)^{1/2}
	+\sqrt n\,\biaserr\E_n[\rate{\mu}]\right)\\
	&\quad=O_\P\!\left(\left\{\sqrt{\frac nK}+\sqrt n\,\biaserr\right\}
	\big(\E_n[\rate{\mu}^2]\big)^{1/2}\right)=o_\P(1).
	\end{split}
	\]
	Here we split $v_{i,s}$ into $\tilde v_{i,s}+\E[v_{i,s}\mid\mathcal F]$ and use boundedness of the outer factors. Conditional Markov's inequality gives the bound for the sum involving $\tilde v_{i,s}$, and Cauchy--Schwarz bounds $\E_n[\rate{\mu}]$. The last equality follows from $\sqrt n/K=o(1)$ and $\sqrt n\big(\biaserr^2+\E_n[\rate{\mu}^2]\big)=o_\P(1)$.
	}

    For the last term in the decomposition of $I_{3,1,1}$, apply the same conditional moment calculation with $\rate{\mu}$ replaced by $\sam^{-1}+\rate{\eta}$, using \eqref{SA eq: projected loading error}. This gives
     \begin{small}
     	\[
     	\begin{split}
		&\quad\;\frac{1}{K\sqrt n}\sum_{i=1}^{n}v_{i,s}\psi_{\tty}^{(1)}(\psi_{\tty}^{-1}(\zeta_{i,t}))
		\widehat{\bpi}_{i,\subi}'\bV_{\subi}^{-1}
		\sum_{j=1}^{n}\frac{\psi_{\tty}^{(1)}(\psi_{\tty}^{-1}(\zeta_{j,t}))}{\bar{V}_{\tty}(\psi_{\tty}^{-1}(\zeta_{j,t}))}(\widehat{\blambda}_{j,\subi}-\blambda_{j,\subi})'\bfrb_{t,\subi}\widehat{\bpi}_{j,\subi}d_j(t)\I(j\in\mathcal{N}_i)\\
		&=\frac{1}{K\sqrt n}
     		\sum_{j=1}^{n}\frac{d_j(t)\psi_{\tty}^{(1)}(\psi_{\tty}^{-1}(\zeta_{j,t}))}{\bar{V}_{\tty}(\psi_{\tty}^{-1}(\zeta_{j,t}))}
     		\sum_{i=1}^{n}v_{i,s}\psi_{\tty}^{(1)}(\psi_{\tty}^{-1}(\zeta_{i,t}))
		\widehat{\bpi}_{i,\subi}'\bV_{\subi}^{-1}\widehat{\bpi}_{j,\subi}\I(j\in\mathcal{N}_i)(\widehat{\blambda}_{j,\subi}-\blambda_{j,\subi})'\bfrb_{t,\subi}\\
		&\lesssim_\P
	\left(\sqrt{\frac nK}+\sqrt n\,\biaserr\right)
	\left(\sam^{-1}+\big(\E_n[\rate{\eta}^2]\big)^{1/2}\right)=o_\P(1).
     	\end{split}
     	\]
     \end{small}

    Then, the proof for part (i) is complete.
	
	Now, consider the variance estimation. Note that
	\[
		\sigma_{t,s}^2=\V[\varphi_{i,t,s}]=\E\Big[\frac{\varrho_{i,s}(\zeta_{i,t}-\theta_{t,s})^2}{\varrho_{s}^2}\Big]+
		\E\Big[\frac{\varrho_{i,s}^2\E[\epsilon_{i,t}^2\mid\balpha_i,\bz_i]}{\varrho_{s}^2\varrho_{i,t}}\Big]=:\mathscr{V}_1+\mathscr{V}_2.
	\]
	The proposed variance estimator is
	\[
	\widehat{\sigma}_{t,s}^2=\,
	\E_n\Big[\frac{d_i(s)(\widehat{\zeta}_{i,t}-\widehat{\theta}_{t,s})^2}{\widehat{\varrho}_{s}^2}\Big]+
	\E_n\Big[\frac{\widehat{\varrho}_{i,s}^2d_i(t)(y_i-\widehat{\zeta}_{i,t})^2}{\widehat{\varrho}_{s}^2\widehat{\varrho}_{i,t}^2}\Big]
	=:\widehat{\mathscr{V}}_1+\widehat{\mathscr{V}}_2.
	\]
	Begin with $\widehat{\mathscr{V}}_1$.
	\begin{align*}
		\widehat{\mathscr{V}}_1=&\E_n\Big[\frac{d_i(s)(\zeta_{i,t}-\theta_{t,s})^2}{\varrho_{s}^2}\Big]+
		\E_n\Big[\frac{d_i(s)(\zeta_{i,t}-\theta_{t,s})^2}{\widehat{\varrho}_{s}^2\varrho_{s}^2}(\varrho_{s}-\widehat{\varrho}_{s})(\varrho_{s}+\widehat{\varrho}_{s})\Big]+\\
		&\E_n\Big[\frac{d_i(s)[(\widehat{\zeta}_{i,t}-\widehat{\theta}_{t,s})^2-(\zeta_{i,t}-\theta_{t,s})^2]}{\widehat{\varrho}_{s}^2}\Big].
	\end{align*}
	The first term is consistent for $\mathscr{V}_1$.
	Since $\widehat{\varrho}_{s}=\varrho_{s}+o_\P(1)$, the second term is $o_\P(1)$ due to Assumption \ref{SA Assumption: ID-counterfactual}. For the third term, note 
	\[
	(\widehat{\zeta}_{i,t}-\widehat{\theta}_{t,s})^2-(\zeta_{i,t}-\theta_{t,s})^2=(\widehat{\zeta}_{i,t}^2-\zeta_{i,t}^2)+(\widehat{\theta}_{t,s}^2-\theta_{t,s}^2)-2(\widehat{\theta}_{t,s}-\theta_{t,s})\widehat{\zeta}_{i,t}-2\theta_{t,s}(\widehat{\zeta}_{i,t}-\zeta_{i,t}).
	\]
	Due to the uniform consistency shown in Theorem \ref{SA thm: QMLE}, the third term is $o_\P(1)$ as well.
	
	Moreover,
	\[
	\begin{split}
		\widehat{\mathscr{V}}_2=\,
		&\E_n\bigg[\frac{\varrho_{i,s}^2d_i(t)\epsilon_{i,t}^2}{\varrho_{s}^2\varrho_{i,t}^2}\bigg]+
		\E_n\bigg[\frac{\varrho_{i,s}^2d_i(t)\epsilon_{i,t}^2}{\varrho_{i,t}^2}\bigg](\widehat{\varrho}_{s}^{-2}-\varrho_{s}^{-2})+
		\E_n\bigg[\frac{\varrho_{i,s}^2d_i(t)\epsilon_{i,t}^2}{\widehat{\varrho}_{s}^2}(\widehat{\varrho}_{i,t}^{-2}-\varrho_{i,t}^{-2})\bigg]+\\
		&\E_n\bigg[\frac{(\widehat{\varrho}_{i,s}^2-\varrho_{i,s}^2)d_i(t)\epsilon_{i,t}^2}{\widehat{\varrho}_{s}^2\widehat{\varrho}_{i,t}^2}\bigg]+
		\E_n\bigg[\frac{\widehat{\varrho}_{i,s}^2d_i(t)((\epsilon_{i,t}+\zeta_{i,t}-\widehat{\zeta}_{i,t})^2-\epsilon_{i,t}^2)}{\widehat{\varrho}_{s}^2\widehat{\varrho}_{i,t}^2}\bigg].
	\end{split}
	\]
		By Theorem \ref{SA thm: QMLE}, the first term is consistent for $\mathscr{V}_2$ and the remainders are all $o_\P(1)$. Since $\sigma_{t,s}^2>0$, the desired result in part (ii) then follows by the consistency of the variance estimator, part (i), and Slutsky's theorem.

\subsection{Proof of Theorem \ref{SA thm: uniform inference}}
\begin{proof}
	By definition, in this context $\zeta_{i,t,\tau}=\P(y_i(t)\leq \tau|\balpha_i,\bz_i)$,  $y_{i,\tau}=\sum_{t=0}^{J}d_i(t)\I(y_i(t)\leq \tau)$, and the expansion for $\widehat{\theta}_{t,s}(\tau)$ given in the proof of Theorem \ref{SA thm: pointwise inference} becomes
	\[
	\begin{split}
		\sqrt{n}(\widehat{\theta}_{t,s}(\tau)-
		\theta_{t,s}(\tau))=&
		\frac{1}{\sqrt{n}}\sum_{i=1}^n\varphi_{i,t,s}(\tau)+
		\frac{1}{\sqrt{n}}\sum_{i=1}^n\mathscr{R}_{i,\tau}+
		\Big(\frac{\varrho_{s}}{\widehat{\varrho}_{s}}-\frac{\varrho_{s}}{\varrho_{s}}\Big)\cdot\frac{1}{\sqrt{n}}\sum_{i=1}^n\mathscr{R}_{i,\tau}+\\
		&\sqrt{n}\theta_{t,s}(\tau)(\varrho_{s}-\widehat{\varrho}_{s})\Big(\frac{1}{\widehat{\varrho}_{s}}-\frac{1}{\varrho_{s}}\Big)+\\
		&\sqrt{n}\frac{\varrho_{s}-\widehat{\varrho}_{s}}{\widehat{\varrho}_{s}}\left(\E_n\left[
		\frac{d_i(s)\zeta_{i,t,\tau}}{\varrho_{s}}+
		\frac{\varrho_{i,s}}{\varrho_{s}}\frac{d_i(t)(y_{i,\tau}-\zeta_{i,t,\tau})}{\varrho_{i,t}}\right]-\theta_{t,s}(\tau)\right).
	\end{split}
	\]
	The definition of $\mathscr{R}_{i,\tau}$ is the same as in the proof of Theorem \ref{SA thm: pointwise inference}, but
	its dependence on $\tau$ is made explicit.

	\medskip
	\noindent\textit{Step 1: Uniform convergence in distribution of the leading term.}
	Regarding the first term, the threshold indicators and conditional distribution functions are bounded and nondecreasing in $\tau$, so the resulting threshold-indexed class has the standard entropy bound for monotone functions. Overlap bounds the propensity weights, which do not depend on $\tau$. Thus, the function class $\{(\balpha_i,\bz_i,t_i,y_i)\mapsto\varphi_{i,t,s}(\tau):\tau\in\mathcal Y\}$ is Donsker, as in Lemma A.2 of \cite{Donald-Hsu_2014_JoE}. By unconfoundedness, $\E[\varphi_{i,t,s}(\tau)]=0$; Donsker's theorem then implies the stated Gaussian limit \citep[][Section 2.8.2]{vandeVaart_book_1996}.

	\medskip
	\noindent\textit{Step 2: Uniform bound for the remainder in the causal-estimand expansion.}
	Then, it remains to show that the remainder of the above expansion is $o_\P(1)$ uniformly over $\tau\in\mathcal{Y}$, denoted by $\tilde{o}_\P(1)$ within this proof. In the fourth term, $\theta_{t,s}(\tau)$ is bounded uniformly over $\tau$ and $\widehat{\varrho}_{s}$ does not depend on $\tau$. Thus, it is $\tilde{o}_\P(1)$. Regarding the last term, $\sqrt{n}(\varrho_{s}-\widehat{\varrho}_{s})/\widehat{\varrho}_{s}=O_\P(1)$, which does not depend on $\tau$, and
	the term within the parentheses is $\tilde{o}_\P(1)$ by uniform law of large numbers. Then, it remains to bound the second term, since the third term must be of smaller order compared to the second one.

	Recall that $\frac{1}{\sqrt{n}}\sum_{i=1}^n\mathscr{R}_{i,\tau}$ can be decomposed into
	\begin{align*}
		&\frac{1}{\sqrt{n}}\sum_{i=1}^n\bigg[\frac{\varrho_{i,s}}{\varrho_{s}}d_i(t)(y_{i,\tau}-\zeta_{i,t,\tau})\Big(
		\frac{1}{\widehat{\varrho}_{i,t}}-\frac{1}{\varrho_{i,t}}\Big)\bigg]+
		\frac{1}{\sqrt{n}}\sum_{i=1}^n\bigg[\Big(\frac{\widehat{\varrho}_{i,s}-\varrho_{i,s}}{\varrho_{s}}\Big)\frac{d_i(t)(y_{i,\tau}-\zeta_{i,t,\tau})}{\widehat{\varrho}_{i,t}}\bigg]\\
		&+\frac{1}{\sqrt{n}}\sum_{i=1}^n\bigg[\Big(\frac{d_i(s)}{\varrho_{s}}-
		\frac{\widehat{\varrho}_{i,s}d_i(t)}{\varrho_{s}\widehat{\varrho}_{i,t}}\Big)(\widehat{\zeta}_{i,t,\tau}-\zeta_{i,t,\tau})\bigg]=:I_1+I_2+I_3.
	\end{align*}

	For $I_1$, first notice that conditional on $\mathcal{H}=\mathcal{F}\vee\sigma(\{\bm{d}_{i}\}_{i\in[n]})$, as defined in the proof of Theorem \ref{SA thm: pointwise inference}, the summand in $I_1$ is independent over $i$ for any $\tau$, and $\widehat{\varrho}_{i,t}$ and $\varrho_{i,t}$ do not depend on $\tau$. Define $\epsilon_{i,t,\tau}=\I(y_i(t)\leq \tau)-\zeta_{i,t,\tau}$ and $\tilde{\epsilon}_{i,t,\tau}=\epsilon_{i,t,\tau}-\E[\epsilon_{i,t,\tau}|\mathcal{H}]$. Then, $\E[\tilde{\epsilon}_{i,t,\tau}|\mathcal{H}]=0$ and $|\E[\epsilon_{i,t,\tau}|\mathcal{H}]|\leq\barbiaserr$.
	For brevity, write $R_n^2=\sam^{-2}+\barbiaserr^2+\E_n[\rate{\eta}^2+\barrate{\mu}^2+\rate{e}^2]$. The theorem's rate condition is $\sqrt{n\log n}R_n^2=o_\P(1)$.
	For the centered component of $I_1$, define
	\[
	\mathcal{T}=\Big\{\ba=(a_1,\cdots, a_n)\in\mathbb{R}^n: a_i=\varrho_{i,s}d_i(t)\tilde{\epsilon}_{i,t,\tau}\Big(\frac{1}{\widehat{\varrho}_{i,t}}-\frac{1}{\varrho_{i,t}}\Big),\tau\in\mathcal Y\Big\}.
	\]
	Use the empirical norm $\|\ba\|_{n,2}^2=\E_n[a_i^2]$, and set $\Theta^2=\E_n[\varrho_{i,s}^2d_i(t)(\widehat\varrho_{i,t}^{-1}-\varrho_{i,t}^{-1})^2]$. Since $\tilde\epsilon_{i,t,\tau}=\I(y_i(t)\leq\tau)-\P(y_i(t)\leq\tau\mid\mathcal H)$ and each component is bounded and nondecreasing in $\tau$, separate covers give $N(\mathcal T,\|\cdot\|_{n,2},\varepsilon\Theta)\lesssim\varepsilon^{-4}$ for $0<\varepsilon<1$. Hence, by conditional symmetrization and Dudley's inequality,
	\[
	\E\Big[\sup_{\tau\in\mathcal Y}\Big|\frac1{\sqrt n}\sum_{i=1}^n a_i\Big|\,\Big|\,\mathcal H\Big]\lesssim\Theta\lesssim_\P R_n.
	\]
	Here Theorem \ref{SA thm: QMLE} bounds the propensity error, including its conditional-mean component, by $R_n$ in empirical $L_2$ norm. The component of $I_1$ involving $\E[\epsilon_{i,t,\tau}\mid\mathcal H]$ is uniformly $O_\P(\sqrt n\barbiaserr R_n)$. Since $I_2$ has the same centered-score and conditional-mean structure, we have
	\[
	\sup_{\tau\in\mathcal Y}(|I_1|+|I_2|)=O_\P\big((1+\sqrt n\barbiaserr)R_n\big)=o_\P(1).
	\]

	For now, suppose that \eqref{eq: asymptotic expansion} holds uniformly over $\tau$ after indexing all outcome quantities by $\tau$, and that the coefficient and comparison-Hessian paths appearing in the expansion have polynomial covering numbers. Denote its remainder by $\mathrm{Rem}_{i,t,\tau}$. Then,
	\begin{align*}
	\sup_{\tau\in\mathcal Y}|\widehat\zeta_{i,t,\tau}-\zeta_{i,t,\tau}|
	&=\bar O_\P(\sam^{-1}+\rate{\eta}+\barrate{\mu}+\barbiaserr),\\
	\sup_{\tau\in\mathcal Y}|\mathrm{Rem}_{i,t,\tau}|
	&=\bar O_\P\big((\sam^{-1}+\rate{\eta}+\barrate{\mu}+\barbiaserr)^2\big),\\
	\sqrt n\,\E_n[\sup_{\tau\in\mathcal Y}|\mathrm{Rem}_{i,t,\tau}|]
	&=O_\P(\sqrt n R_n^2)=o_\P(1).
	\end{align*}
	We verify this uniform version of \eqref{eq: asymptotic expansion} after completing the argument for $I_3$.

	Regarding $I_3$, recall that
	\begin{align*}
		\varrho_{s}I_3&=\frac{1}{\sqrt{n}}\sum_{i=1}^n
		\Big[(\widehat{\zeta}_{i,t,\tau}-\zeta_{i,t,\tau})\Big(d_i(s)-\frac{\varrho_{i,s}d_i(t)}{\varrho_{i,t}}\Big)\Big]\\
		&\qquad+\frac{1}{\sqrt{n}}\sum_{i=1}^n\Big[(\widehat{\zeta}_{i,t,\tau}-\zeta_{i,t,\tau})\Big(\frac{\varrho_{i,s}d_i(t)}{\varrho_{i,t}}-\frac{\widehat{\varrho}_{i,s}d_i(t)}{\widehat{\varrho}_{i,t}}\Big)\Big]\\
		&=:I_{3,1}+I_{3,2}.
	\end{align*}
	By Cauchy--Schwarz and the uniform nuisance bound, $\sup_{\tau\in\mathcal Y}|I_{3,2}|=O_\P(\sqrt n R_n^2)=o_\P(1)$. For $I_{3,1}$, use the decomposition $I_{3,1,1}-I_{3,1,2}$ from the proof of Theorem \ref{SA thm: pointwise inference}. We detail the uniform bounds for $I_{3,1,1}$; the bounded factor $\varrho_{i,s}/\varrho_{i,t}$ in $I_{3,1,2}$ does not depend on $\tau$.

	For the first double sum in the decomposition of $I_{3,1,1}$ in the proof of Theorem \ref{SA thm: pointwise inference}, write
	\[
	a_{j,\tau}=\frac1K\sum_{i=1}^n v_{i,s}
	\psi_{\tty}^{(1)}(\psi_{\tty}^{-1}(\zeta_{i,t,\tau}))
	\widehat{\bpi}_{i,\subi}'\bV_{\subi,\tau}^{-1}
	\widehat{\bpi}_{j,\subi}\I(j\in\mathcal N_i).
	\]
	All coefficients multiplying $v_{i,s}$ are measurable on $\mathcal F$. With $D_j=\sum_i\I(j\in\mathcal N_i)$, the centered weights have conditional second moments of order $D_j\log n/K^2$ by conditional symmetrization over $\tau$, and their conditional means are bounded by $D_j\barbiaserr/K$. Since $\sum_jD_j=nK$ and $\sum_jD_j^2\leq n^2K$, as established in the proof of Theorem \ref{SA thm: pointwise inference}, we have
	\begin{align*}
	\E_n[\sup_{\tau\in\mathcal Y}a_{j,\tau}^2]
	&=O_\P\big(\log n/K+n\barbiaserr^2/K\big),\\
	\E_n[\sup_{\tau\in\mathcal Y}|a_{j,\tau}|]
	&=O_\P\big(\sqrt{\log n/K}+\barbiaserr\big).
	\end{align*}
	Since the weights are fixed conditional on $\mathcal H$ and $\tilde\epsilon_{j,t,\tau}$ is a bounded centered threshold-indexed process, by the conditional maximal inequality, we have
	\[
	\sup_{\tau\in\mathcal Y}\bigg|\frac1{\sqrt n}\sum_{j=1}^n
	\frac{d_j(t)\tilde\epsilon_{j,t,\tau}a_{j,\tau}}
	{\bar V_{\tty}(\psi_{\tty}^{-1}(\zeta_{j,t,\tau}))}\bigg|
	=O_\P\bigg(\frac{\log n}{\sqrt K}+\sqrt{\frac{n\log n}{K}}\barbiaserr\bigg).
	\]
	Since $|\E[\epsilon_{j,t,\tau}\mid\mathcal H]|\leq\barbiaserr$, by the first-moment bound for $a_{j,\tau}$, we have
	\[
	\sup_{\tau\in\mathcal Y}\bigg|\frac1{\sqrt n}\sum_{j=1}^n
	\frac{d_j(t)\E[\epsilon_{j,t,\tau}\mid\mathcal H]a_{j,\tau}}
	{\bar V_{\tty}(\psi_{\tty}^{-1}(\zeta_{j,t,\tau}))}\bigg|
	=O_\P\big(\sqrt n\barbiaserr\{\sqrt{\log n/K}+\barbiaserr\}\big).
	\]
	These bounds include the within-unit terms and do not require disjoint neighborhoods.

	For the approximation and loading terms in the decomposition of $I_{3,1,1}$ in the proof of Theorem \ref{SA thm: pointwise inference}, let $c_{i,j,\tau}$ denote the corresponding coefficient. Since these coefficients are measurable on $\mathcal F$, their envelope is bounded by $(\sam^{-1}+\rate{\eta}+\barrate{\mu})\I(j\in\mathcal N_i)$, and the centered treatment residuals are conditionally independent, the conditional maximal inequality, the conditional first-moment calculation in the proof of Theorem \ref{SA thm: pointwise inference}, and Jensen's inequality imply
	\[
	\frac1{K\sqrt n}\sum_{j=1}^n
	\E\Big[\sup_{\tau\in\mathcal Y}\Big|\sum_{i=1}^n
	(v_{i,s}-\E[v_{i,s}\mid\mathcal F])c_{i,j,\tau}\Big|\,\Big|\,\mathcal F\Big]
	\lesssim_\P\sqrt{\frac{n\log n}{K}}R_n.
	\]
	For the conditional-mean components of these approximation and loading terms, $|\E[v_{i,s}\mid\mathcal F]|\leq\barbiaserr$ and the envelope for $c_{i,j,\tau}$ imply
	\[
	\sup_{\tau\in\mathcal Y}\frac1{K\sqrt n}\sum_{j=1}^n
	\bigg|\sum_{i=1}^n\E[v_{i,s}\mid\mathcal F]c_{i,j,\tau}\bigg|
	\lesssim_\P \sqrt n\barbiaserr\,
	\E_n[\sam^{-1}+\rate{\eta}+\barrate{\mu}]
	=O_\P(\sqrt n\barbiaserr R_n),
	\]
	where the inequality uses $\sum_j\I(j\in\mathcal N_i)=K$, and the final equality follows by Cauchy--Schwarz. Write $\kappa_{i,\tau}$ and $\widehat\kappa_{i,\tau}$ for the threshold-indexed versions of $\kappa_i$ and $\widehat\kappa_i$ in the proof of Theorem \ref{SA thm: pointwise inference}. Since the centered treatment residuals are conditionally independent given $\mathcal F$ and the two summand classes have envelopes bounded by $R_n$, by the conditional maximal inequality, their centered components are $O_\P(\sqrt{\log n}R_n)$. Since $|\E[v_{i,s}\mid\mathcal F]|\leq\barbiaserr$, by the approximation bounds, their conditional-mean components are $O_\P(\sqrt n\barbiaserr R_n)$. Therefore,
	\[
	\begin{split}
	&\sup_{\tau\in\mathcal Y}\bigg|\frac1{\sqrt n}\sum_{i=1}^n v_{i,s}
	\{\zeta_{i,t,\tau}-\psi_{\tty}(\kappa_{i,\tau})\}\bigg|\\
	&\quad+\sup_{\tau\in\mathcal Y}\bigg|\frac1{\sqrt n}\sum_{i=1}^n v_{i,s}
	\{\psi_{\tty}(\widehat\kappa_{i,\tau})-\psi_{\tty}(\kappa_{i,\tau})\}\bigg|
	=O_\P\big((\sqrt{\log n}+\sqrt n\barbiaserr)R_n\big).
	\end{split}
	\]
	Since the second derivative of $\psi_{\tty}$, the treatment residual, and the estimated regressors are bounded, and $\sup_{\tau\in\mathcal Y}\|\widehat{\bb}^{\circ}_{\tau}\|=\bar O_\P(R_n)$, the quadratic Taylor remainder satisfies
	\[
	\sup_{\tau\in\mathcal Y}\frac1{\sqrt n}\sum_{i=1}^n
	|v_{i,s}|(\widehat{\bpi}_{i,\subi}'\widehat{\bb}^{\circ}_{\tau})^2
	=O_\P(\sqrt n R_n^2).
	\]
	Consequently,
	\[
	\sup_{\tau\in\mathcal Y}(|I_{3,1,1}|+|I_{3,1,2}|)
	=O_\P\big(\sqrt{\log n}R_n+\sqrt n R_n^2\big)=o_\P(1).
	\]
	Here $\sqrt{\log n/K}\leq\sam^{-1}\leq R_n$ and $\barbiaserr\leq R_n$ absorb the double-sum bounds, while the theorem's rate condition makes both remaining terms vanish. Therefore,
	\[
	\sup_{\tau\in\mathcal Y}\bigg|\frac1{\sqrt n}\sum_{i=1}^n\mathscr R_{i,\tau}\bigg|
	\leq\sup_{\tau\in\mathcal Y}(|I_1|+|I_2|+|I_3|)=o_\P(1),
	\]
	Conditional on the uniform expansion invoked above, this result, together with the bounds for the other terms and the leading empirical process, establishes the stated weak convergence. It remains to verify that expansion.

	\medskip
	\noindent\textit{Step 3: Verification of the uniform version of \eqref{eq: asymptotic expansion}.}

	\smallskip
	\noindent\textit{Step 3.1: Approximation and index paths.}
	We use the notation from the proof of Theorem \ref{SA thm: QMLE}, adding $\tau$ to the outcome quantities. The true indices $\psi_{\tty}^{-1}(\zeta_{j,t,\tau})$ are bounded and monotone in $\tau$, so their smooth transformations have the standard covering bounds for monotone functions.

	Let $\bfrb^0_{t,\tau,\subi}$ denote coefficients satisfying Assumption \ref{SA Assumption: QMLE, uniform}(b). On events of arbitrarily high probability, they lie in a fixed ball $\mathcal B_M$. On these events, choose
	\[
	\bfrb_{t,\tau,\subi}\in\argmin_{\bb\in\mathcal B_M}
	\frac1K\sum_{j\in\mathcal N_i}
	\{\mu_{t,\tau}(\balpha_j)-\blambda_{j,\subi}'\bb\}^2.
	\]
	Since $\bfrb^0_{t,\tau,\subi}$ is feasible, optimality, the loading normalization, and the bounded normalized loading rows imply
	\[
	K^{-1/2}\|\bLambda_{\subi}(\bfrb_{t,\tau,\subi}-\bfrb^0_{t,\tau,\subi})\|
	\leq 2\barrate{\mu},
	\qquad
	\max_{j\in\mathcal N_i}|\mu_{t,\tau}(\balpha_j)-\blambda_{j,\subi}'\bfrb_{t,\tau,\subi}|
	\lesssim_\P\barrate{\mu}.
	\]
	Thus this choice preserves the approximation rate in Assumption \ref{SA Assumption: QMLE, uniform}(b). Moreover, constrained least squares projects the vector of local conditional means onto the convex set $\{\bLambda_{\subi}\bb:\bb\in\mathcal B_M\}$. By Assumption \ref{SA Assumption: QMLE, uniform}(a),
	\begin{align*}
	\|\bUpsilon_{\subi}(\bfrb_{t,\tau,\subi}-\bfrb_{t,\tau',\subi})\|
	&\leq\bigg\{\frac1K\sum_{j\in\mathcal N_i}
	[\mu_{t,\tau}(\balpha_j)-\mu_{t,\tau'}(\balpha_j)]^2\bigg\}^{1/2}
	\lesssim|\tau-\tau'|,\\
	\|\bfrb_{t,\tau,\subi}-\bfrb_{t,\tau',\subi}\|
	&\lesssim_\P\upsilon_{\ttd_i,\subi}^{-1}|\tau-\tau'|,
	\qquad \max_i\upsilon_{\ttd_i,\subi}^{-1}=O_\P(\sam)=O_\P(\sqrt n).
	\end{align*}
	Since the normalized true and estimated loading rows are bounded, the preceding Lipschitz bound and Assumption \ref{SA Assumption: QMLE, uniform}(a) imply that the index-error path
	\[
	\widehat\kappa_{j,\tau}-\psi_{\tty}^{-1}(\zeta_{j,t,\tau})
	=(\widehat\blambda_{j,\subi}-\blambda_{j,\subi})'\bfrb_{t,\tau,\subi}
	-r_{t,\tau,\subi}(\balpha_j)
	=\widehat\blambda_{j,\subi}'\bfrb_{t,\tau,\subi}-\mu_{t,\tau}(\balpha_j)
	\]
	is Lipschitz in $\tau$ uniformly over $i$ and $j$. By \eqref{SA eq: projected loading error} and the approximation bound, it is $\bar O_\P(\sam^{-1}+\rate{\eta}+\barrate{\mu})$. After normalization by this rate, its Lipschitz constant is $O_\P(\sam)=O_\P(\sqrt n)$, so its covering numbers are polynomial in $n$ and the inverse covering radius. Smooth transformations of these paths have the same property.

	\smallskip
	\noindent\textit{Step 3.2: Score and Hessian.}
	Since the threshold-error class is the difference of two bounded monotone classes, conditional on $\mathcal H$ we have
	\[
	\sup_{\tau\in\mathcal Y}\max_i
	\bigg\|\frac1K\sum_{j\in\mathcal N_{i,t}}
	\frac{\widehat{\bpi}_{j,\subi}\tilde\epsilon_{j,t,\tau}}
	{\bar V_{\tty}(\psi_{\tty}^{-1}(\zeta_{j,t,\tau}))}\bigg\|
	=O_\P\big(\sqrt{\log n/K}\big).
	\]
	Since the conditional means are bounded by $\barbiaserr$, the approximation error is bounded by $\barrate{\mu}$, and \eqref{SA eq: projected loading error} holds uniformly over $\tau$, the decomposition in the proof of Lemma \ref{lem: An expansion} gives
	\[
	\sup_{\tau\in\mathcal Y}\|\bA_{n,\tau}\|
	=\bar O_\P(\sam^{-1}+\rate{\eta}+\barrate{\mu}+\barbiaserr).
	\]
	Next, consider the ``Hessian'' class $\mathcal T$ defined in the proof of Lemma \ref{lem: bound on coefficients}, with $\widehat\kappa_j$ and $\epsilon_{j,t}^-$ indexed by $\tau$ and $\tau\in\mathcal Y$ added to the index set. The monotone-class covering bound and Step 3.1 imply that this enlarged class has polynomial covering numbers; the corresponding comparison-Hessian path has the same property. Since the centered treatment--outcome pairs are conditionally independent, the conditional maximal inequality makes the empirical Hessian fluctuation uniformly negligible. The local Gram and curvature bounds in the proof of Lemma \ref{lem: bound on coefficients} keep the comparison Hessian uniformly nonsingular. Hence,
	\[
	\inf_{\tau\in\mathcal Y}\min_i\inf_{\bb^\circ\in\mathcal B}
	s_{\min}\{-\bQ_{n,\tau}(\bb^\circ)\}\gtrsim_\P1.
	\]
	Since the score has the rate above and the local criterion is concave, the argument in the proof of Lemma \ref{lem: bound on coefficients} now implies
	\[
	\sup_{\tau\in\mathcal Y}\|\widehat\bb_\tau^\circ\|
	=\bar O_\P(\sam^{-1}+\rate{\eta}+\barrate{\mu}+\barbiaserr).
	\]
	Let $\bV_{\subi,\tau}$ be the comparison Hessian conditional on $\mathcal F$, with the treatment indicator inside the expectation and the estimated regressors retained. Since the preceding coefficient bound localizes the intermediate index and Assumption \ref{SA Assumption: QMLE, uniform}(c) controls the change in the Hessian, the Hessian calculation in the proof of Theorem \ref{SA thm: QMLE} gives
	\begin{align*}
	\sup_{\tau\in\mathcal Y}\|\bV_{n,\tau}-\bV_{\subi,\tau}\|
	&=\bar O_\P(\sam^{-1}+\rate{\eta}+\barrate{\mu}+\barbiaserr),\\
	\sup_{\tau\in\mathcal Y}\{\|\bV_{n,\tau}^{-1}\|+\|\bV_{\subi,\tau}^{-1}\|\}
	&=\bar O_\P(1).
	\end{align*}
	Therefore, by the matrix-inverse identity,
	\[
	\sup_{\tau\in\mathcal Y}\|\bV_{n,\tau}^{-1}-\bV_{\subi,\tau}^{-1}\|
	=\bar O_\P(\sam^{-1}+\rate{\eta}+\barrate{\mu}+\barbiaserr).
	\]

	\smallskip
	\noindent\textit{Step 3.3: Expansion remainder.}
	It remains to control the terms omitted from the leading expansion in the proof of Theorem \ref{SA thm: QMLE}. First, the term linear in both $\epsilon_{j,t,\tau}$ and $\widehat\kappa_{j,\tau}-\psi_{\tty}^{-1}(\zeta_{j,t,\tau})$ is indexed by the product of the threshold-error class and the normalized index-error class. The threshold-error class has the standard covering bound for monotone functions, while Step 3.1 gives a polynomial covering bound for the normalized index-error class. Conditional centering and the maximal inequality therefore give
	\begin{align*}
	&\sup_{\tau\in\mathcal Y}\max_i\left\|\frac1K\sum_{j\in\mathcal N_{i,t}}
	\frac{\epsilon_{j,t,\tau}\bar V_{\tty}^{(1)}(\psi_{\tty}^{-1}(\zeta_{j,t,\tau}))}
	{\bar V_{\tty}^2(\psi_{\tty}^{-1}(\zeta_{j,t,\tau}))}
	\{\widehat\kappa_{j,\tau}-\psi_{\tty}^{-1}(\zeta_{j,t,\tau})\}
	\widehat{\bpi}_{j,\subi}\right\|\\
	&\qquad=\bar O_\P\big((\sam^{-1}+\rate{\eta}+\barrate{\mu})
	(\sam^{-1}+\barbiaserr)\big).
	\end{align*}
	Second, Assumption \ref{SA Assumption: QMLE, uniform}(c) and the uniform coefficient bound imply that the Taylor remainders are $\bar O_\P((\sam^{-1}+\rate{\eta}+\barrate{\mu}+\barbiaserr)^2)$. Finally, replacing $\bV_{n,\tau}^{-1}$ by $\bV_{\subi,\tau}^{-1}$ produces a product of the score and inverse-Hessian errors and therefore has the same quadratic order. Consequently, \eqref{eq: asymptotic expansion} holds uniformly over $\tau$, with
	\[
	\sup_{\tau\in\mathcal Y}|\mathrm{Rem}_{i,t,\tau}|
	=\bar O_\P\big((\sam^{-1}+\rate{\eta}+\barrate{\mu}+\barbiaserr)^2\big).
	\]
	This verifies the uniform version of \eqref{eq: asymptotic expansion} and completes the proof.
	\end{proof}

\subsection{Proof of Corollary \ref{SA coro: functional delta method}}
\begin{proof}
	By the functional delta method \citep[e.g.,][Theorem 3.9.4]{vandeVaart_book_1996} and the
	linearity of the Hadamard derivative, the weak convergence to a Gaussian process is implied.
\end{proof}

\subsection{Proof of Corollary \ref{SA coro: multiplier bootstrap}}
\begin{proof}
	The result follows by the conditional multiplier central limit theorem, as in \cite{Donald-Hsu_2014_JoE}. Note that  the proposed estimator of $\varphi_{i,t,s}(\cdot)$ is uniformly consistent  by Theorem \ref{SA thm: QMLE} and the argument given in the proof of Theorem \ref{SA thm: uniform inference}.
\end{proof}

\newpage
\newcounter{model}
\forloop{model}{1}{\value{model} < 7}{
	
	\begin{table}[h]
		\renewcommand{\arraystretch}{1.25}
		\centering
		\scriptsize
		\caption{Simulation results, Model \themodel, heteroskedastic error}
		\label{table:simul \themodel}
		\resizebox{.85\textwidth}{1.17\height}{\scriptsize \input{input/Table_Pointwise_SA_\themodel.txt}}
		\flushleft\footnotesize{\textbf{Notes}:
			RMSE = root MSE of point estimator, CR = coverage rate of $95\%$ nominal confidence intervals, AL = average interval length of $95\%$ nominal confidence intervals. $\widehat{K}_{\mathtt{CV}}$= cross-validation choice of $K$.
		}\newline
	\end{table}

}

\begin{table}[h]
	\renewcommand{\arraystretch}{1.25}
	\centering
	\scriptsize
	\caption{Simulation results, $n=500$, $p=250$, i.i.d. error}
	\label{table:simul iid}
	\resizebox{.85\textwidth}{1.17\height}{\scriptsize \begin{tabular}{lccccccccc}
\hline\hline
\multicolumn{1}{l}{\bfseries }&\multicolumn{5}{c}{\bfseries Local PCA, $K=$}&\multicolumn{1}{c}{\bfseries }&\multicolumn{1}{c}{\bfseries DR}&\multicolumn{1}{c}{\bfseries }&\multicolumn{1}{c}{\bfseries DR-DL}\tabularnewline
\cline{2-6} \cline{8-8} \cline{10-10}
\multicolumn{1}{l}{}&\multicolumn{1}{c}{32}&\multicolumn{1}{c}{63}&\multicolumn{1}{c}{95}&\multicolumn{1}{c}{126}&\multicolumn{1}{c}{$\widehat{K}_{\mathtt{CV}}$}&\multicolumn{1}{c}{}&\multicolumn{1}{c}{}&\multicolumn{1}{c}{}&\multicolumn{1}{c}{}\tabularnewline
\hline
{\bfseries Model 1}&&&&&&&&&\tabularnewline
~~RMSE&$0.113$&$0.113$&$0.114$&$0.116$&$0.114$&&$0.197$&&$0.203$\tabularnewline
~~CR&$0.916$&$0.920$&$0.919$&$0.912$&$0.919$&&$0.823$&&$0.808$\tabularnewline
~~AL&$0.396$&$0.394$&$0.394$&$0.395$&$0.395$&&$0.520$&&$0.522$\tabularnewline
\hline
{\bfseries Model 2}&&&&&&&&&\tabularnewline
~~RMSE&$0.100$&$0.097$&$0.097$&$0.097$&$0.098$&&$0.156$&&$0.150$\tabularnewline
~~CR&$0.943$&$0.947$&$0.947$&$0.946$&$0.943$&&$0.902$&&$0.899$\tabularnewline
~~AL&$0.376$&$0.373$&$0.373$&$0.374$&$0.373$&&$0.500$&&$0.480$\tabularnewline
\hline
{\bfseries Model 3}&&&&&&&&&\tabularnewline
~~RMSE&$0.101$&$0.099$&$0.100$&$0.102$&$0.101$&&$0.186$&&$0.202$\tabularnewline
~~CR&$0.935$&$0.942$&$0.938$&$0.941$&$0.937$&&$0.844$&&$0.808$\tabularnewline
~~AL&$0.377$&$0.376$&$0.379$&$0.381$&$0.378$&&$0.515$&&$0.521$\tabularnewline
\hline
{\bfseries Model 4}&&&&&&&&&\tabularnewline
~~RMSE&$0.131$&$0.119$&$0.117$&$0.120$&$0.123$&&$0.442$&&$0.220$\tabularnewline
~~CR&$0.932$&$0.951$&$0.965$&$0.970$&$0.947$&&$0.842$&&$0.934$\tabularnewline
~~AL&$0.482$&$0.466$&$0.485$&$0.519$&$0.475$&&$1.217$&&$0.801$\tabularnewline
\hline
{\bfseries Model 5}&&&&&&&&&\tabularnewline
~~RMSE&$0.203$&$0.218$&$0.238$&$0.262$&$0.227$&&$0.515$&&$0.348$\tabularnewline
~~CR&$0.902$&$0.885$&$0.853$&$0.820$&$0.866$&&$0.803$&&$0.703$\tabularnewline
~~AL&$0.672$&$0.679$&$0.698$&$0.723$&$0.688$&&$1.334$&&$0.773$\tabularnewline
\hline
{\bfseries Model 6}&&&&&&&&&\tabularnewline
~~RMSE&$0.266$&$0.274$&$0.289$&$0.304$&$0.285$&&$0.515$&&$0.315$\tabularnewline
~~CR&$0.873$&$0.867$&$0.865$&$0.858$&$0.869$&&$0.810$&&$0.817$\tabularnewline
~~AL&$0.822$&$0.824$&$0.852$&$0.885$&$0.847$&&$1.354$&&$0.855$\tabularnewline
\hline
\end{tabular}
}
	\flushleft\footnotesize{\textbf{Notes}:
		RMSE = root MSE of point estimator, CR = coverage rate of $95\%$ nominal confidence intervals, AL = average interval length of $95\%$ nominal confidence intervals. $\widehat{K}_{\mathtt{CV}}$= cross-validation choice of $K$.
	}\newline
\end{table}

\begin{table}[h]
	\renewcommand{\arraystretch}{1.25}
	\centering
	\scriptsize
	\caption{Simulation results, $n=500$, $p=250$, AR(1) error}
	\label{table:simul ar1}
	\resizebox{.85\textwidth}{1.17\height}{\scriptsize \begin{tabular}{lccccccccc}
\hline\hline
\multicolumn{1}{l}{\bfseries }&\multicolumn{5}{c}{\bfseries Local PCA, $K=$}&\multicolumn{1}{c}{\bfseries }&\multicolumn{1}{c}{\bfseries DR}&\multicolumn{1}{c}{\bfseries }&\multicolumn{1}{c}{\bfseries DR-DL}\tabularnewline
\cline{2-6} \cline{8-8} \cline{10-10}
\multicolumn{1}{l}{}&\multicolumn{1}{c}{32}&\multicolumn{1}{c}{63}&\multicolumn{1}{c}{95}&\multicolumn{1}{c}{126}&\multicolumn{1}{c}{$\widehat{K}_{\mathtt{CV}}$}&\multicolumn{1}{c}{}&\multicolumn{1}{c}{}&\multicolumn{1}{c}{}&\multicolumn{1}{c}{}\tabularnewline
\hline
{\bfseries Model 1}&&&&&&&&&\tabularnewline
~~RMSE&$0.117$&$0.117$&$0.120$&$0.124$&$0.118$&&$0.191$&&$0.201$\tabularnewline
~~CR&$0.911$&$0.915$&$0.908$&$0.895$&$0.911$&&$0.826$&&$0.803$\tabularnewline
~~AL&$0.400$&$0.397$&$0.398$&$0.400$&$0.398$&&$0.506$&&$0.519$\tabularnewline
\hline
{\bfseries Model 2}&&&&&&&&&\tabularnewline
~~RMSE&$0.100$&$0.098$&$0.098$&$0.099$&$0.100$&&$0.159$&&$0.140$\tabularnewline
~~CR&$0.941$&$0.942$&$0.946$&$0.947$&$0.940$&&$0.902$&&$0.925$\tabularnewline
~~AL&$0.376$&$0.373$&$0.373$&$0.375$&$0.374$&&$0.492$&&$0.471$\tabularnewline
\hline
{\bfseries Model 3}&&&&&&&&&\tabularnewline
~~RMSE&$0.102$&$0.101$&$0.101$&$0.104$&$0.101$&&$0.185$&&$0.196$\tabularnewline
~~CR&$0.937$&$0.937$&$0.937$&$0.938$&$0.941$&&$0.854$&&$0.816$\tabularnewline
~~AL&$0.377$&$0.377$&$0.379$&$0.382$&$0.378$&&$0.506$&&$0.515$\tabularnewline
\hline
{\bfseries Model 4}&&&&&&&&&\tabularnewline
~~RMSE&$0.133$&$0.119$&$0.117$&$0.120$&$0.124$&&$0.440$&&$0.222$\tabularnewline
~~CR&$0.934$&$0.947$&$0.961$&$0.969$&$0.945$&&$0.840$&&$0.936$\tabularnewline
~~AL&$0.483$&$0.468$&$0.487$&$0.520$&$0.477$&&$1.215$&&$0.803$\tabularnewline
\hline
{\bfseries Model 5}&&&&&&&&&\tabularnewline
~~RMSE&$0.218$&$0.232$&$0.252$&$0.274$&$0.241$&&$0.513$&&$0.344$\tabularnewline
~~CR&$0.890$&$0.873$&$0.844$&$0.815$&$0.854$&&$0.796$&&$0.705$\tabularnewline
~~AL&$0.697$&$0.701$&$0.719$&$0.744$&$0.708$&&$1.324$&&$0.769$\tabularnewline
\hline
{\bfseries Model 6}&&&&&&&&&\tabularnewline
~~RMSE&$0.282$&$0.292$&$0.306$&$0.322$&$0.301$&&$0.512$&&$0.311$\tabularnewline
~~CR&$0.871$&$0.864$&$0.861$&$0.858$&$0.865$&&$0.806$&&$0.825$\tabularnewline
~~AL&$0.869$&$0.875$&$0.904$&$0.936$&$0.894$&&$1.344$&&$0.852$\tabularnewline
\hline
\end{tabular}
}
	\flushleft\footnotesize{\textbf{Notes}:
		RMSE = root MSE of point estimator, CR = coverage rate of $95\%$ nominal confidence intervals, AL = average interval length of $95\%$ nominal confidence intervals. $\widehat{K}_{\mathtt{CV}}$= cross-validation choice of $K$.
	}\newline
\end{table}

\begin{table}[h]
	\renewcommand{\arraystretch}{1.25}
	\centering
	\scriptsize
	\caption{Simulation Results, $n=500$, $p=250$, heteroskedastic error}
	\label{table:simul loc const}
	\resizebox{.85\textwidth}{!}{\scriptsize \begin{tabular}{lccccccccc}
\hline\hline
\multicolumn{1}{l}{\bfseries }&\multicolumn{5}{c}{\bfseries Local constant, $K=$}&\multicolumn{1}{c}{\bfseries }&\multicolumn{1}{c}{\bfseries DR}&\multicolumn{1}{c}{\bfseries }&\multicolumn{1}{c}{\bfseries DR-DL}\tabularnewline
\cline{2-6} \cline{8-8} \cline{10-10}
\multicolumn{1}{l}{}&\multicolumn{1}{c}{32}&\multicolumn{1}{c}{63}&\multicolumn{1}{c}{95}&\multicolumn{1}{c}{126}&\multicolumn{1}{c}{$\widehat{K}_{\mathtt{CV}}$}&\multicolumn{1}{c}{}&\multicolumn{1}{c}{}&\multicolumn{1}{c}{}&\multicolumn{1}{c}{}\tabularnewline
\hline
{\bfseries Model 1}&&&&&&&&&\tabularnewline
~~RMSE&$0.136$&$0.148$&$0.160$&$0.170$&$0.157$&&$0.198$&&$0.203$\tabularnewline
~~CR&$0.846$&$0.819$&$0.784$&$0.756$&$0.790$&&$0.822$&&$0.808$\tabularnewline
~~AL&$0.397$&$0.402$&$0.407$&$0.412$&$0.406$&&$0.520$&&$0.522$\tabularnewline
\hline
{\bfseries Model 2}&&&&&&&&&\tabularnewline
~~RMSE&$0.097$&$0.097$&$0.101$&$0.114$&$0.099$&&$0.156$&&$0.150$\tabularnewline
~~CR&$0.942$&$0.945$&$0.937$&$0.914$&$0.936$&&$0.902$&&$0.901$\tabularnewline
~~AL&$0.372$&$0.371$&$0.377$&$0.389$&$0.373$&&$0.500$&&$0.480$\tabularnewline
\hline
{\bfseries Model 3}&&&&&&&&&\tabularnewline
~~RMSE&$0.099$&$0.101$&$0.112$&$0.130$&$0.109$&&$0.186$&&$0.202$\tabularnewline
~~CR&$0.936$&$0.936$&$0.915$&$0.879$&$0.916$&&$0.844$&&$0.808$\tabularnewline
~~AL&$0.374$&$0.378$&$0.387$&$0.399$&$0.383$&&$0.515$&&$0.521$\tabularnewline
\hline
{\bfseries Model 4}&&&&&&&&&\tabularnewline
~~RMSE&$0.135$&$0.161$&$0.194$&$0.223$&$0.160$&&$0.442$&&$0.218$\tabularnewline
~~CR&$0.947$&$0.936$&$0.910$&$0.897$&$0.932$&&$0.842$&&$0.935$\tabularnewline
~~AL&$0.532$&$0.601$&$0.667$&$0.723$&$0.590$&&$1.215$&&$0.799$\tabularnewline
\hline
{\bfseries Model 5}&&&&&&&&&\tabularnewline
~~RMSE&$0.214$&$0.238$&$0.267$&$0.293$&$0.249$&&$0.515$&&$0.356$\tabularnewline
~~CR&$0.879$&$0.847$&$0.813$&$0.792$&$0.840$&&$0.800$&&$0.681$\tabularnewline
~~AL&$0.663$&$0.686$&$0.719$&$0.758$&$0.699$&&$1.333$&&$0.776$\tabularnewline
\hline
{\bfseries Model 6}&&&&&&&&&\tabularnewline
~~RMSE&$0.273$&$0.282$&$0.295$&$0.311$&$0.293$&&$0.514$&&$0.323$\tabularnewline
~~CR&$0.858$&$0.858$&$0.858$&$0.849$&$0.856$&&$0.811$&&$0.809$\tabularnewline
~~AL&$0.806$&$0.824$&$0.857$&$0.893$&$0.848$&&$1.353$&&$0.861$\tabularnewline
\hline
\end{tabular}
}
	\flushleft\footnotesize{\textbf{Notes}:
		RMSE = root MSE of point estimator, CR = coverage rate of $95\%$ nominal confidence intervals, AL = average interval length of $95\%$ nominal confidence intervals. $\widehat{K}_{\mathtt{CV}}$= cross-validation choice of $K$.
	}\newline
\end{table}

\end{document}